\documentclass[11pt,a4paper]{article}
\usepackage{placeins,setup}
\usepackage{jheppub}
\usepackage{graphicx,xcolor,float,afterpage,booktabs}
\usepackage{amssymb,amsmath,multirow,booktabs,multirow,tabularx}

\newcolumntype{C}[1]{>{\centering\arraybackslash}p{#1}}

\newcommand{\be}{\begin{equation}}
\newcommand{\ee}{\end{equation}}
\newcommand{\bea}{\begin{eqnarray}}
\newcommand{\eea}{\end{eqnarray}}
\newcommand{\bi}{\begin{itemize}}
\newcommand{\ei}{\end{itemize}}
\newcommand{\ben}{\begin{enumerate}}
\newcommand{\een}{\end{enumerate}}

\newcommand{\lp}{\left(}
\newcommand{\rp}{\right)}

\def\frac#1#2{{{#1}\over {#2}}}
\def\gsim{\gtrsim}
\def\lsim{\lesssim}

\newcommand{\draft}[1]{}

\def\beq{\begin{equation}}  
\def\eeq{\end{equation}}

\def\({\left(}
\def\){\right)}
\def\[{\left[}
\def\]{\right]}
\let\originalleft\left
\let\originalright\right
\renewcommand{\left}{\mathopen{}\mathclose\bgroup\originalleft}
\renewcommand{\right}{\aftergroup\egroup\originalright}

\numberwithin{equation}{section}
\numberwithin{figure}{section}
\numberwithin{table}{section}

\let\oldsubsection\subsection
\renewcommand\subsection[2][\subsectiontoc]{%
  \def\subsectiontoc{#2}%
  \oldsubsection[#1]{\boldmath #2}%
}

\let\oldsubsubsection\subsubsection
\renewcommand\subsubsection[2][\subsubsectiontoc]{%
  \def\subsubsectiontoc{#2}%
  \oldsubsubsection[#1]{\boldmath #2}%
}

\subheader{\small Nikhef/2018-017}

\title{\boldmath Neutrino Telescopes as QCD Microscopes}

\author[a]{Valerio~Bertone,}
\author[b]{Rhorry Gauld,}
\author[a]{and Juan Rojo}

\affiliation[a]{Department of Physics and Astronomy, VU University, NL-1081 HV Amsterdam,\\
and Nikhef Theory Group, Science Park 105, 1098 XG Amsterdam, The Netherlands}
\affiliation[b]{Institute for Theoretical Physics, ETH, CH-8093 Z\"urich, Switzerland}

\emailAdd{v.bertone@vu.nl}
\emailAdd{rgauld@phys.ethz.ch}
\emailAdd{j.rojo@vu.nl}

\abstract{
We present state-of-the-art predictions for the ultra-high energy (UHE)
neutrino-nucleus cross-sections in charged- and neutral-current scattering.
The calculation is performed in the framework of collinear factorisation at NNLO, 
extended to include the resummation of small-$x$ BFKL effects.
Further improvements are made by accounting for the free-nucleon PDF
constraints provided by $D$-meson data from LHCb and assessing the
impact of nuclear corrections and heavy-quark mass effects.
The calculations presented here should play an important role in the
interpretation of future data from neutrino telescopes such as IceCube 
and KM3NET, and highlight the opportunities that astroparticle
experiments offer to study the strong interactions.
}

\begin{document}
\maketitle
\flushbottom


\section{Introduction}
\label{sec:introduction}

Ultra-high energy (UHE) neutrinos represent a unique messenger
for a wide variety of astrophysical and cosmological phenomena~\cite{Gaisser:1994yf,Halzen:2010yj,Halzen:2002pg}.
Being charge-neutral, neutrinos are not deflected by galactic magnetic fields,
while being weakly interacting implies that they are not attenuated {\it e.g.} by dust.
Therefore, neutrinos point directly to their original source of production,
offering crucial information which would be not available
from other messengers such as electromagnetic waves or cosmic rays.
This potential is illustrated by the recent report of the detection of UHE neutrinos
from the same direction as a known Blazar~\cite{IceCube:2018cha}.

UHE neutrinos also represent a very promising probe
of physics beyond the Standard Model~\cite{Anchordoqui:2013dnh},
for instance testing scenarios where PeV neutrinos arise
from heavy dark matter decays~\cite{Murase:2015gea} or those where neutrinos
have non-standard interactions (NSI)~\cite{Salvado:2016uqu}.
Moreover, neutrino-nucleon scattering
at very high energies also provides unique
opportunities to test Quantum Chromodynamics (QCD) in an
extreme kinematic regime far from those accessible
at colliders~\cite{Gandhi:1998ri}.

At very high energies, the astrophysical neutrino flux $\Phi_{\nu}$
decreases steeply with the neutrino energy $E_{\nu}$ with a power-law of approximately
$E_{\nu}^{-2}$~\cite{Aartsen:2014gkd}.
The fact that the incoming UHE neutrino flux is so small, combined
with the very feeble interactions of neutrinos with matter, makes the detection 
of UHE neutrinos extremely challenging.
To bypass these limitations, neutrino telescopes such
as IceCube~\cite{Achterberg:2006md},
KM3NET~\cite{Adrian-Martinez:2016fdl}, and Baikal~\cite{Belolaptikov:1997ry}
instrument large volumes of ice or
water, which then act as detectors with
an effective volume of up to 1~km$^3$.
This way, by accumulating data taken during several years, neutrino telescopes
are expected to be able to detect neutrinos with energies
as high as $E_{\nu}\simeq 10^{12}$~GeV.
In addition, UHE neutrinos can also be studied by means of an
array of radio antennas~\cite{Martineau-Huynh:2017bpw} or
with balloon experiments such as ANITA~\cite{Gorham:2010kv}.

Irrespective of the specific detection method adopted, the physical interpretation 
of the UHE neutrino event rates relies on an accurate theoretical prediction for the 
relevant signal process.
In the case of the neutrino telescopes described above, the signal arises from the interaction~\cite{Gandhi:1998ri} 
of a highly energetic neutrino with a nucleon from the target material, typically an H$_2$O molecule. 
The prediction for this scattering process is sensitive to both perturbative and non-perturbative
QCD effects including the quark and gluon content of the target nucleon~\cite{Rojo:2015acz,Gao:2017yyd},
the treatment of heavy-quark mass effects~\cite{Forte:2010ta}, and the stability of the perturbative 
expansion at small values of the partonic momentum fraction $x$~\cite{Altarelli:2008aj}.
The sensitivity of the prediction to QCD dynamics represents both a challenge and an 
opportunity, given that the strong interactions must be probed in an unexplored kinematic regime.
It is the main goal of this work to present a state-of-the-art calculation, taking into account recent developments 
in our understanding of perturbative and non-perturbative QCD, for the UHE neutrino-nucleon cross-section 
relevant for signal detection at neutrino telescopes.
While a number of calculations have been provided by different groups in the past, both in the framework of 
collinear DGLAP factorisation~\cite{Gluck:1998js,CooperSarkar:2007cv,Gandhi:1995tf,Connolly:2011vc,CooperSarkar:2011pa,Gandhi:1998ri,Dicus:2001kb} and beyond it~\cite{Fiore:2005wf,Albacete:2015zra,Goncalves:2015fua,Arguelles:2015wba,Block:2013nia,JalilianMarian:2003wf}, there are now several significant motivations to revisit
this calculation. The improvements made in this work, and their motivation, are detailed below.

Firstly, we account for the effects of small-$x$ (BFKL)
resummation up to the next-to-leading logarithm (NLL$x$) accuracy in
the calculation of deep-inelastic scattering (DIS) structure functions
and in the collinear evolution of parton distribution functions (PDFs). Specifically, the fixed-order
calculations at next-to-next-to-leading order (NNLO) are consistently
matched to the corresponding resummed
results~\cite{Bonvini:2016wki,Bonvini:2017ogt}.
Small-$x$ corrections to the DIS structure functions are provided in the
framework of the FONLL general-mass scheme~\cite{Forte:2010ta,Ball:2015tna}.
Recently, evidence for the onset of small-$x$ resummation has been
found~\cite{Ball:2017otu,Abdolmaleki:2018jln} in the inclusive and
charm HERA data following the approach proposed in Refs.~\cite{Caola:2009iy,Caola:2010cy}.
It is thus of the utmost importance to include these effects also in
the computation of the UHE cross-sections, which probe much smaller
values of $x$ than at HERA, down to $x\simeq 10^{-8}$.

Secondly, we include the constraints on PDFs arising from the 
LHCb $D$-meson production measurements in $pp$ collisions at 5, 7, and 
13~TeV~\cite{Aaij:2016jht,Aaij:2015bpa,Aaij:2013mga} following the approach 
of Ref.~\cite{Gauld:2016kpd}. As demonstrated in this work, and in the 
previous studies~\cite{Zenaiev:2015rfa,Gauld:2015yia,Cacciari:2015fta}, the LHCb data
provides important constraints at small-$x$ which are in turn relevant for the UHE cross-section
predictions.
Here we revisit the study of Ref.~\cite{Gauld:2016kpd} to account for the impact that this 
data has on the PDFs extracted from a global analysis where small-$x$ resummation 
effects have been included~\cite{Ball:2017otu}.
As will be shown, including the LHCb $D$-meson measurements leads to 
a significant reduction of the uncertainties on the UHE cross-sections
associated to PDFs.
It is worth mentioning that the same PDF constraints are also an important ingredient
in the prediction of the so-called prompt neutrino flux~\cite{Enberg:2008te}, induced by the decays 
of $D$ mesons produced in cosmic ray collisions in the atmosphere. The neutrinos produced
in this process are a dominant background source in the search for UHE neutrinos of astrophysical origin.
See Refs.~\cite{Gauld:2015yia,Gauld:2015kvh,Garzelli:2016xmx} for more details as well as a number of related studies~\cite{Gelmini:1999ve,Martin:2003us,Bhattacharya:2015jpa,Bhattacharya:2016jce,Benzke:2017yjn,Halzen:2016thi,Garzelli:2015psa}.

Finally, as the signal process of interest is neutrino scattering on an H$_2$O target (as opposed to a free nucleon), we 
account for nuclear modifications in the UHE cross-section predictions. 
There have been a number of studies quantifying the impact that nuclear modifications have on the distribution of 
quarks and gluons within bound nuclei in Refs.~\cite{deFlorian:2011fp,Eskola:2016oht,Kovarik:2015cma,Khanpour:2016pph}, 
which can be used to study the impact that nuclear corrections have on neutrino-nucleon scattering. These modifications 
(and the associated uncertainty) are absent from many of benchmark UHE cross-section calculations.

Our results represent a significant improvement over previous calculations of
the UHE cross-sections both in terms of the perturbative (NNLO accuracy, small-$x$
resummation) and non-perturbative (PDF constraints from $D$-meson data, nuclear corrections)
content.
This allows us to provide a more reliable comparison to the very recent direct measurements 
of the neutrino-nucleus interaction cross-sections at high energies~\cite{Aartsen:2017kpd,Bustamante:2017xuy} 
extracted from the IceCube data, and to extrapolate the calculation to as of yet unexplored energies.
Our calculations are made publicly available at the level of the total cross-section 
for a large range of (anti)neutrino values up to $10^{12}$~GeV. In addition, we also
provide predictions for the DIS structure functions in the form numerical grids. 
These grids (with the relevant interpolation routines) may be used to construct a 
double-differential cross-section and can also be integrated numerically to obtain the total cross-section.

The outline of this paper is as follows.
In Sect.~\ref{sec:theoryformalism} we present the theoretical
formalism adopted for the calculation of the cross-sections.
In Sect.~\ref{sec:lhcb} we revisit the impact of the LHCb $D$-meson data
on the small-$x$ PDFs once the effects of small-$x$ resummation are accounted for.
The main results of this work are presented in Sect.~\ref{sec:results}, where
we also discuss the various sources of theoretical uncertainties associated
to our calculation.
We conclude in Sect.~\ref{sec:summary} and outline
possible future developments.
Appendix~\ref{sec:UHExsec} contains a tabulated version
of our calculation for the UHE cross-sections
with the corresponding uncertainties, and also includes a description
of the structure function grids and how they may be used to construct UHE
cross-section predictions at both differential and integrated levels.

\section{Theoretical formalism}
\label{sec:theoryformalism}

In this section we present an overview of the theoretical settings used 
throughout this work to compute the UHE neutrino-nucleus cross-sections.
We first review the basic ingredients required to evaluate
predictions for neutrino-induced DIS cross-sections, focussing on the
kinematic region relevant to UHE neutrino-nucleus scattering.
We then provide details on the theoretical accuracy and implementation
of our calculation.
Finally, we discuss the role played by heavy-quark mass effects and
the treatment of nuclear corrections.

\subsection{Neutrino-nucleon deep-inelastic scattering}
\label{sec:DISkin}

\begin{figure}[t]
\centering
\includegraphics[width=.50\linewidth]{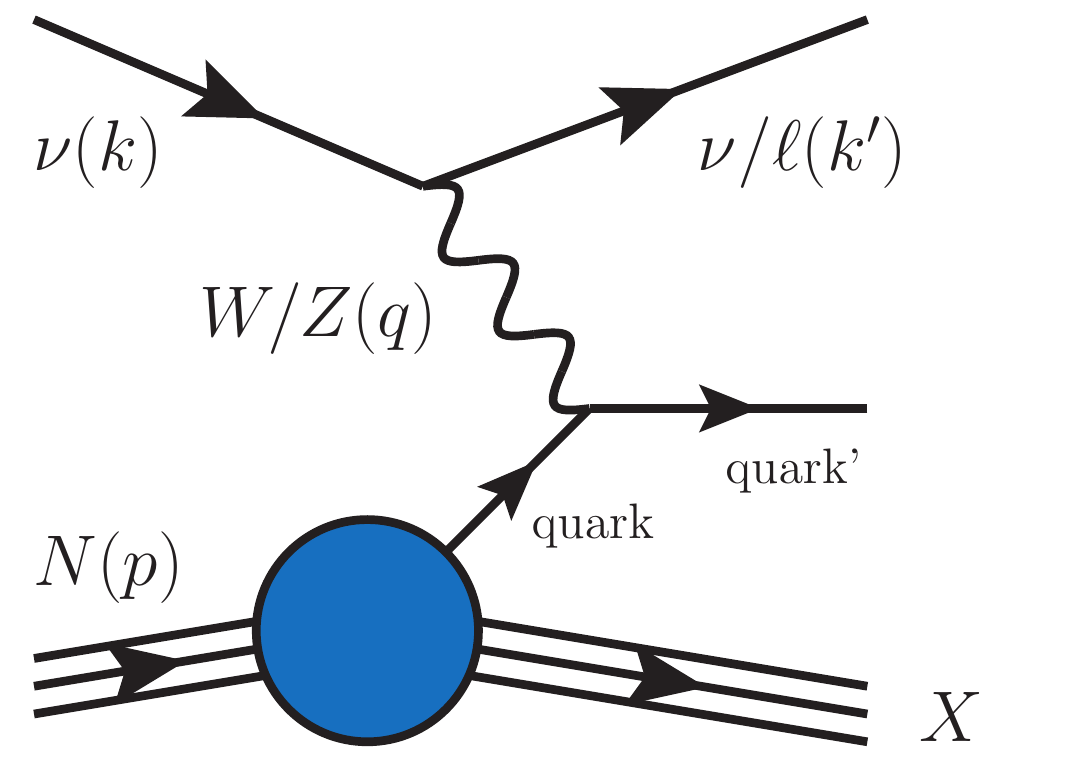}
\caption{\small 
Diagram for deep-inelastic neutrino-nucleon scattering. The process
proceeds through the exchange of an electroweak gauge boson either
neutral ($Z$) or charged ($W^\pm$).}
\label{fig:UHEdiagram}
\end{figure}

As illustrated in Fig.~\ref{fig:UHEdiagram}, a high-energy neutrino
interacts with a nucleon from the target through the exchange of an
electroweak gauge-boson, either neutral ($Z$) or charged ($W^\pm$).
The energy of the incoming neutrino can then be reconstructed by the
energy deposited in the detector by either the outgoing lepton, in
charged-current (CC) scattering, or the recoiling hadronic system in
the case of neutral-current (NC) scattering.

The formalism for describing both NC and CC processes is equivalent, 
and in the following we focus on the CC case, $\nu(k) + N (p) \to \ell^{\pm}(k^{\prime}) + X(W)$. 
Details on the implementation of the NC case are discussed in Appendix~\ref{sec:UHExsec}.
The differential cross-section for this process can be written in
terms of DIS structure functions as follows:
\begin{align}
  \label{eq:d2sig}
  \nonumber
\frac{\rd^2 \sigma^{\rm CC}_{\nu(\bar{\nu})N} (x,Q^2,E_{\nu})}{\rd x\,\rd Q^2} = &\frac{G_{F}^2 M_W^{4} }{4\pi x (Q^2+M_W^2)^2}  \\
	&\times  \Bigg(Y_+ F_{2,{\rm CC}}^{\nu(\bar{\nu})N}(x,Q^2)   \mp Y_- x F_{3,{\rm CC}}^{\nu(\bar{\nu})N}(x,Q^2) - y^2 F_{L,{\rm CC}}^{\nu(\bar{\nu})N}(x,Q^2)   \Bigg) \,,
\end{align}
where the sign of the $xF_3$ term is positive (negative) for
(anti)neutrino scattering, and $N$ represents the struck
nucleon.\footnote{In the case of (anti)neutrino scattering on an
H$_2$O molecule, the total structure function can be obtained as the
combination of that for a free proton (hydrogen) and that for bound
nucleons within an oxygen nucleus. We will return to this point in
Sect.~\ref{sec:nuclearPDFs}.}
In Eq.~\eqref{eq:d2sig} we have defined $Y_\pm = 1\pm(1-y)^2$ and the
DIS kinematic variables are given by
\begin{align}  \nonumber
Q^2 	&= -q^2\,, \qquad
y 	= \frac{q \cdot p}{k \cdot p} = 1 - \frac{E^{\prime}}{E_{\nu}} \,, \qquad
x 	= \frac{Q^2}{2\,q \cdot p} = \frac{Q^2}{2\,m_N y E_{\nu} } \,, \\[2mm]
W^2 &= (p+q)^2 = m_N^2 + Q^2 \left(\frac{1}{x} -1\right)\,, \qquad
s = (k+p)^2 = m_N^2 + 2 E_{\nu} m_N\,, \label{eq:DISkinematics}
\end{align}
where some of the Lorentz-invariant quantities are also given in the
laboratory centre-of-mass frame where the target nucleon is at rest.
In these expressions, $m_N$ is the nucleon mass, $E_\nu$ and
$E^{\prime}$ are the incoming- and outgoing-lepton energies, and $k$,
$q$, and $p$ are the four-momenta of the neutrino, the gauge boson,
and the nucleon (see Fig.~\ref{fig:UHEdiagram}).

The main ingredients for the theoretical prediction of the
double-differential cross-section defined in Eq.~\eqref{eq:d2sig} are
the structure functions $F^{\nu(\bar{\nu})N}_{i = 2,3,L}(x,Q^2)$,
which describe the underlying QCD dynamics of the scattering
process. Structure functions factorise as follows:
\begin{align} \label{eq:SF}
F_i (x, Q^2) = \sum_{a = g, q} \int_x^1 \frac{\rd z}{z} C_{i,a} \left( \frac{x}{z}, Q^2 \right) f_a\left(z,Q^2\right) \,,
\end{align}
corresponding to the convolution of the universal PDFs ($f_a$) with
the process-dependent coefficient functions ($C_{i,a}$).
Typically, coefficient functions are computed in perturbation theory
as a truncated expansion in powers of the strong coupling $\alpha_s$.

Before discussing the theoretical accuracy of the structure function
predictions, it is useful to illustrate the
kinematic coverage of the UHE neutrino-nucleon cross-section for some
representative values of the incoming neutrino energy $E_\nu$.
To do so, we compute the total CC neutrino-nucleon cross-section
integrating Eq.~(\ref{eq:d2sig}) over the relevant kinematic
region. The computation is performed at NLO using the NLO fixed-order
NNPDF3.1sx central PDF set~\cite{Ball:2017otu} assuming an isoscalar target.\footnote{That is, a nucleus assumed to be composed of equal numbers of free protons and neutrons and where nuclear effects are 
neglected. The PDFs for such a nucleon can be obtained by assuming
isospin symmetry to connect protons and neutrons.}

\begin{figure}[t]
\centering
\includegraphics[width=.49\linewidth]{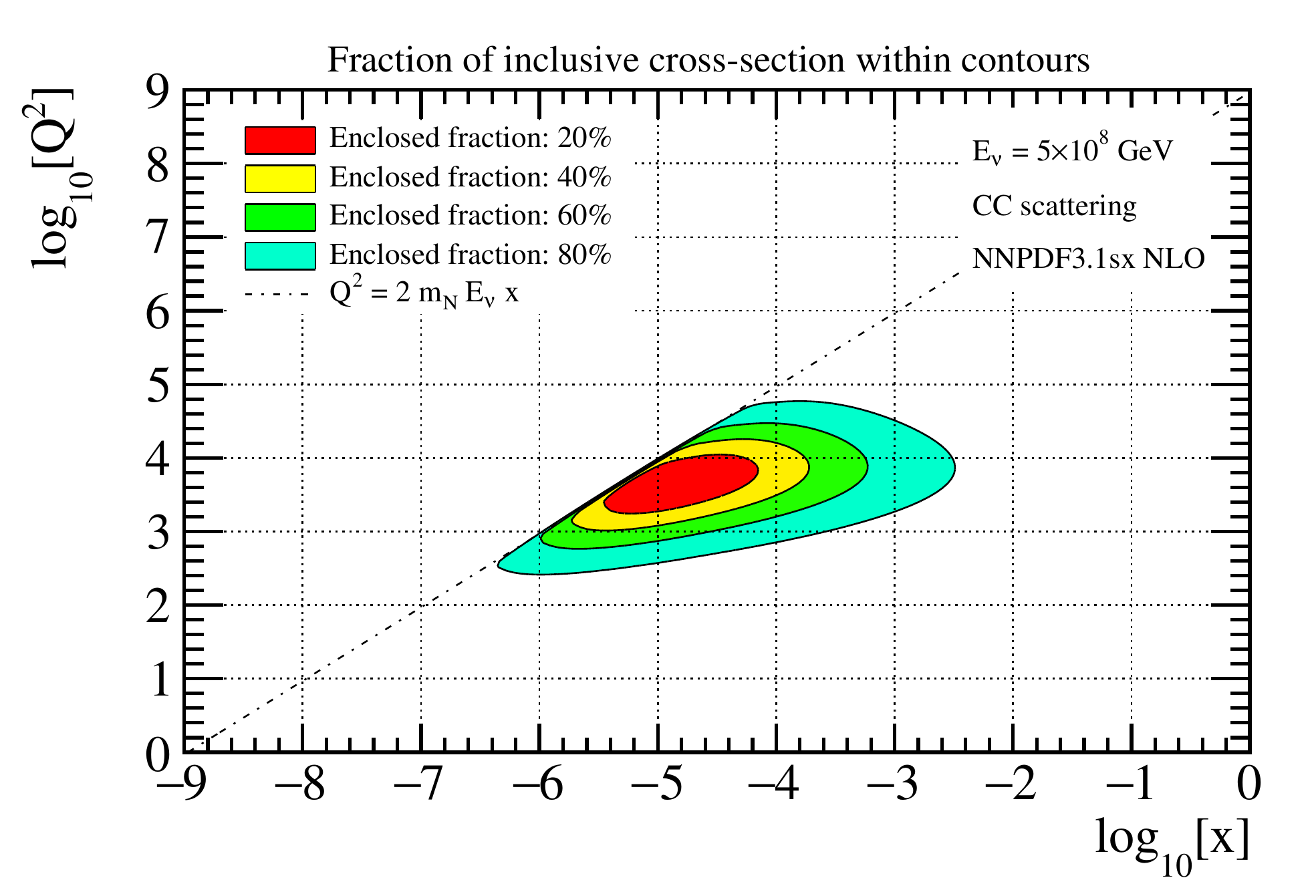}
\includegraphics[width=.49\linewidth]{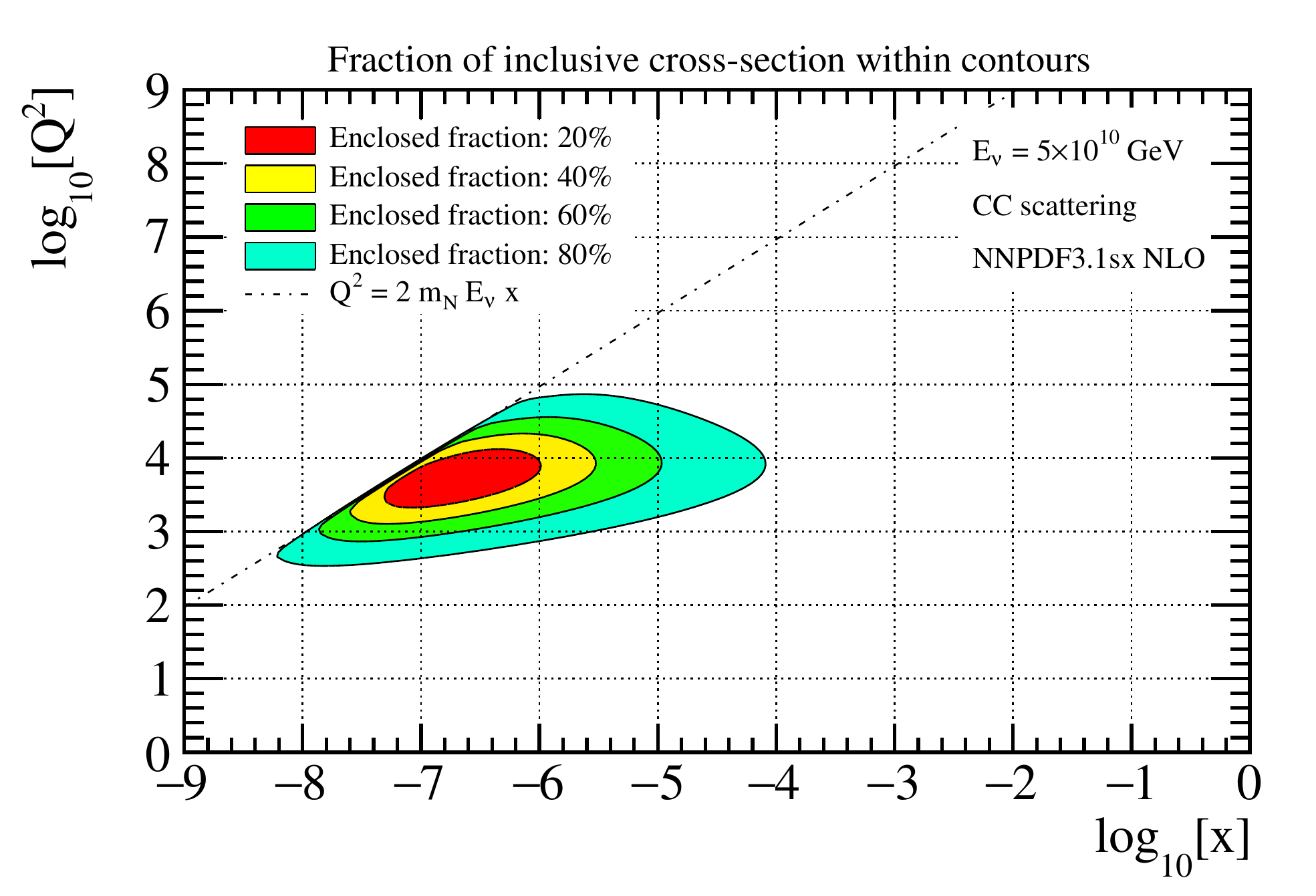}
\hfill
\caption{\small The percentage of the total UHE neutrino-isoscalar cross-section corresponding
to the contribution of different kinematic regions in the $(x,Q^2)$ plane in CC
scattering.
Predictions are computed in the FONLL scheme at NLO using the
NNPDF3.1sx NLO PDF set.
Results are shown for $E_{\nu}=5\times 10^8$~GeV (left) and
$E_{\nu}=5\times 10^{10}$~GeV (right).
The dot-dashed diagonal lines correspond to the kinematic limit
$Q^2_{\rm max}(x)=2m_N E_{\nu} x$.
}
\label{fig:contour}
\end{figure}

Fig.~\ref{fig:contour} shows the integration regions on the
$(\mbox{log}_{10}x$ \textit{vs.}  $\mbox{log}_{10}Q^2)$ plane that
contribute most (20\%, 40\%, 60\%, and 80\%) to the total
cross-section for $E_{\nu}=5\times 10^8$~GeV (left) and
$E_{\nu}=5\times10^{10}$~GeV (right).
The dot-dashed diagonal line indicates the upper bound of the
integration of $Q^2$ as a function of $x$, that is given by
$Q^2_{\rm max}(x)=2m_N E_{\nu} x$.
For $E_{\nu}=5\times 10^8~(5\times 10^{10})$~GeV, the main
contribution to the cross-section arises from the region with
$Q^2 \simeq M_W^2$ and $x\simeq 2\times 10^{-5}~(2\times 10^{-7})$,
with sensitivity to the small-$x$ region down to
$x\simeq 10^{-6}~(10^{-8})$.
This pattern is the consequence of two separate effects. For
$Q^2\gg M_W^2$ the contribution to the total cross-section is
power-suppressed by the $W$-boson propagator, see
Eq.~(\ref{eq:d2sig}). For $Q^2\ll M_W^2$, instead, the suppression of
the cross-section is driven by the decrease of PDFs as $Q^2$
decreases.

In the kinematic regions highlighted in Fig.~\ref{fig:contour},
fixed-order calculations are affected by the presence of large
logaritms of $x$ that spoil the perturbative convergence, see
Refs.~\cite{Bonvini:2016wki,Bonvini:2017ogt,Bonvini:2018iwt} and
references therein.
However, it has been shown that fixed-order calculations can be
complemented by small-$x$ resummation effects to extend their validity
down to very small values of $x$. Remarkably, this leads to a marked
improvement in the description of the precise HERA 1+2 combined
data~\cite{Ball:2017otu,Abdolmaleki:2018jln}.
It is therefore important to account for these effects when computing
predictions for the UHE neutrino-nucleus cross-section. As discussed
below, predictions for the DIS structure functions in this work are
accurate to NNLO+NLL$x$.

\subsection{Numerical implementation} \label{sec:numerics}
The DIS structure functions appearing in Eq.~\eqref{eq:SF}
are computed at fixed-order with the {\tt APFEL} program~\cite{Bertone:2013vaa}. {\tt APFEL} has
been interfaced to the {\tt HELL} code~\cite{Bonvini:2016wki,Bonvini:2017ogt,Bonvini:2018iwt} that
provides small-$x$ resummation corrections to the DIS coefficient
functions and the DGLAP splitting functions.
The computation of the structure functions is performed with the NNPDF3.1sx sets obtained in the
corresponding global PDF analysis~\cite{Ball:2017otu}.
Heavy-quark mass effects are included in the calculation of the
structure functions using the FONLL general-mass variable flavour
number scheme~\cite{Forte:2010ta,Ball:2015tna}.  In the NC
case, the FONLL-B and -C variants have been used for the NLO and NNLO
computations, respectively.
In the case of CC interactions, NNLO corrections to the massive
coefficient functions have recently been
computed~\cite{Berger:2016inr} (see also Ref.~\cite{Gao:2017kkx}).
However, the results of this calculation have not been made publicly
available yet in a format suitable for the present calculation.
Mass effects to the CC structure functions are therefore taken into
account only up to NLO accuracy~\cite{Ball:2011mu}.

The numerical values of the heavy quark masses are the following:
$m_c = 1.51$~GeV, $m_b = 4.92$~GeV, and $m_t = 172.5$~GeV.
As for the electroweak parameters appearing in Eq.~\eqref{eq:d2sig},
we take: $M_W = 80.385$~GeV, $M_Z = 91.1876$~GeV,
$G_{F} = 1.663787\cdot10^{-5}$~GeV$^{-2}$. For the CC interactions, we
use a non-diagonal CKM matrix with the following absolute values:
$|V_{ud}| = 0.97427$, $|V_{us}| = 0.22536$, $|V_{ub}| = 0.0035$,
$|V_{cd}| = 0.22522$, $|V_{cs}| = 0.97343$, $|V_{cb}| = 0.04140$,
$|V_{td}| = 0.00886$, $|V_{ts}| = 0.04050$, $|V_{tb}| = 0.99914$.
This setup is consistent with that adopted in the NNPDF3.1sx analysis.

For a number of phenomenological applications, such as the calculation of
event rates at neutrino telescopes using Monte Carlo simulations, the
double-differential distribution in Eq.~\eqref{eq:d2sig} is the most
relevant physical quantity. To this end, predictions for both CC and
NC structure functions differential in $x$ and $Q^2$ are made publicly
available in the form of interpolation grids (see Appendix~\ref{sec:UHExsec}).

However, the total neutrino-nucleon cross-section is also a highly relevant 
quantity for a number of analyses at neutrino telescopes.
To construct the theoretical predictions for this quantity, it is
necessary to integrate Eq.~\eqref{eq:d2sig} over the
allowed kinematic range in $x$ and $Q^2$.
The total cross-section is therefore computed as
\beq
\label{eq:integrate}
\sigma(E_{\nu}) = \int_{Q^2_{\rm min}}^{Q^2_{\rm max}} \rd Q^2 \bigg[
\int_{x_0(Q^2)}^{1} \rd x \frac{\rd^2 \sigma }{\rd x\,\rd
  Q^2}(x,Q^2,E_{\nu}) \bigg] \,.  \eeq
This integral is evaluated numerically using Gauss quadrature as implemented in
{\tt GSL} with relative precision set to $10^{-4}$.
Results were cross-checked by comparing to those obtained by
performing a two-dimensional Monte Carlo integration with the {\tt VEGAS} 
algorithm (as implemented in the {\tt CUBA} library~\cite{Hahn:2004fe}) 
and requiring a relative precision of $5\times10^{-3}$.

The integration limits of the integrals in Eq.~\eqref{eq:integrate}
are obtained by inspection of the DIS kinematics in
Eq.~\eqref{eq:DISkinematics}.
The inelasticity $y$ is bounded within the range of $y\in [0,1]$, and
consequently the maximum value that $Q^2$ can take for a given
neutrino projectile energy is given by $Q^2_{\rm max}=2m_N E_\nu$.
While the integration in $Q^2$ in Eq.~\eqref{eq:DISkinematics} should
extend all the way down to zero, it is in practice necessary to impose
a cut so that the structure functions are evaluated in the region of
validity of perturbation theory.
We choose $Q_{\rm min}^2 = Q_0^2$, where $Q_0$ is the scale at which
the input PDFs are parameterised. In the case of NNPDF3.1sx,
$Q_0 = 1.64$~GeV.
As far as the integration over $x$ is concerned, the upper integration
bound is naturally set to one, while the lower bound for each
particular value of $Q^2$ should be set to $x_0(Q^2)= Q^2/(2 m_N E_{\nu})$.  
However, when $E_{\nu}$ becomes very
large and for values of $Q^2$ close to $Q^2_{\rm min}$, $x_0$ can
potentially become very small. Since we cannot access PDFs down to
indefinitely small values of $x$, we need to impose a cut on the
minimum value that $x_0$ can take. To this end, we replace the lower
bound of the integration in $x$ in Eq.~\eqref{eq:integrate} with
$x_0(Q^2)= \mbox{max}\left[Q^2/(2 m_N E_{\nu}), x_{\rm min}\right]$.
For our nominal results we choose $x_{\rm min} = 10^{-9}$.

\begin{figure}[t]
\centering
\includegraphics[width=.49\linewidth]{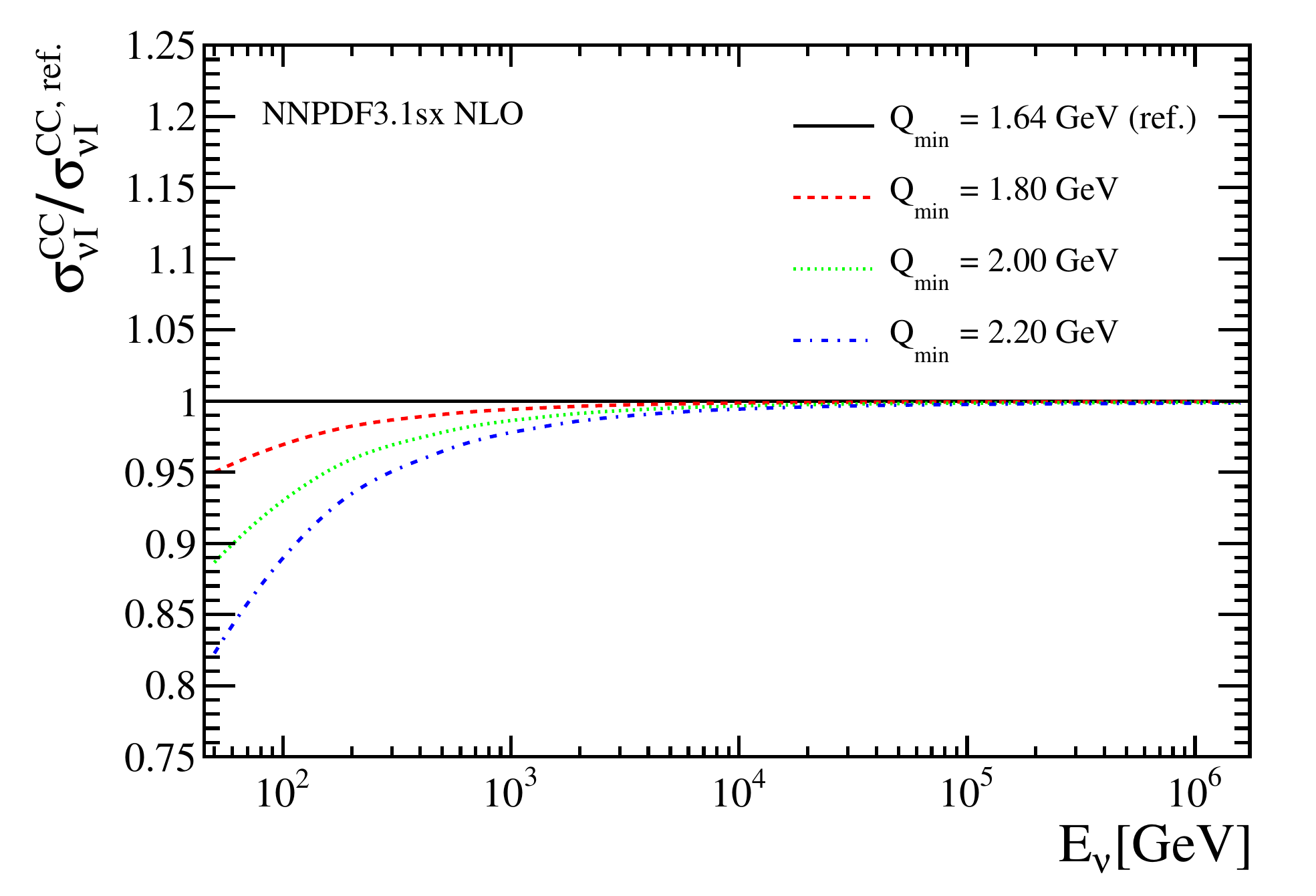} \hfill
\includegraphics[width=.49\linewidth]{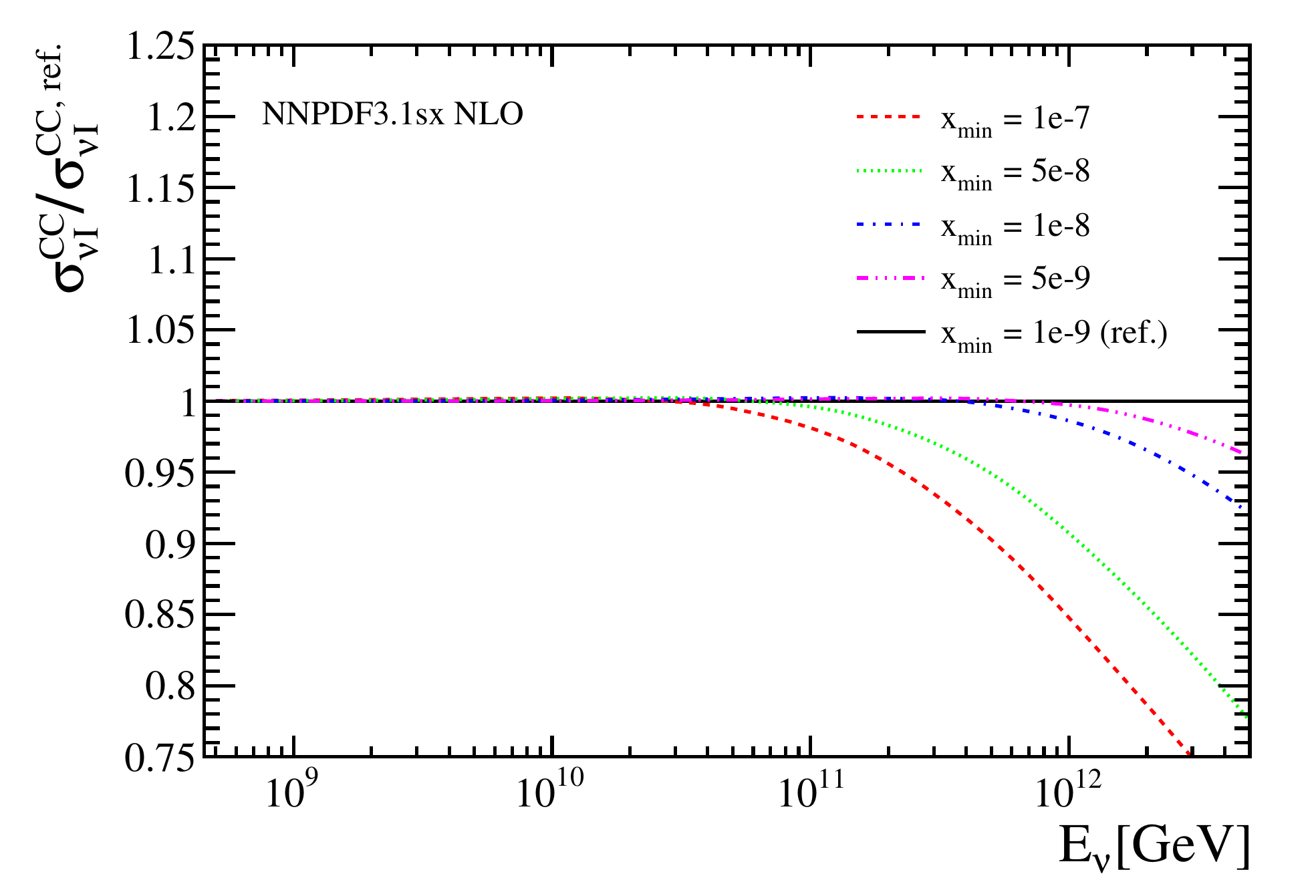}
\caption{\small The dependence of the total CC neutrino-isoscalar
  cross-section, Eq.~(\ref{eq:integrate}), with respect to variations
  of the lower integration limits $Q_{\rm min}^2$ (left) and
  $x_{\rm min}$ (right) presented as ratios to the default (ref.) values.}
\label{fig:limits}
\end{figure}

In order to investigate the dependence of the results on the
integration limits $Q_{\rm min}^2$ and $x_{\rm min}$, we have computed
the total CC neutrino cross-section on an isoscalar target as a function of incoming neutrino 
energy $E_\nu$ for different values of $Q_{\rm min}^2$ and $x_{\rm min}$.
The results are shown in Fig.~\ref{fig:limits} where the
cross-section is computed at NLO for different values of $Q^2_{\rm min}$
(left) and $x_{\rm min}$ (right), presented as a ratio to that
obtained with the nominal values of $Q_{\rm min}=1.64$~GeV and
$x_{\rm min}=10^{-9}$.
Regarding the dependence on the value of $x_{\rm min}$, we find that the
total cross-section becomes sensitive to the precise value of $x_{\rm min}$
only for extremely large energies ($E_{\nu}\gsim 10^{10}$~GeV).
For instance, at $E_{\nu}=10^{12}$~GeV the total cross-section is
reduced by around $15\%$ if the integration is restricted to
$x_{\rm min}=10^{-7}$ as compared to the nominal value
$x_{\rm min}=10^{-9}$.
With respect to the choice of $Q_{\rm min}$, we find that the total
cross-section receives non-negligible contributions from the region
$Q \in [1.64,2.2]$~GeV for neutrino energies below
$E_{\nu}\lsim 5\times 10^{3}$~GeV. For instance, raising $Q_{\rm min}$
from the nominal value of 1.64~GeV up to 2.2~GeV leads to a
suppression of the total cross-section by around $15\%$ at $E_{\nu}=100$~GeV.
The predictions provided in this work can thus be considered reliable
for neutrino energies $E_{\nu}\gsim 5\times10^{3}$~GeV.
Accurate predictions in the region $E_{\nu}\lsim 5\times10^{3}$~GeV
could be achieved by matching the perturbative QCD calculation of the
structure functions to a parameterisation of low-$Q^2$ neutrino
structure functions~\cite{Formaggio:2013kya}, following for instance
the approach developed in Ref.~\cite{DelDebbio:2004qj}.

\subsection{Heavy-quark mass effects}
\label{sec:heavyquarks}

Previous calculations of the UHE neutrino-nucleon cross-sections have
been performed in the zero-mass variable-flavour-number scheme
(ZM-VFNS), where finite mass effects in the computation of the partonic
cross-sections are neglected.
Nonetheless, it is now well known that heavy-quark mass effects are
required in order to provide an adequate description of the HERA data
in the low-$Q^2$ region.
Since the computation of the total neutrino cross-sections requires
integrating over a wide range in $Q^2$ that includes the regions
close to the charm- and bottom-quark masses, heavy-quark mass corrections can be relevant.
As mentioned above, our predictions for the DIS structure functions
are computed using the FONLL scheme that accounts for heavy-quark mass
effects.
In the following, we assess the importance of including these effects
in the UHE cross-section predictions. We also discuss the relevance of
top-quark mass effects.

\begin{figure}
\centering
\includegraphics[width=.49\linewidth]{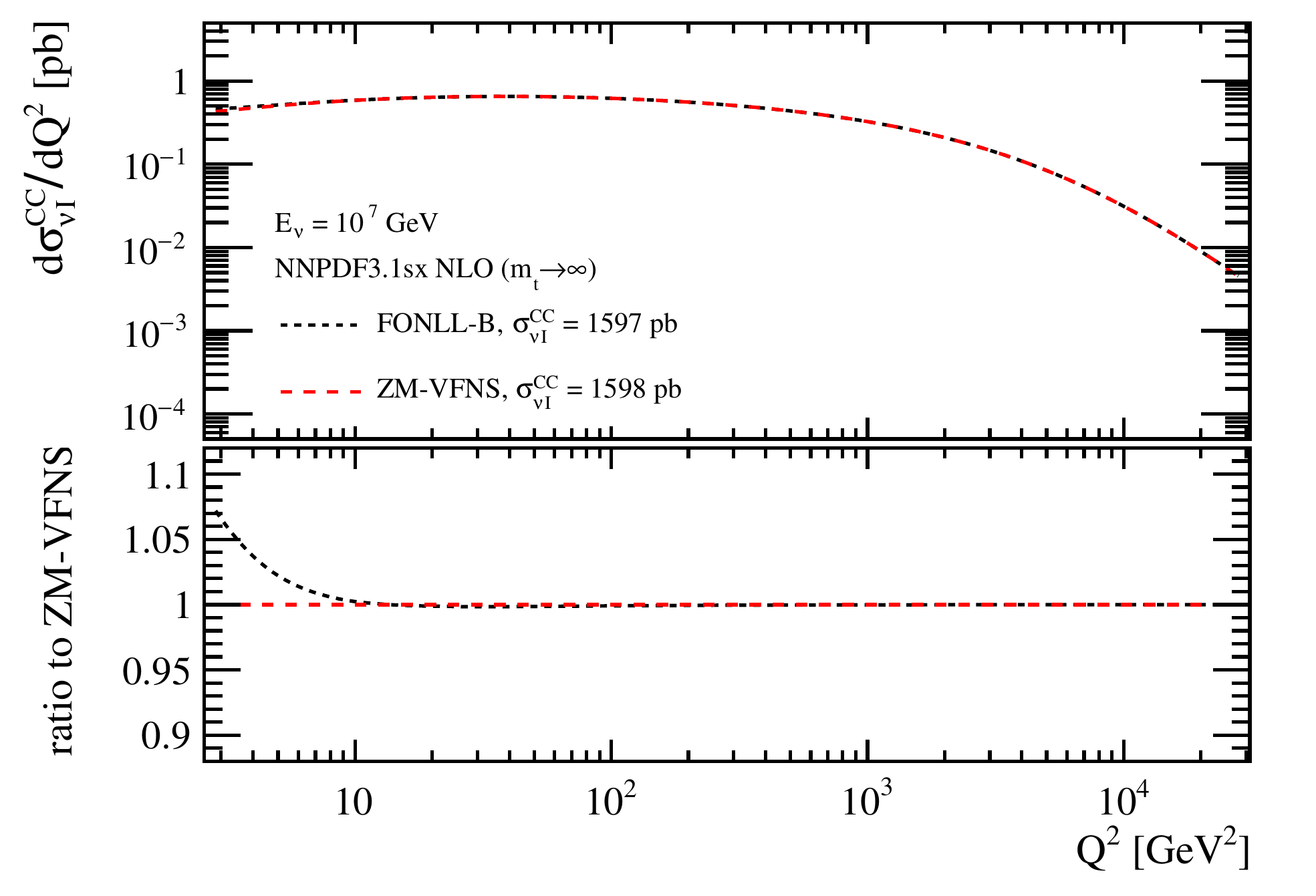} 
\includegraphics[width=.49\linewidth]{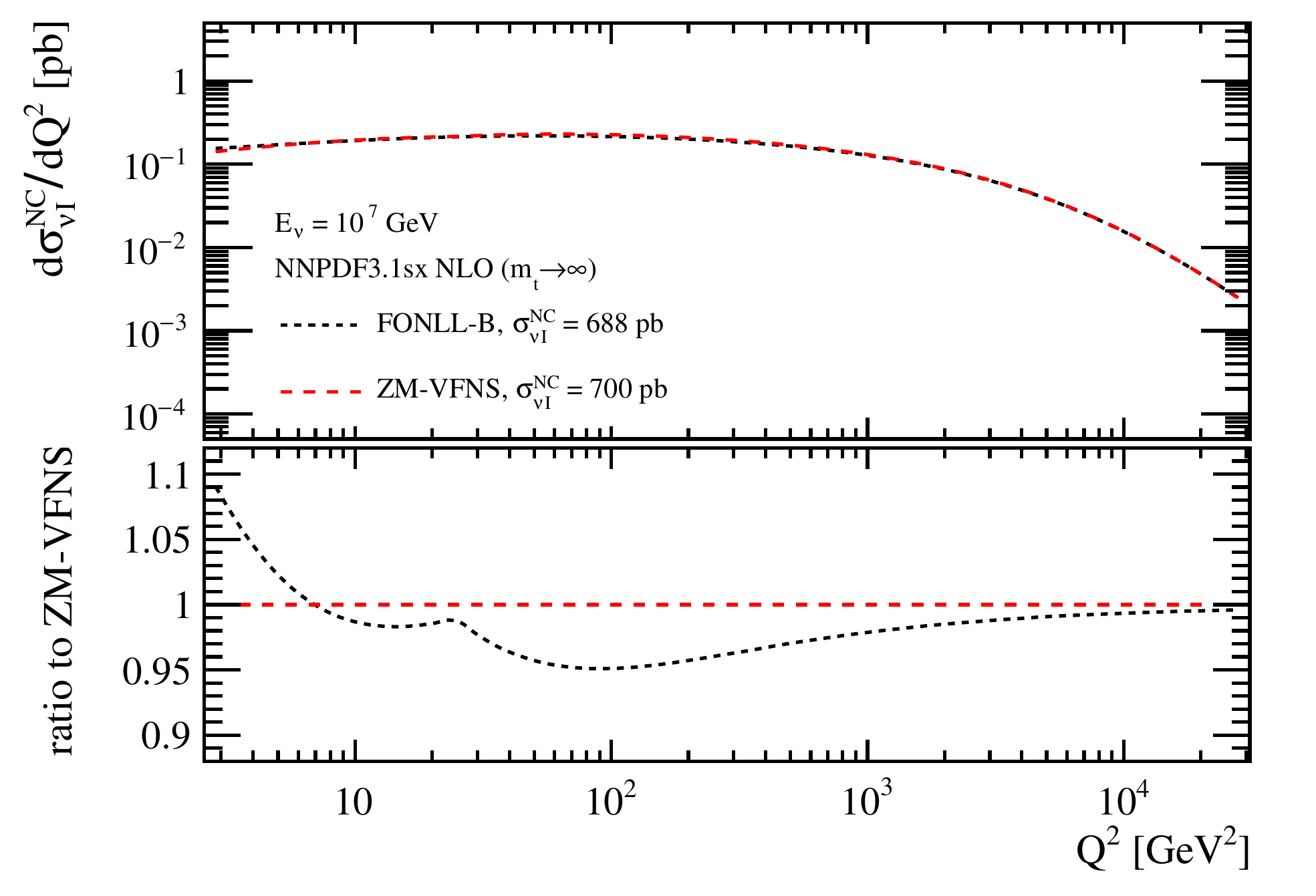}
\caption{\small Comparison between the FONLL and ZM-VFN heavy-quark
  mass schemes at NLO for the calculation of the single-differential
  cross-section $\rd\sigma_{\nu I}/\rd Q^2$,
  Eq.~(\ref{eq:integrate2}), for CC (left) and NC (right) scattering
  at $E_\nu=10^7$~GeV.  In both cases the limit $m_t\to \infty$ is
  taken to decouple the effects of the top quark.}
\label{fig:masseffects}
\end{figure}

We first consider the impact of charm- and bottom-quark mass effects on both
the CC and NC calculations at the level of the single-differential
cross-section $\rd\sigma/\rd Q^2$. This is obtained by integrating
Eq.~(\ref{eq:d2sig}) over the allowed kinematic range of $x$, namely
\beq
\label{eq:integrate2}
\frac{\rd\sigma(Q^2,E_{\nu})}{\rd Q^2} =  \int_{x_0(Q^2)}^{1} \rd x \frac{\rd^2 \sigma }{\rd x\,\rd Q^2}(x,Q^2,E_{\nu}) \,.
\eeq
In Fig.~\ref{fig:masseffects} we consider the predictions for the
differential DIS cross-section in Eq.~(\ref{eq:integrate2}) in the CC
(left) and in the NC (right) case. Predictions are computed as
functions of $Q^2$ at $E_{\nu} = 10^{7}$~GeV at NLO in both the FONLL scheme and the
ZM-VFNS. The upper panels display the absolute distributions while the
lower panels show the ratio to the ZM-VFNS. In these plots the limit
$m_t\to \infty$ is taken to decouple the effect of the top
quark.\footnote{This is done for illustration purposes only. Finite
$m_t$ effects are included in our calculation, see the
discussion below.}
In the upper panel of these plots, the total cross-section obtained by
integrating over the allowed range in $Q^2$ is also reported.

In the case of CC scattering, the inclusion of charm-mass effects
introduces a positive shift of up to 5\% in the region of
$Q^2 \lsim 10{\rm~GeV}^2$ but has a negligible impact on the total
cross-section.
The impact of bottom-mass effects is entirely negligible. This is due
to the strong suppression for the subprocesses involving the CKM
matrix elements $V_{ub}$ and $V_{cb}$.
Heavy-quark mass corrections in NC case display a similar behaviour at low
energies but introduce a negative correction in the intermediate
energy range of $Q^{2} \in [20,500]{\rm~GeV}^2$.
The net effect is a reduction of the total cross-section by
(1-2)\%. While not shown here, similar results are also obtained for
neutrino energies in the range $E_{\nu} \in [10^3,10^{12}]{\rm~GeV}$.
From this comparison, we conclude that charm- and bottom-quark mass
effects are negligibly small at the level of the total cross-section,
but may still be relevant for the calculation of the
double-differential distributions in Eq.~(\ref{eq:d2sig}).
These effects are included in our predictions unless specifically stated.

In addition to charm- and bottom-quark mass effects, it is also
important to consider the mass effects related to the top quark (see
also Ref.~\cite{Barge:2016uzn}).
The Born processes for massive top-quark production in neutrino-nucleon scattering are
\beq
{\rm CC:} \quad \nu + q_d \to \ell^{-}+ t \,, \qquad
{\rm NC:} \quad \nu + g \to \ell^{-}+ t +\bar{t} \,, 
\eeq
where in the CC process $q_d$ denotes a down-type quark.
Kinematically, the CC (NC) process may only proceed if the
\textit{physical} threshold for the production of a top quark (pair)
is reached, namely
\be\label{eq:w2def}
W^2=\frac{Q^2(1-x)}{x} > m_t^2 \quad \lp W^2 > 4m_t^2 \rp\,,
\ee
where the target nucleon mass is neglected. An accurate prediction
for these processes relies on the inclusion of mass effects in the partonic
cross-section.
We account for these effects by adopting the FONLL scheme with a maximum of $n_f = 6$
active flavours ($n_f = 6$ scheme, in short). In this scheme the top quark is treated 
as a massive parton for energies close to the top-quark mass, and becomes effectively massless
at much larger energies, $Q^2\gg m_t^2$. This has the consequence of introducing top-quark PDFs 
whose evolution resums potentially large logarithms of the form $\ln(Q^2/m_t^2)$.
As discussed in Sect.~\ref{sec:numerics}, heavy-quark mass effects are accounted for at 
$\mathcal{O}\lp \alpha_s^2\rp$ for NC scattering and $\mathcal{O}\lp \alpha_s\rp$ for CC scattering.
We note that the scheme choice in this work contrasts with the approach commonly 
used in the extraction of PDFs from experimental data, such as the NNPDF3.1sx sets used in this
work, in which the $n_f = 5$ scheme is preferred. 
In order to employ the $n_f=6$ scheme, we have generated variants of
the NNPDF3.1sx sets in which the top-quark PDFs are dynamically
generated at $Q^2=m_t^2$~\cite{Bertone:2017ehk}. These sets coincide
with the original ones for $Q^2<m_t^2$ and are made publicly available
(see Appendix~\ref{sec:UHExsec}).

In Fig.~\ref{fig:UHE_topquarkmass} we assess the impact of top-quark
production on the total cross-section for both CC (left) and NC
(right) scattering.
To do so, cross-sections are computed at NLO in the $m_{t}\to \infty$
limit and compared to the default $n_f=6$ FONLL calculation with
$m_t=172.5$~GeV. Taking the limit of $m_{t}\to \infty$ effectively
increases the production threshold for top production such that it
becomes kinematically inaccessible. Comparing these predictions is
therefore useful to single out the relative contribution for the
top-quark production to the total cross-section.

For CC scattering (left plot of Fig.~\ref{fig:UHE_topquarkmass}), it
is found that the top-quark contribution becomes relevant for neutrino
energies of $E_{\nu} \gsim 10^{7}$~GeV, while for NC scattering (right
plot) the contribution is negligible.
Inspection of Eq.~\eqref{eq:w2def} reveals that the 
production of a single top-quark in CC scattering is 
kinematically possible for $Q^2/x \gsim m_t^2$.
Therefore, for neutrino energies of $E_{\nu} \gsim 10^{7}$~GeV top-quark 
production becomes accessible at relatively small values of $Q^2$ in the low-$x$ region.
This process is dominated by the partonic subprocess $W + b \to t$ as a consequence of
the steep growth of the $b$-quark PDF (generated at $Q = m_b$) at low-$x$, and 
of the fact that $|V_{tb}| \approx 1$.
The combination of these effects results in a non-negligible
contribution to the total CC cross-section which represent up to 
$5\%$ of the total cross-section at $E_{\nu} \sim 10^{12}$~GeV.

\begin{figure}
\centering
\includegraphics[width=.49\linewidth]{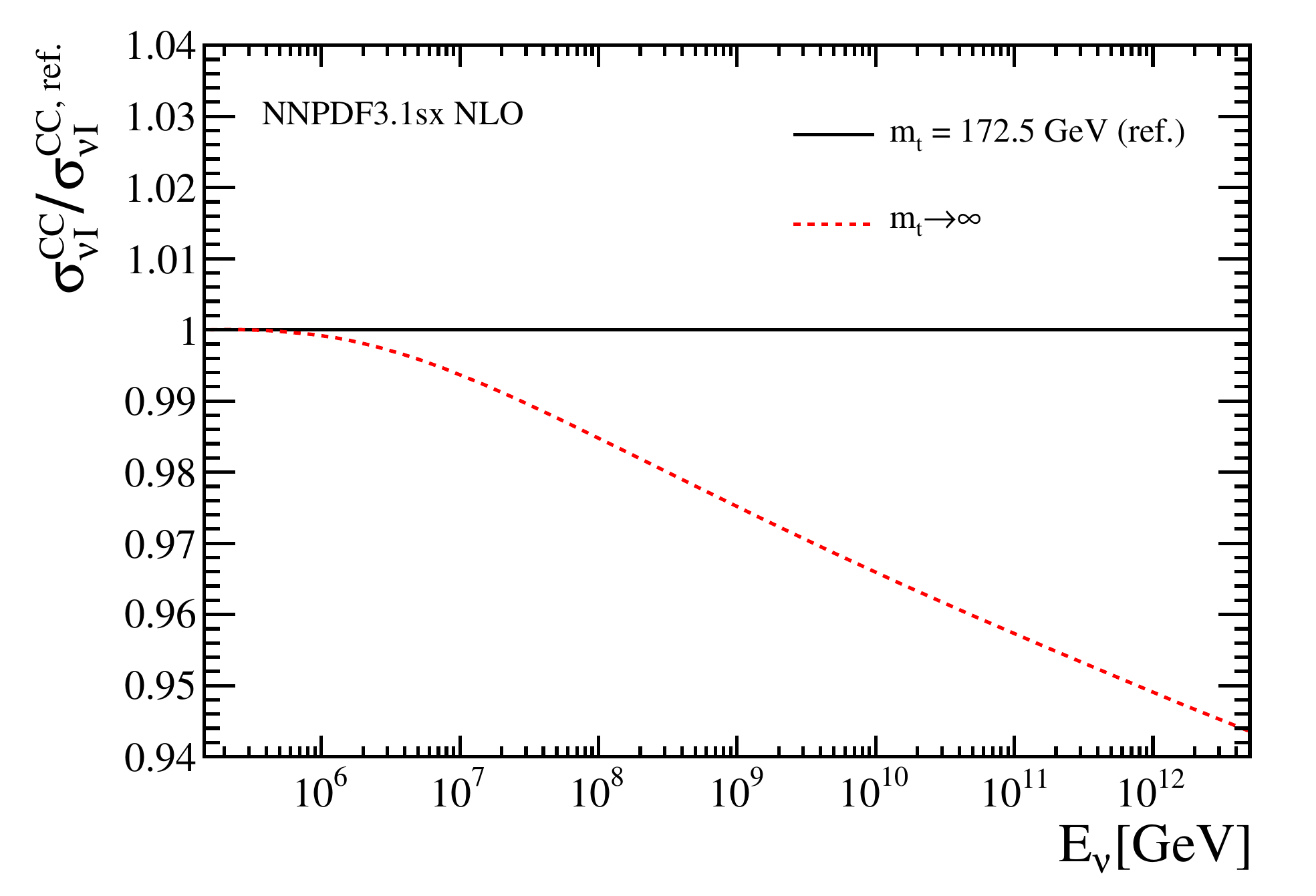}
\includegraphics[width=.49\linewidth]{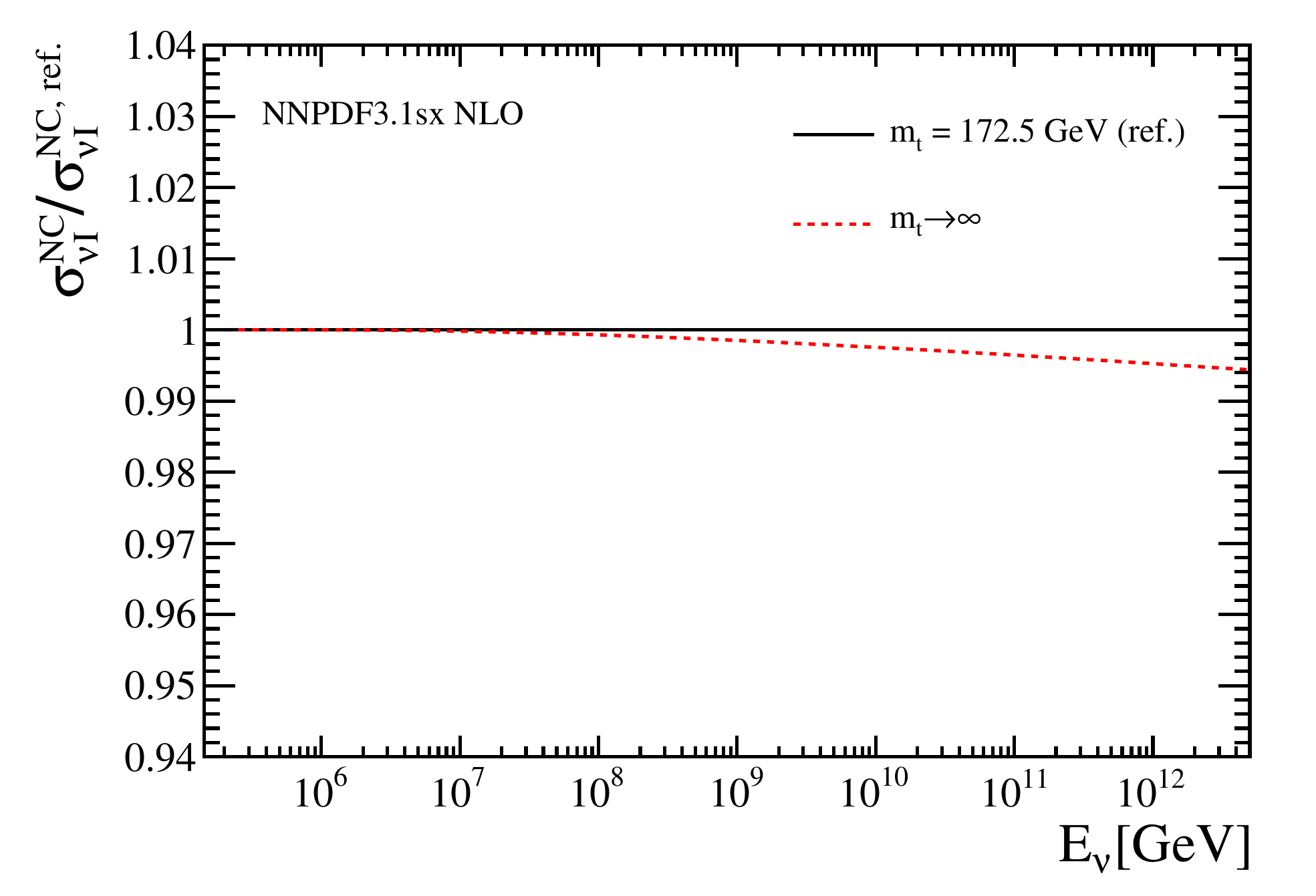}
\caption{\small The ratio of the CC (left) and NC (right)
  neutrino-isoscalar cross-section at NLO computed in the
  $m_{t}\to \infty$ limit compared with our default (ref.) $n_f=6$
  FONLL calculation with $m_t=172.5$~GeV.}
\label{fig:UHE_topquarkmass}
\end{figure}

\subsection{Impact of nuclear modifications}
\label{sec:nuclearPDFs}

At neutrino telescopes, the incoming UHE neutrinos scatter upon
nucleons which may be bound within a nucleus of the target material.
Therefore, in principle it is necessary to account for the presence of
nuclear-binding effects.
The most relevant of these effects for the calculation of the UHE
cross-sections is that of shadowing~\cite{Frankfurt:2011cs}, namely
the depletion of nuclear structure functions as compared to their
free-nucleon counterparts.
In experiments such as IceCube, KM3NET, and Baikal the nuclear target
is a molecule of H$_{2}$O, and the scattering is dominated by bound
nucleons inside the oxygen nuclei with a small admixture of the free
protons from the hydrogen.

The mass-averaged structure function for a water/ice target may thus
be written as
\begin{align} \label{eq:Nuc}
F^{\rm{H}_2\rm{O}} = \frac{1}{2+A}\left(2 F^p + Z F^{p,A} + N F^{n,A}\right)\,,
\end{align}
where $F^p$ is the structure function of a free proton, and
$F^{p(n),A}$ represents that of a proton (neutron) which is bound
within a nucleus with mass number $A$ containing $Z$ protons and $N$
neutrons, with $A=Z+N$.
In this case, the bound nucleus is that of oxygen which is an
isoscalar target ($N = Z$) with $A = 16$.
Predictions for these structure functions can be computed according to
Eq.~\eqref{eq:SF}, where the PDFs correspond to those of the (bound)
nucleon.
This implies that one should account for the effects that quark and
gluon PDFs of a nucleon experience inside a heavy nucleus.  These
corrections have been quantified in a number of analyses of nuclear
parton distribution functions
(nPDFs)~\cite{deFlorian:2011fp,Eskola:2016oht,Kovarik:2015cma,Khanpour:2016pph}.

\begin{figure}[t]
\centering
\includegraphics[width=.49\linewidth]{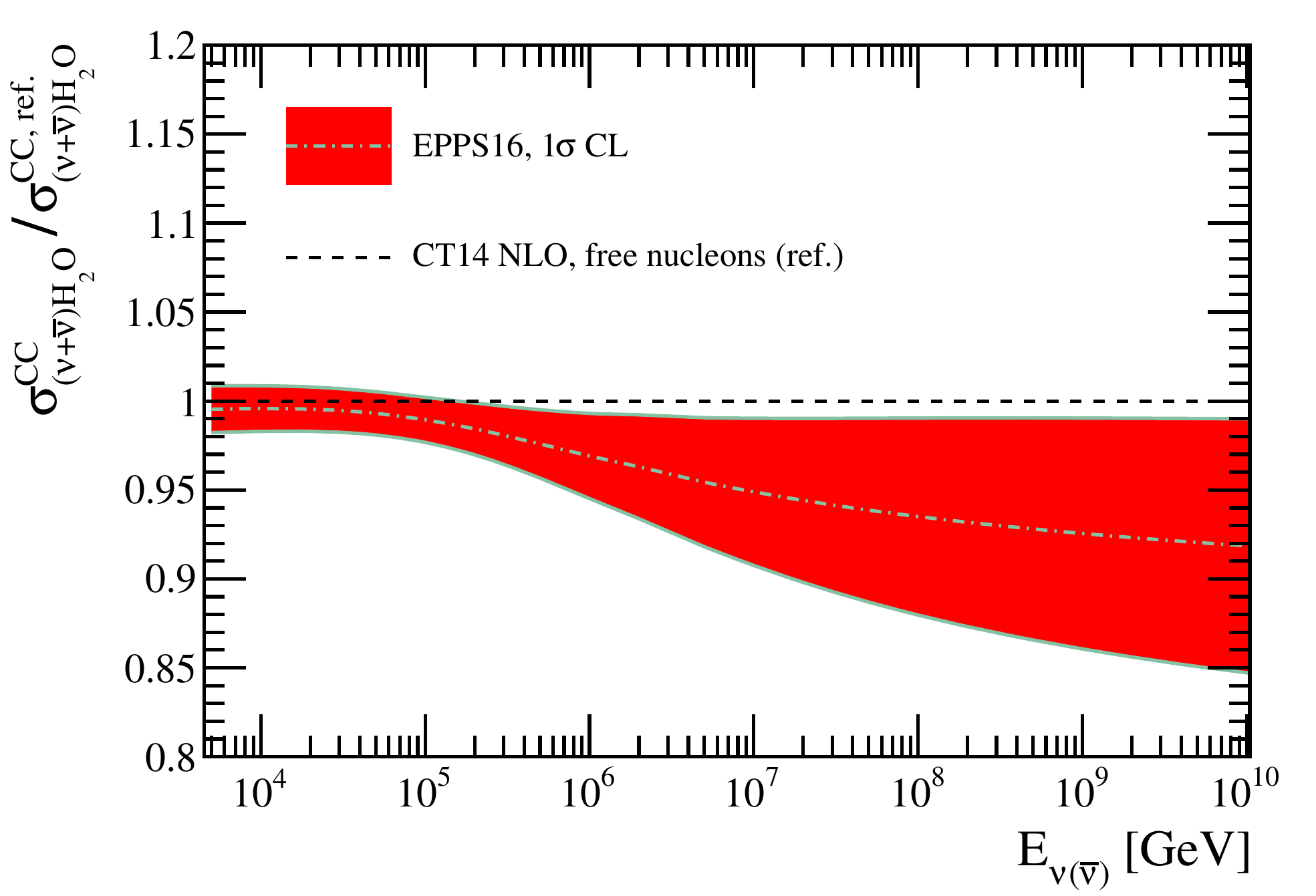}
\includegraphics[width=.49\linewidth]{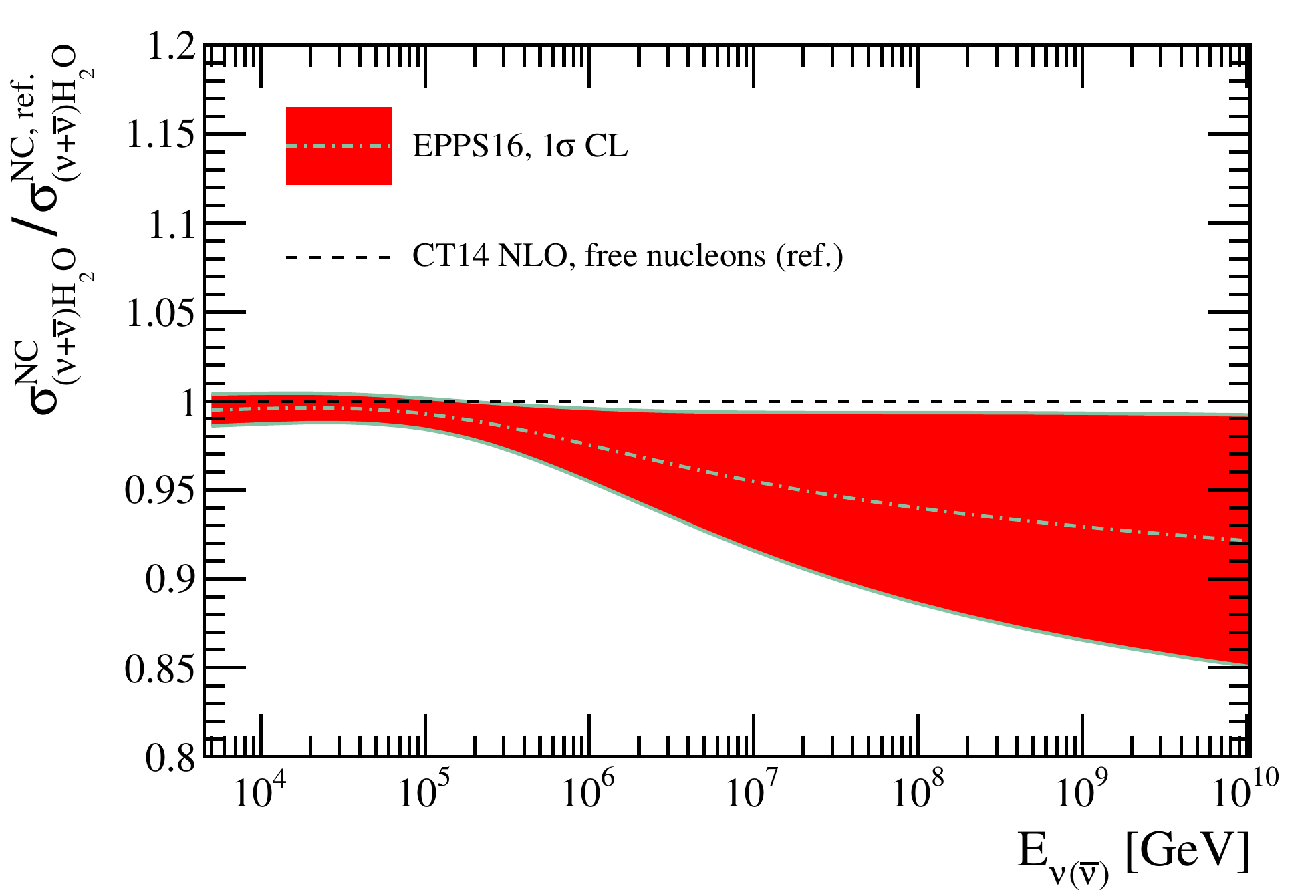}
\caption{\small The CC (left) and NC (right)
neutrino-nucleon cross-sections (adding the contributions
from neutrinos and antineutrinos) for an H$_2$O molecule computed 
with the EPPS16 nPDF set and presented as a function of the
neutrino projectile energy $E_{\nu({\bar{\nu}})}$.
The quoted 1$\sigma$ confidence-level (CL) uncertainty bands include only the uncertainty from the nuclear PDF fit,
and have been evaluated using the asymmetric Hessian prescription.
Each distribution has been normalised with respect to the baseline free-nucleon prediction.
\label{fig:F2_nuclearffects}
}
\end{figure}

To assess the impact of nuclear corrections on the UHE
neutrino-nucleon cross-sections, we have computed the mass-averaged
total cross-section for an H$_{2}$O molecule using the nPDFs from the
EPPS16 global analysis~\cite{Eskola:2016oht}.
These predictions are obtained by first computing the cross-section
for both a free proton and an oxygen target, and then performing a
combination according to
\beq \label{eq:SigH2O}
\sigma_{\nu{\rm H}_2{\rm O}}(E_{\nu}) = \frac{1}{18} \left( 2 \sigma_{\nu p}(E_{\nu}) + \sigma_{\nu O}(E_{\nu}) \right) \,.
\eeq
The EPPS16 fit is constructed taking the CT14 NLO free-nucleon
PDFs~\cite{Dulat:2015mca} as a baseline. Therefore, we use the
central value of this PDF set for the free-nucleon
predictions.\footnote{In the numerical computation, we also adjust the
  values of $Q_{\rm min}$ and $x_{\rm min}$ values to match those of
  EPPS16.}
Results for both CC and NC scattering cross-sections (for the sum of
neutrino- and antineutrino-induced processes) are shown in
Fig.~\ref{fig:F2_nuclearffects}. Distributions are presented as
normalised to those of the corresponding free-nucleon predictions.
The quoted uncertainty bands represent the 1$\sigma$ uncertainty of the 
EPPS16 set (excluding free-nucleon uncertainties) evaluated using the asymmetric Hessian prescription.

We find a suppression of the cross-section for (anti)neutrino energies
$E_{\nu({\bar{\nu}})} \gsim 10^6$~GeV due to the shadowing
effect present in the nPDFs at small-$x$.
The central value is reduced by 3\% for $E_{\nu({\bar{\nu}})} = 10^6$~GeV,
and by as much as 10\% for $E_{\nu({\bar{\nu}})} = 10^{10}$~GeV.
However, it should be noted that while a suppression of the
cross-section is preferred, the uncertainty of the nuclear corrections
are almost as large as the shift of the central value.
Therefore, using EPPS16, the significance of nuclear modifications is
mostly within the 1$\sigma$ level.
At lower energies, instead, the impact of the nuclear corrections
becomes less important.
The results of Fig.~\ref{fig:F2_nuclearffects} indicate that nuclear
corrections represent a large source of theoretical uncertainty in the
predictions of the UHE neutrino-nucleon cross-section. Therefore, it
is necessary to account for such effects to provide reliable
predictions.

In Sect.~\ref{sec:results}, where predictions are provided for the
total UHE cross-sections, the impact of nuclear corrections is
accounted for in a factorised form.
To illustrate this procedure, we consider here the construction of the
cross-section for the neutrino-induced scattering on an oxygen
nucleus.
First, the nuclear modification factor $R_{\nu O}(E_{\nu})$ is
computed with the EPPS16 nPDFs as follows:
\beq
\label{eq:REPPS16}
R_{\nu O}(E_{\nu}) \equiv \left(  \frac{\sigma^{\rm EPPS16}_{\nu O}(E_{\nu})}{\sigma^{\rm free}_{\nu I}(E_{\nu})}\right) \,,
\eeq
where $\sigma^{\rm free}_{\nu I}(E_{\nu})$ is the cross-section for an isoscalar
target computed with the central CT14 NLO free-nucleon PDFs, and the
normalisation is such that $R_{\nu O}(E_{\nu}) \to A=16$
in the limit of vanishing nuclear effects.
Note that the flavour symmetry of PDFs at small-$x$ implies that
Eq.~(\ref{eq:REPPS16}) gives essentially the same results for both NC
and CC scattering, as also seen from Fig.~\ref{fig:F2_nuclearffects}.

This modification factor is then applied to the cross-section for an
isoscalar target computed with a different set of free-nucleon PDFs
according to \beq
\label{eq:NucMod}
\widetilde{\sigma}_{\nu O}(E_{\nu})  = R_{\nu O}(E_{\nu})
\widetilde{\sigma}_{\nu I}(E_{\nu}) \,,
\eeq
for either NC or CC scattering.
The mass-averaged cross-section for neutrino scattering on a molecule
of H$_2$O is obtained according to Eq.~\eqref{eq:SigH2O}.
In fact, for the highest neutrino energies, any departures from
non-isoscalarity can be ignored and the following approximation can be
made
\beq
\label{eq:nPDF_UHE}
\sigma_{\nu{\rm H}_2{\rm O}}(E_{\nu}) \simeq \frac{\sigma_{\nu I}(E_{\nu})}{18} \left( 2 + R_{\nu O}(E_{\nu}) \right)\,.
\eeq
The total uncertainty in Eq.~(\ref{eq:nPDF_UHE}) can then be computed
by adding in quadrature the free-nucleon PDF uncertainties (arising
from $\sigma_{\nu I}$) with those of the nPDFs (from
$R_{\nu O}$).

The factorised expression in Eq.~(\ref{eq:nPDF_UHE}) makes it
straightforward to improve the prediction of the total
cross-section when a more precise determination of
$R_{\nu O}(E_{\nu})$ becomes available.\footnote{Another benefit of this factorised expression is that it is straightforward
to calculate the nuclear correction for other nuclear targets. This may be relevant
for the modelling of neutrino absorption within the Earth as neutrinos may scatter via the
NC process on an Fe target~\cite{Aartsen:2017kpd}.
}
The large uncertainties associated to $R_{\nu O}(E_{\nu})$ in the
current calculation can be related to the lack of experimental data
used to determine nPDFs (and in particular those sensitive to the
gluon) in the region of $x\lsim 10^{-2}$.
Indeed, to compute $R_{\nu O}(E_{\nu})$ at large $E_{\nu}$ values it
is necessary to extrapolate nPDFs to small values of $x$ by several
orders of magnitude.
In the extrapolation region results are driven to a large extent by
the particular methodological choices made to extract nPDFs. Examples
are the parameterisation (\textit{i.e.} the functional form assumed to
parameterise the $x$ and $A$ dependence) or $\chi^2$ tolerances that
define the 1$\sigma$ PDF uncertainties.\footnote{We have also repeated
the analysis above using the nCTEQ15 nuclear PDF
sets~\cite{Kovarik:2015cma} and obtained consistent results.
However, it was found that the 1$\sigma$ uncertainties were
considerably smaller as compared to the EPPS16 results. Therefore,
we quote the EPPS16 results which provide a more conservative
uncertainty estimate.}

As will become apparent in Sect.~\ref{sec:results}, the uncertainty
associated to the nuclear corrections is a limiting factor in the
calculation of the UHE neutrino-nucleon cross-section. This provides a
strong motivation to improve global fits of nPDFs by extending the
kinematic coverage of the input data set.
One possibility is to include LHC data collected in $p$+$Pb$
collisions which are sensitive to the small-$x$ region. The nPDFs are
typically parameterised as continuous functions of the nucleus mass
number $A$. Therefore, constraints obtained for nucleons bound within
a $Pb$ nucleus ($A = 208$) are relevant also for lighter
nuclei such as oxygen.
Progress in this direction may be possible by studying forward
$D$-meson production in $p$+$Pb$ collisions~\cite{Aaij:2017gcy}, but
these data have not yet been included in any nPDF fits (see
Ref.~\cite{Kusina:2017gkz} for initial work in this direction).
In the longer term, stringent constraints would also be provided by
possible future lepton-ion colliders such as the
EIC~\cite{Boer:2011fh} and the LHeC~\cite{AbelleiraFernandez:2012cc}.

\section{Constraining small-$x$ resummed PDFs with $D$-meson data}
\label{sec:lhcb}

The analysis of Ref.~\cite{Gauld:2016kpd} quantified the impact of
the LHCb $D$-meson cross-section measurements on the NNPDF3.0 NLO PDFs at
small $x$.
Here we revisit this analysis, applying it to the NNPDF3.1sx
sets~\cite{Ball:2017nwa,Ball:2017otu} which have been extracted either with 
or without including the effects of small-$x$ resummation.
In this section, we review the fit settings and discuss the
experimental inputs along with the corresponding theoretical
calculations used to include the LHCb data into the fit.
We then present the fit results and describe the tests performed to
assess their robustness.

\subsection{Fit settings, experimental data, and theory calculations}

The kinematic coverage of the UHE neutrino-nucleon cross-section as
shown in Fig.~\ref{fig:contour} illustrates the sensitivity of this observable to 
PDFs in a region of extremely small $x$ ($x\simeq 10^{-8}$).
This region is outside the coverage of the input datasets used in
the current PDF fits. In particular, it is beyond the coverage of
the HERA data~\cite{Abramowicz:2015mha} that only reach values of $x$
as small as $x\simeq 2\times 10^{-5}$ for $Q^2 \gsim 1$ GeV$^2$.
However, recent work has demonstrated that smaller values of $x$ can
be probed using $D$-meson production measurements as provided by the LHCb experiment.
Several groups have shown~\cite{Zenaiev:2015rfa,Gauld:2015yia,Cacciari:2015fta} that
forward $D$-meson cross-section measurements can provide important information on both
the normalisation and the shape of the small-$x$ PDFs, especially that of the gluon.
In particular, in Ref.~\cite{Gauld:2016kpd} it has been shown that the
combination of the LHCb measurements at $\sqrt{s}=$5, 7, and 13~TeV
allows to substantially reduce the uncertainties of the gluon PDF at small $x$.
              
\begin{figure}
\centering
\includegraphics[width=.55\linewidth]{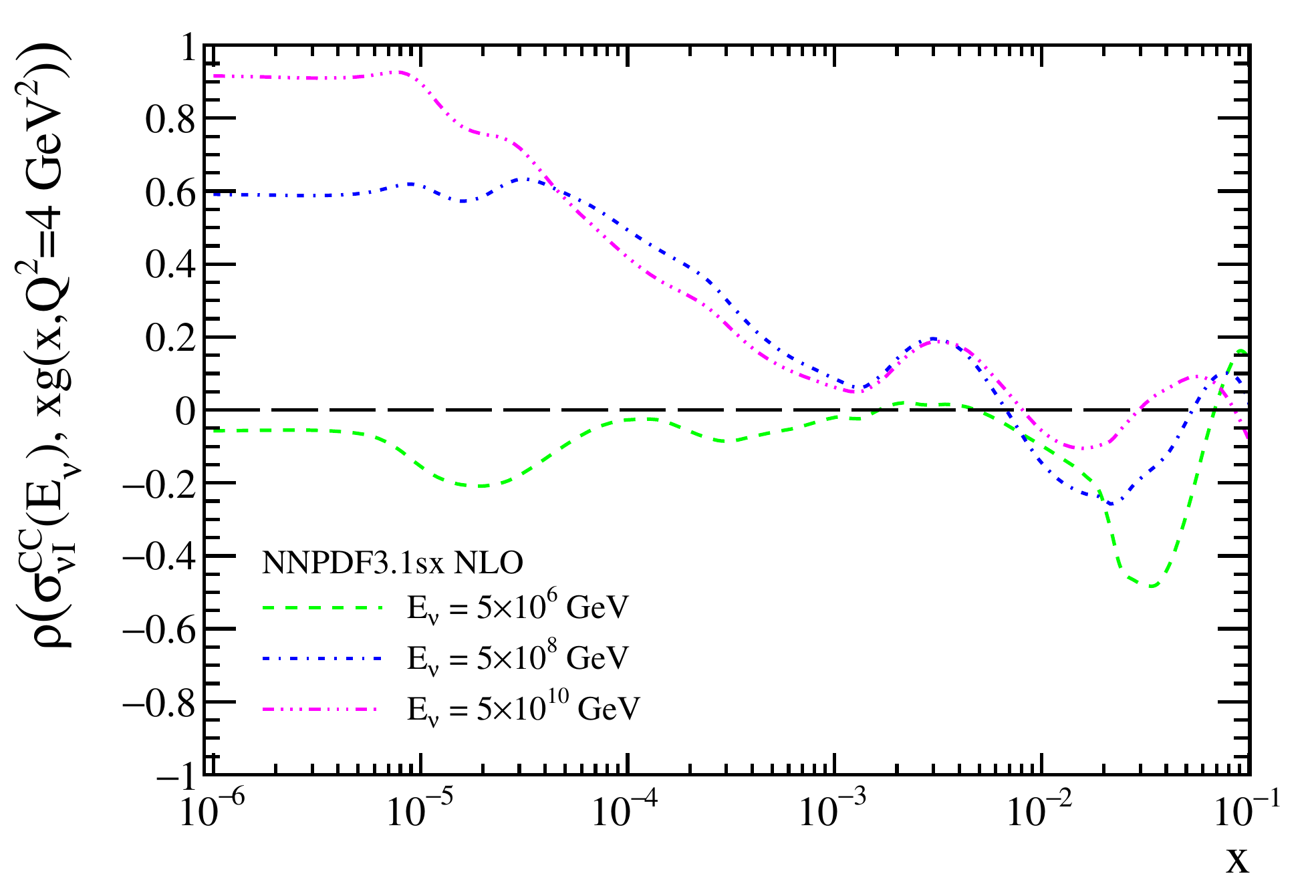}
\caption{\small The correlation coefficient $\rho$ between the gluon PDF at $Q^2=4$~GeV$^2$
and the CC neutrino-isoscalar cross-section evaluated at specific values of $E_{\nu}$, presented
as a function of the momentum fraction $x$ carried by the gluon.
\label{fig:F2correl}
}
\end{figure}
In order to illustrate the interplay between the small-$x$ gluon PDF
and the UHE neutrino-nucleon cross-sections, in
Fig.~\ref{fig:F2correl} we show the correlation coefficient
$\rho$~\cite{Guffanti:2010yu} evaluated between the gluon PDF at
$Q^2=4$~GeV$^2$ and the CC neutrino-isoscalar cross-section
$\sigma_{\nu I}^{CC}(E_{\nu})$ for fixed values of $E_{\nu}$. The correlation
is shown as a function of the momentum fraction $x$ carried by the gluon.
Note that although there is no direct coupling between the electroweak
bosons and the gluon, this correlation mostly arises from the impact of the
latter on the sea quarks by means of the DGLAP evolution.
From this comparison, we find a large correlation ($\rho\gsim 0.6$) 
in the small-$x$ region ($x\lesssim 10^{-4}$) for the cross-section evaluated 
at large values of the neutrino energy $E_{\nu} \gsim 5\times10^8$~GeV.
Therefore, a better understanding of the small-$x$ gluon PDF is necessary to obtain
reliable predictions for the UHE neutrino-nucleon cross-sections in this energy range.

In this work we aim at providing predictions for the UHE
neutrino-nucleon cross-sections using a state-of-the-art calculation
based on structure functions accurate at NNLO+NLL$x$. This requires
using PDFs exatracted with the same accuracy.
The main limitation in achieving this goal is that theoretical
predictions for $D$-meson production are not available at this
accuracy.
Firstly, because NNLO differential cross-sections for heavy-quark pair
production are only available for top quarks~\cite{Czakon:2015owf}.
Secondly, because small-$x$ resummed hard
cross-sections~\cite{Ball:2001pq} are not yet available in a form
suitable for phenomenology.

Following Ref.~\cite{Gauld:2016kpd}, the impact of the LHCb $D$-meson
production data on PDFs at small-$x$ can be quantified by means of
the following observables:
\begin{align} \nonumber
  \label{eq:Obs}
N_X^{ij} &=  \frac{d^2\sigma({\rm X~TeV})}{dy_i^D d (p_T^D)_j} \bigg{/} \frac{d^2\sigma({\rm X~TeV})}{dy_{\rm ref}^D d (p_T^D)_j} \, , \\[1mm]
R_{13/X}^{ij} &= \frac{d^2\sigma({\rm 13~TeV})}{dy_i^D d (p_T^D)_j} \bigg{/} \frac{d^2\sigma({\rm X~TeV})}{dy_{\rm i}^D d (p_T^D)_j} \, ,
\end{align}
where $X=5{\rm~or~}7$, $p_T^D$ and $y^D$ are the transverse momentum and
rapidity of the $D$ mesons, and $y_{\rm ref}^D$ denotes a reference
rapidity bin.
Eq.~(\ref{eq:Obs}) implies that we use $D$-meson differential
distributions always in a normalised form, taking as a reference
either a given rapidity bin or data taken at a different
centre-of-mass energy.
An important implication of this approach is that missing higher-order
corrections to partonic cross-sections partially cancel in the ratios,
reducing the sensitivity of these observables to such corrections.
Crucially, numerators and denominators in Eq.~(\ref{eq:Obs}) probe
different regions in $x$, and therefore these observables still provide 
constraints on the PDFs at small $x$.

We apply Bayesian reweighting~\cite{Ball:2010gb,Ball:2011gg} to 
prior PDF sets which have previously been extracted with different theory settings.
Specifically, we use the NNPDF3.1sx sets based on: NLO, small-$x$
resummation matched to NLO (\textit{i.e.} NLO+NLL$x$), and small-$x$
resummation matched to NNLO (\textit{i.e.} NNLO+NLL$x$).
In all cases, the partonic cross-sections for $D$-meson production are
computed at NLO. We discuss below the stability of our results with respect to this choice.

For the PDF reweighting, we exploit the LHCb measurements of $D$-meson
cross-section measurements in $pp$ collisions at $\sqrt{s}=5$, 7, and
13~TeV~\cite{Aaij:2016jht,Aaij:2013mga,Aaij:2015bpa}.
These data are provided double differentially with respect to $p_T^D$ 
and $y^D$ for different types of $D$ meson.
The kinematic coverage of these cross-sections is approximately
$p_T^D \in [0,8]$~GeV and $y^{D} \in [2,4.5]$, with small differences
depending on the specific $D$-meson species and the hadronic
centre-of-mass energy.
From these measurements we construct the two normalised observables
defined in Eq.~(\ref{eq:Obs}) for the $D^0$, $D^+$, and $D_s$ species
and the corresponding anti-particles.

In what follows we adopt as a baseline the results of the fit with the
normalised cross-sections $N_X^{ij}$, denoted by $N_{5+7+13}$.
The LHCb data is restricted to the region $p_T^D \in [1,8]$~GeV. The
reason for this cut is that in our calculation the factorisation and
renormalisation scales are set equal to the transverse mass of the
outgoing heavy quark.
Since the NNPDF3.1sx fits are determined at the input scale of $Q_0 = 1.64$~GeV, 
restricting to the region $p_T^D \in [1,8]$~GeV avoids sampling the PDFs outside their validity range.
Practically, we set $\mu_{\rm min}  = 1.64$~GeV in the calculation which is relevant for a very small fraction
of events with $p_T^c \simeq 0$~GeV and $p_T^{D} > 1$~GeV.
As summarised in Table~\ref{tab:chi2}, with this kinematic cut, a
total of 78, 72, and 119 data points at 5, 7, and 13~TeV,
respectively, are included in the fit.

Applying the reweighting procedure requires computing the theory
predictions for the observables in Eq.~\eqref{eq:Obs} using the
$N_{\rm rep}=100$ replicas of the NNPDF3.1sx sets.
These predictions are obtained at {\sc\small NLO$+$PS} accuracy using
{\sc\small
  POWHEG}~\cite{Nason:1987xz,Nason:2004rx,Frixione:2007vw,Alioli:2010xd}
to match the fixed-order calculation~\cite{Frixione:2007nw} to the
{\sc\small Pythia8} shower~\cite{Sjostrand:2007gs,Sjostrand:2014zea}
with the default {\sc\small Monash 2013} tune~\cite{Skands:2014pea}.
The calculation is performed with an input value for the charm-quark
mass of $m_c = 1.5$~GeV.

\subsection{Results and validation}

\begin{figure}[t]
\centering
\includegraphics[width=.49\linewidth]{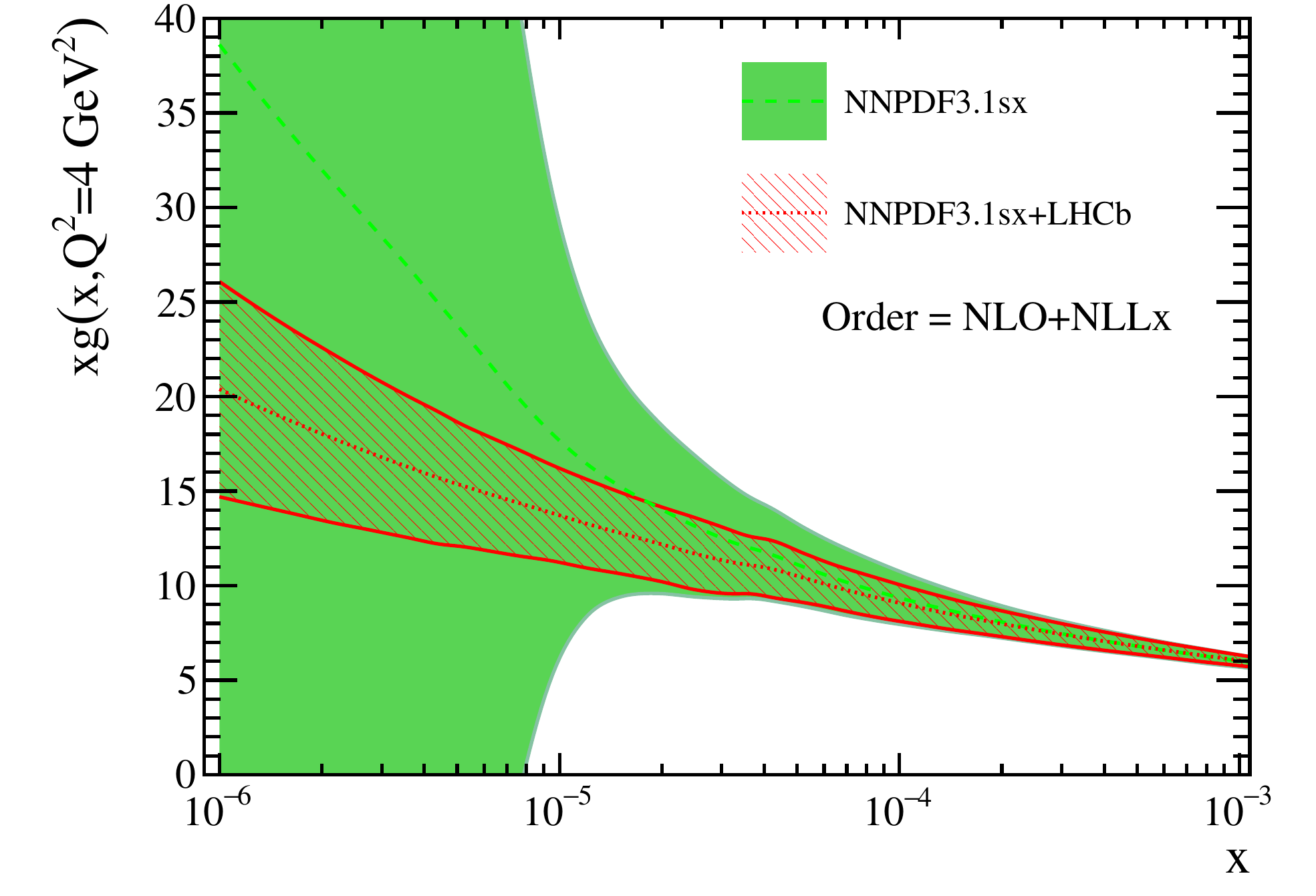}
\includegraphics[width=.49\linewidth]{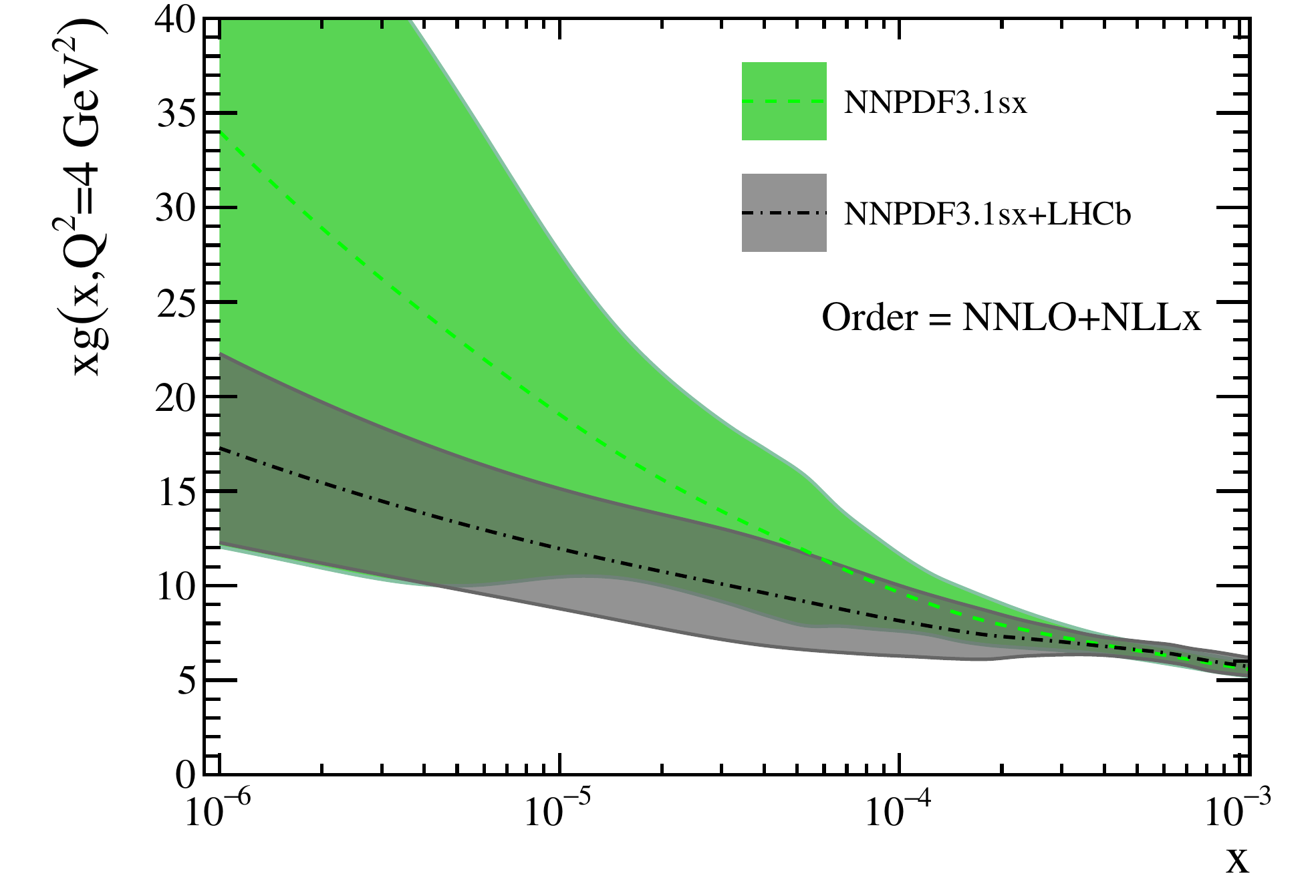} \\
\includegraphics[width=.49\linewidth]{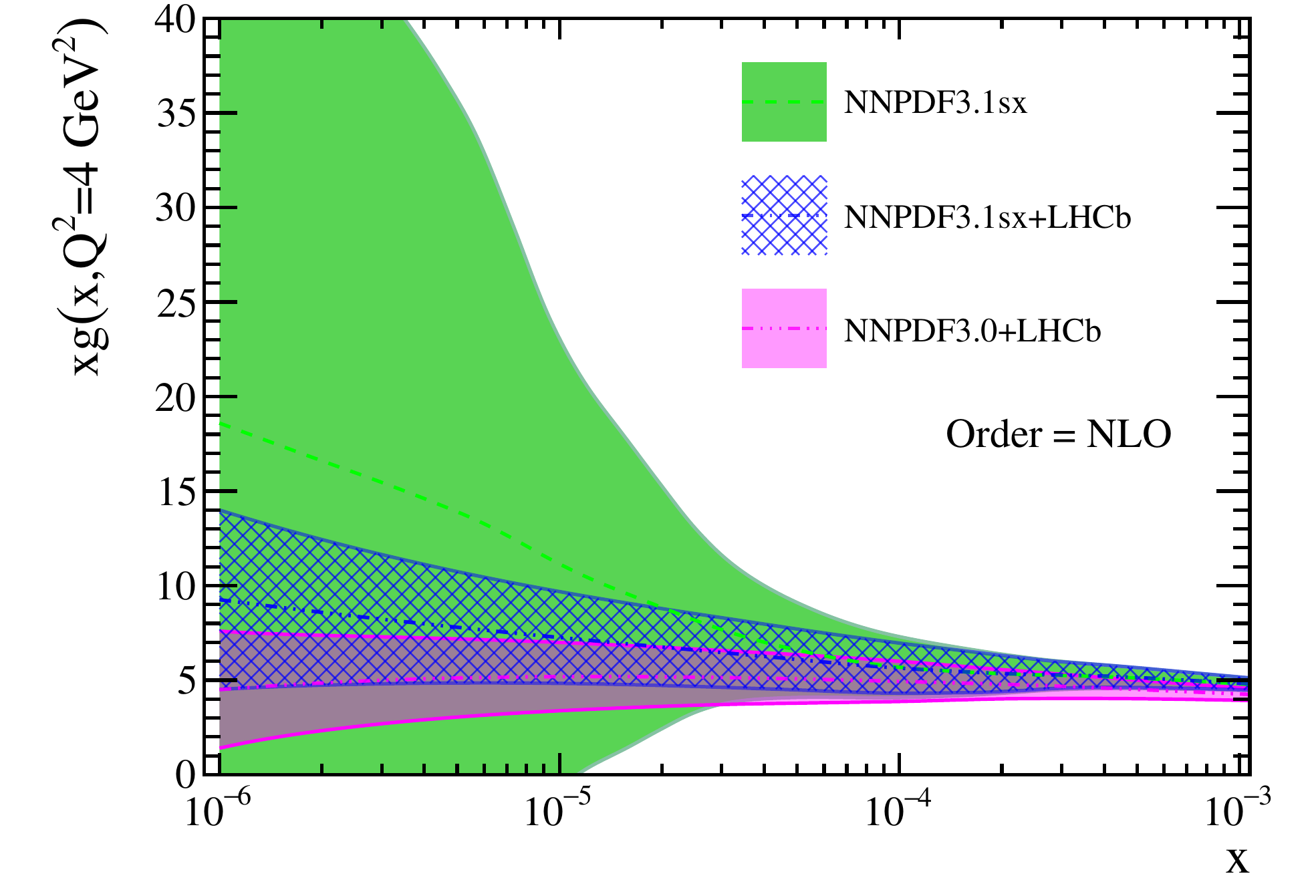}
\includegraphics[width=.49\linewidth]{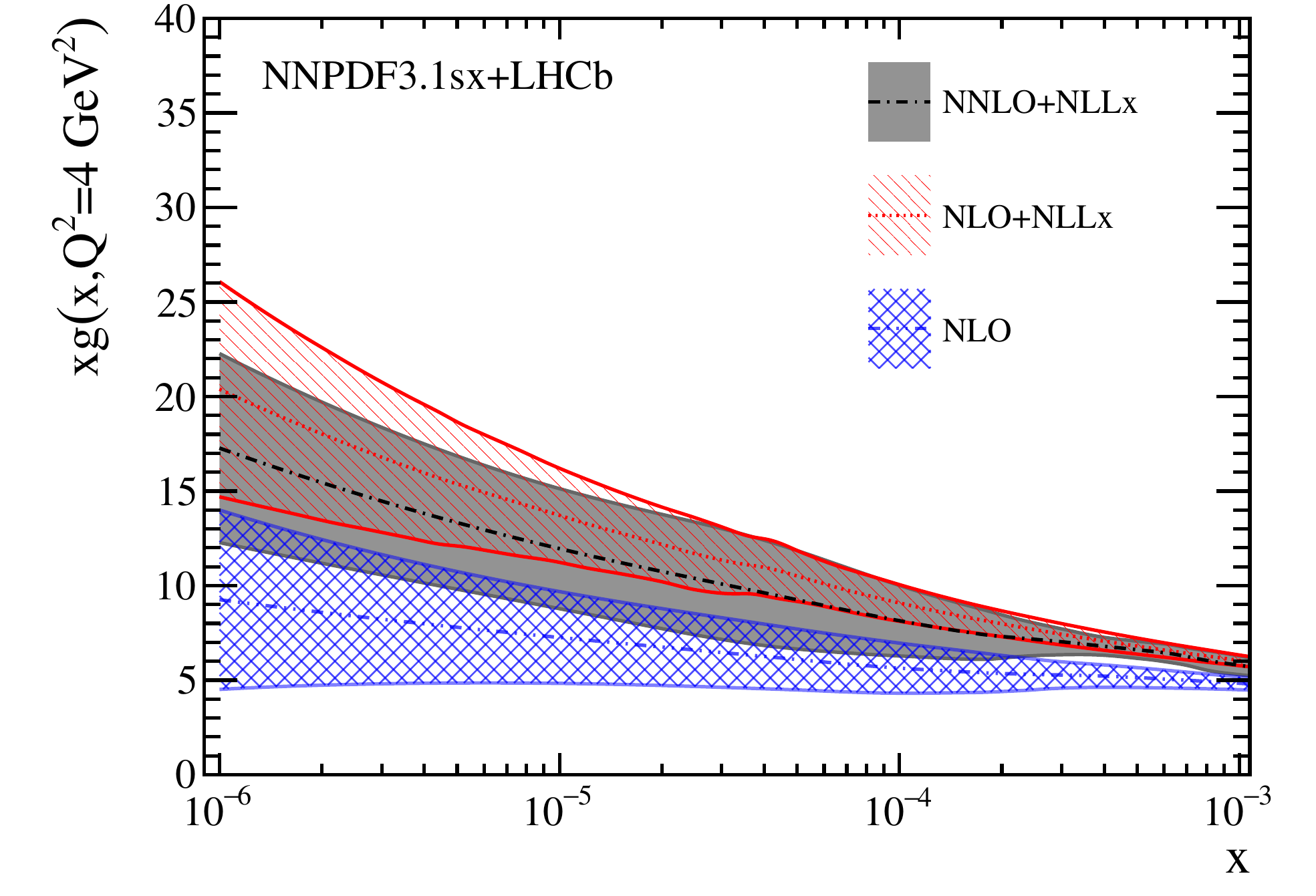}
\caption{\small Comparison of the gluon PDF at $Q^2=4$ GeV$^2$ from
  the different NNPDF3.1sx prior sets 
with the corresponding results once the LHCb $D$-meson cross-sections
have been included in the fit.
  The results are shown for three different theoretical settings:
  using resummed NLO+NLL$x$ and NNLO+NLL$x$ theory (upper plots)
  and using fixed-order NLO theory (lower-left plot).
  In addition, a comparison of the fit
  results obtained with different theory settings is also shown in the lower right plot.
  For completeness, in the NLO comparison we also show the NNPDF3.0+LHCb
  results of Ref.~\cite{Gauld:2016kpd}.
  }
\label{fig:gPDF_rwgt}
\end{figure}

Following the procedure outlined above, we have produced three
variants of the NNPDF3.1sx+LHCb fits, based on the NLO, NLO+NLL$x$,
and NNLO+NLL$x$ theory settings.
In all cases, the matrix-elements for the partonic process are evaluated 
at NLO~\cite{Frixione:2007nw}.
In Fig.~\ref{fig:gPDF_rwgt} we compare the gluon PDF from the three
prior NNPDF3.1sx sets at $Q^2=4$ GeV$^2$ with the corresponding
results after the LHCb $D$-meson cross-sections have been included in
the fit.
For completeness, we also compare the NLO results which have been
obtained with the NNPDF3.0+LHCb set~\cite{Gauld:2016kpd}.
PDF uncertainties are computed as 1$\sigma$ intervals.

From the comparisons in Fig.~\ref{fig:gPDF_rwgt} we find a significant
reduction of the PDF uncertainties due to the inclusion of the LHCb
$D$-meson cross-section data.
The magnitude of this reduction turns out to be similar for all the
three theory settings considered.
We also observe that at NLO, consistent results are obtained for the 
two different priors, NNPDF3.0 and NNPDF3.1sx.
This stability is reassuring taking into account the differences
between the two fits in terms of input dataset, treatment of the charm
quark PDF, and the values of the heavy-quark masses.

In the lower-right plot of Fig.~\ref{fig:gPDF_rwgt} we also display the comparison of the
three NNPDF3.1sx+LHCb sets based on NLO, NLO+NLL$x$, and NNLO+NLL$x$
theory.
The stability of the perturbative expansion once small-$x$ resummation
effects are accounted for is evident.
Indeed, the differences between the central values of the NLO+NLL$x$
and NNLO+NLL$x$ gluon PDFs are always smaller than the corresponding
uncertainties in the entire range of $x$ considered.
We also observe that the central value of the small-$x$ gluon PDF is
larger in the NNL$x$ case as compared to fixed-order and that the
effective behaviour at small $x$ is a moderate rise rather than a
constant behaviour as at NLO.

In Table~\ref{tab:chi2} we report the values of $\chi^2/{N_{\rm dat}}$
for each of the LHCb $D$-meson datasets considered in this analysis.
For each of the three theory settings considered, we show the results
both before ($\chi^2_{\rm orig}$) and after ($\chi^2_{\rm new}$)
adding the LHCb data into the fit.
In the first column of this table, we also indicate the values of
$N_{\rm dat}$ for each dataset.
The results for the $N_{5+7+13}$ combination, corresponding to
$N_{\rm dat}=269$ data points, represent the baseline fit of this work.
For completeness, we also provide the $\chi^2$ values for the ratio
between the cross-sections at 13 and 5 TeV, $R_{13/5}$.

From Table~\ref{tab:chi2} one finds that an excellent description is
obtained for the normalised $D$-meson cross-sections, with similar
values of the $\chi^2/{N_{\rm dat}}$ for the three different theory
settings, and with the NLO+NLL$x$ fit leading to the smallest
$\chi^2_{\rm new}$.
As discussed in Ref.~\cite{Gauld:2016kpd}, in the calculation of these
$\chi^2$ values the experimental bin-by-bin correlation matrices are
included for $R_{13/5}$, while for the normalised cross-section data
these correlations (which are only available for $N_5$ and $N_{13}$)
are not included.

\renewcommand*{\arraystretch}{1.5}
\begin{table}[t]
	\centering
	\begin{tabular}{ r | c c | c c | c c @{}}
	 		& \multicolumn{2}{c|}{NLO} & \multicolumn{2}{c|}{NLO+NLL$x$} & \multicolumn{2}{c}{NNLO+NLL$x$} \\ \toprule
	 Dataset ($N_{\rm dat}$) &  $\chi^2_{\rm orig}/{N_{\rm dat}}$	& $\chi^2_{\rm new}/{N_{\rm dat}}$	& $\chi^2_{\rm orig}/{N_{\rm dat}}$	& $\chi^2_{\rm new}/{N_{\rm dat}}$	& $\chi^2_{\rm orig}/{N_{\rm dat}}$	& $\chi^2_{\rm new}/{N_{\rm dat}}$ \\
	 \hline
        $N_{5}~~(78)$ & 1.0 & 0.71 & 1.11 & 0.78 & 1.61 & 0.84\\
        $N_{7}~~(72)$ & 0.8 & 0.69 & 0.84 & 0.72 & 0.96 & 0.75\\
        $N_{13}~~(119)$ & 1.51 & 1.13 & 1.6 & 1.16 & 2.0 & 1.22\\ \midrule
        $N_{5+7+13}~~(269)$ & 1.17 & 0.89 & 1.25 & 0.93 & 1.61 & 0.98\\ \midrule
	 $R_{13/5}~~(99)$ & 1.64 & 1.66 & 1.87 & 1.79 & 1.83 & 1.74\\
         \bottomrule
	\end{tabular}
        \vspace{2.6mm}
        \caption{\small The values of the $\chi^2$ per
          data point, $\chi^2/{N_{\rm dat}}$, for each of the
          LHCb $D$-meson production datasets considered
          in this analysis.
          Each centre-of-mass energy contains the results of all the $D$-meson species
          considered.
          For each of the three theory settings,
          we show the results both before ($\chi^2_{\rm orig}$) and after ($\chi^2_{\rm new}$)
          adding the LHCb
          data into the fit.
          In the first column we also indicate the values
          of $N_{\rm dat}$ for each dataset.
          In this work the results based on the  $N_{5+7+13}$
          dataset are taken as the baseline.
        } \label{tab:chi2}
\end{table}

In this analysis the partonic cross-sections for charm-quark production,
convoluted with NLO and (N)NLO+NLL$x$ accurate PDFs, are accurate to
NLO in all cases.
In order to assess the robustness of these results with respect to this approximation, the analysis of 
the $N_{5+7+13}$ normalised cross-section dataset has been repeated in the following way.
A modified $\chi^2$ is introduced to quantify the agreement between
theory and data, defined as
\begin{align} \label{eq:chi2mod} 
\chi^2_{\rm mod} &= \sum_i
  \frac{\left( \, O_{i}^{\rm exp} - O_{i}^{\rm th}\right)^2}{(\delta
    O_i^{\rm exp})^2 + (\delta O_i^{\rm th})^2}\,,
\end{align}
where $O_i$ corresponds to the $i$-th bin value of the observable $O$
and $\delta O_i$ to its uncertainty.
Note that in our baseline analysis the theoretical uncertainty
$\delta O_i^{\rm th}$ is not accounted for.
This estimator is then used to repeat the NLO+NLL$x$ analysis, and in
this case we define the theory error $\delta O_i^{\rm th}$ to be the
following shift
\begin{align}
\delta O_i^{\rm th} \equiv O^{\rm evNLO}_i - O_i^{\rm evNLO+NLLx} \,,
\end{align}
where the cross-section $O^{\rm evNLO}_i$ is evaluated by evolving
upwards the NLO+NLL$x$ NNPDF3.1sx set from $Q_0 = 1.64$~GeV to
larger-$Q$ values using fixed-order NLO evolution.
The cross-section $O_i^{\rm evNLO+NLLx}$ is instead evaluated using
NLO+NLL$x$ settings for the evolution.
The same strategy can be applied to the NNLO+NLL$x$ case, where 
Eq.~(\ref{eq:chi2mod}) is written in terms of $O^{\rm evNNLO}_i$ and
$O_i^{\rm evNNLO+NLLx}$.

This additional source of theoretical uncertainty
$\delta O_i^{\rm th}$ is introduced in the $\chi^2$ in
Eq.~(\ref{eq:chi2mod}) in order to estimate the possible impact of the
missing contributions in the evaluation of the hard cross-sections for
charm production.
It effectively reduces the contribution to the $\chi^2$ due to those
bins that are most sensitive to the difference between (N)NLO+NLL$x$
and (N)NLO PDF evolution. Consequently, the weight of these bins is
reduced in the reweighting procedure.

\begin{figure}[t]
\centering
\includegraphics[width=.49\linewidth]{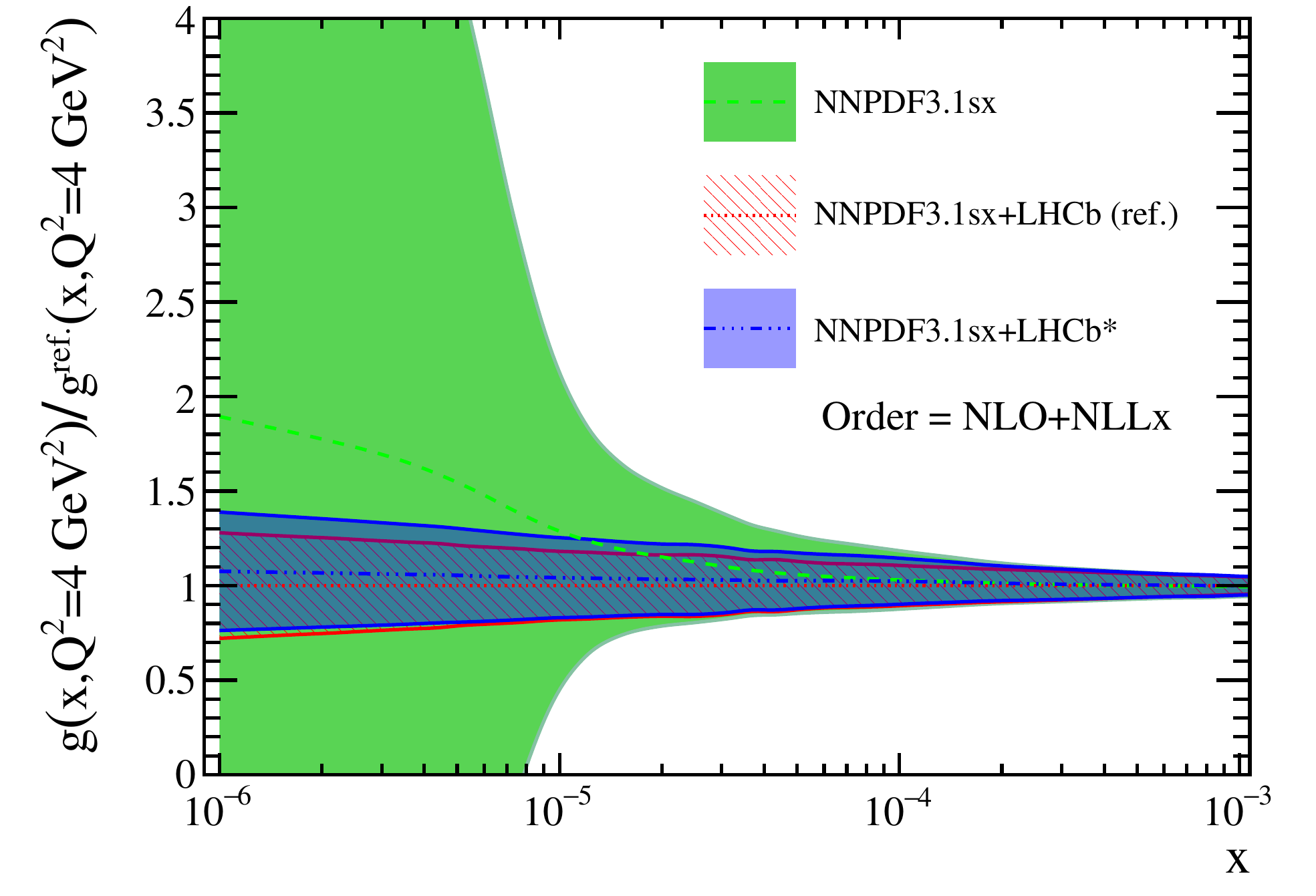}
\includegraphics[width=.49\linewidth]{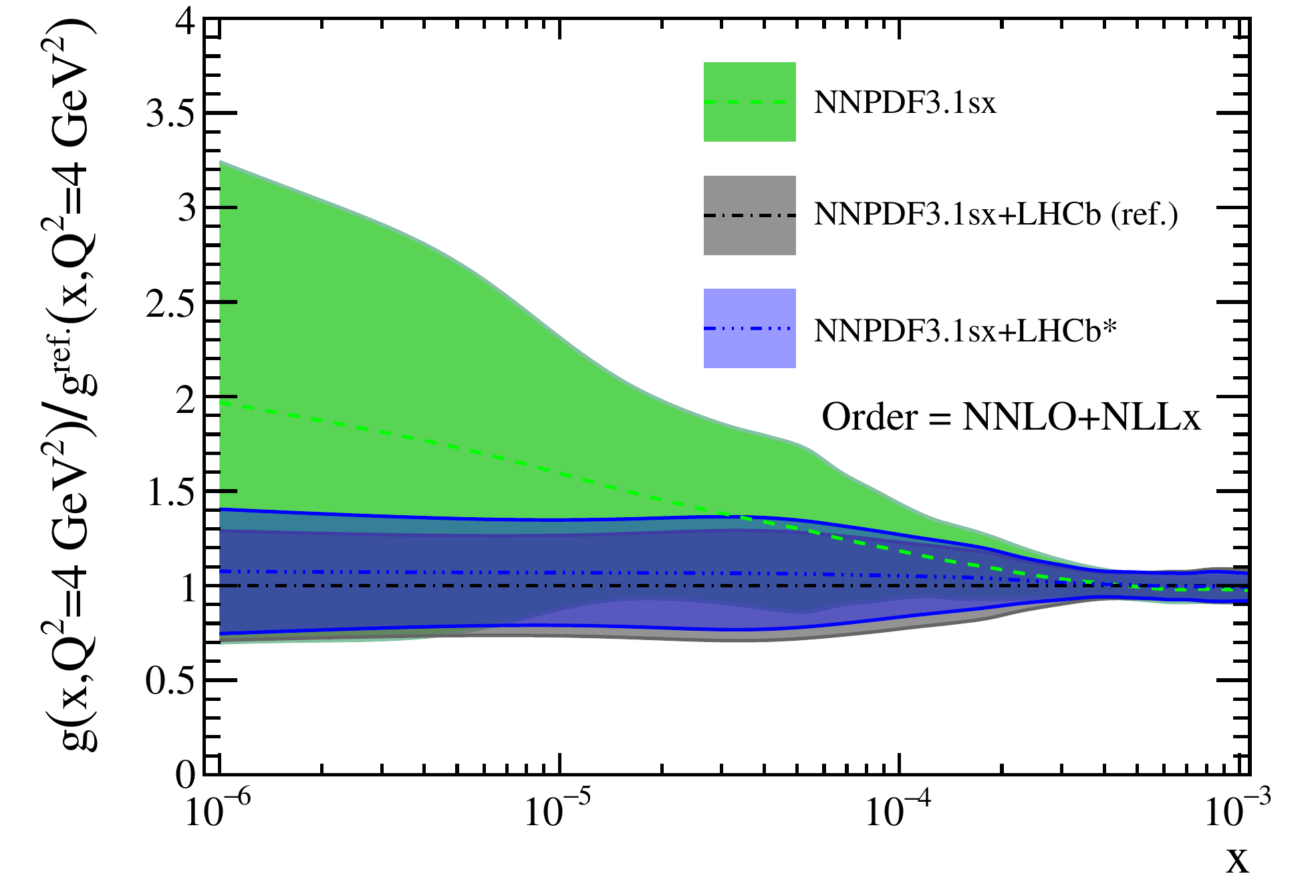} \\
\caption{\small The same comparison as in the upper
  plots in Fig.~\ref{fig:gPDF_rwgt}, now normalised to the central value
  of the NNPDF3.1sx+LHCb baseline result and adding the results
  (indicated by a {\bf *}) of the fits obtained using the
  modified definition for the $\chi^2$ in Eq.~\eqref{eq:chi2mod}.}
\label{fig:gPDF_rel_rwgt}
\end{figure}

In Fig.~\ref{fig:gPDF_rel_rwgt} we show the same comparison as in the
upper plots of Fig.~\ref{fig:gPDF_rwgt}, now normalised to the central
value of the NNPDF3.1sx+LHCb baseline result and adding the results
(indicated by a {\bf *}) of the fits obtained using the modified
definition for the $\chi^2$ in Eq.~\eqref{eq:chi2mod}.
One finds that adding the additional theory uncertainty in the
$\chi^2$ leads to slightly larger PDF uncertainties together with a
small positive shift of the central value, both at NLO+NLL$x$ and at
NNLO+NLL$x$.
These results suggest that missing NLL$x$ corrections in the prediction of the $D$-meson
production cross-sections can be neglected as compared to the
gluon PDF uncertainties, thus justifying the
approximations used in this work.

\begin{figure}[t]
\centering
\includegraphics[width=.49\linewidth]{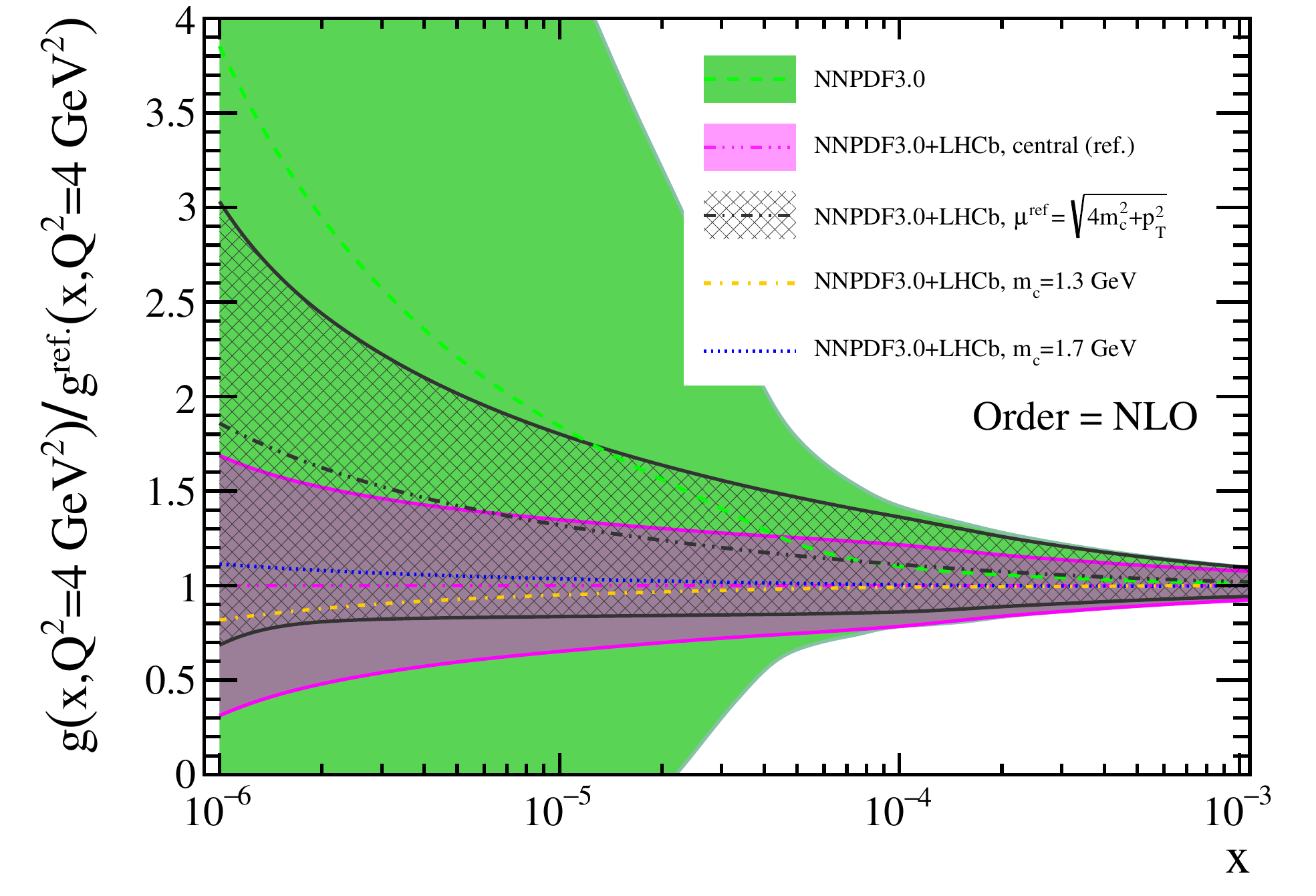}
\caption{\small Same as Fig.~\ref{fig:gPDF_rel_rwgt} for the
  NNPDF3.0+LHCb fit, comparing the baseline results with those
  obtained varying the scale $\mu^{(\rm ref)}$ and the value of the
  charm mass $m_c$.  }
\label{fig:gPDF_rel_rwgt_th}
\end{figure}

The results of the reweighting analysis are affected by further
theoretical uncertainties, such as the choice of renormalisation and
factorisation scales (set equal in this analysis) and of the value of
the charm-quark mass used in the calculation of Ref.~\cite{Gauld:2016kpd}.
For illustration purposes, we show here the impact of performing
scale and charm-quark mass variations in the determination of the gluon
PDF at small $x$.
As a baseline we take the NNPDF3.0+LHCb set of
Ref.~\cite{Gauld:2016kpd} restricting the input data to within
$p_T \in [1.0,8.0]$~GeV to match the current analysis.
This comparison is shown in Fig.~\ref{fig:gPDF_rel_rwgt_th}.
Specifically, we assess how the fit results change if the scale is
varied from the default $\mu^{\rm (ref)}=\sqrt{m_c^2+p_T^2}$ to
$\mu=\sqrt{4m_c^2+p_T^2}$ and if the charm-quark mass is varied from
the default $m_c=1.5$~GeV to 1.3 and 1.7~GeV.

From the comparison in Fig.~\ref{fig:gPDF_rel_rwgt_th} one observes
that in all cases the results of the fits varying either the scale or
the value of $m_c$ are consistent with the baseline within
uncertainties.
In particular, the effects of charm-quark mass variations are much smaller
than the PDF uncertainties, since they partially cancel out in the
ratios in Eq.~(\ref{eq:Obs}).
The impact of varying $\mu^{(\rm ref)}$ is more significant and
reaches the 1$\sigma$~level at $x\lsim 10^{-5}$.

\section{The neutrino-nucleon cross-section at ultra-high energies}
\label{sec:results}

In this section we present the main results of this work, namely the
predictions for the UHE (anti)neutrino-nucleon cross-section for CC
and NC scattering.
Firstly, the perturbative stability of our calculation is assessed by
studying the convergence of results obtained with different
theoretical accuracies.
We then provide a comparison of our baseline predictions to previous
calculations and discuss the origin of differences and similarities.
Finally, we compare our predictions to recent measurements from
IceCube~\cite{Aartsen:2017kpd} and assess the impact of nuclear
corrections following the strategy outlined in
Sect.~\ref{sec:nuclearPDFs}.

\subsection{Impact of theory settings and perturbative stability}

\begin{figure}[t]
\centering
\includegraphics[width=.49\linewidth]{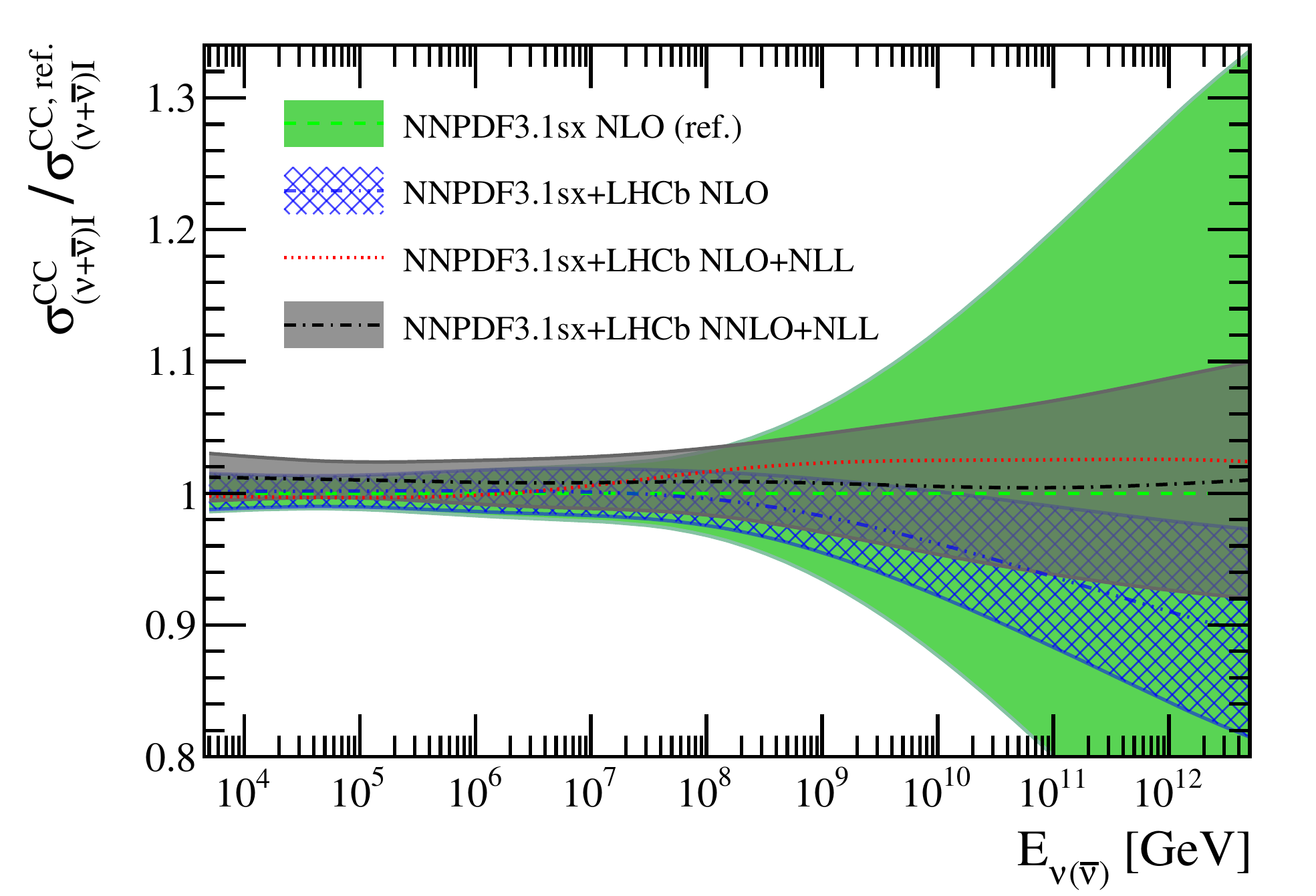}
\includegraphics[width=.49\linewidth]{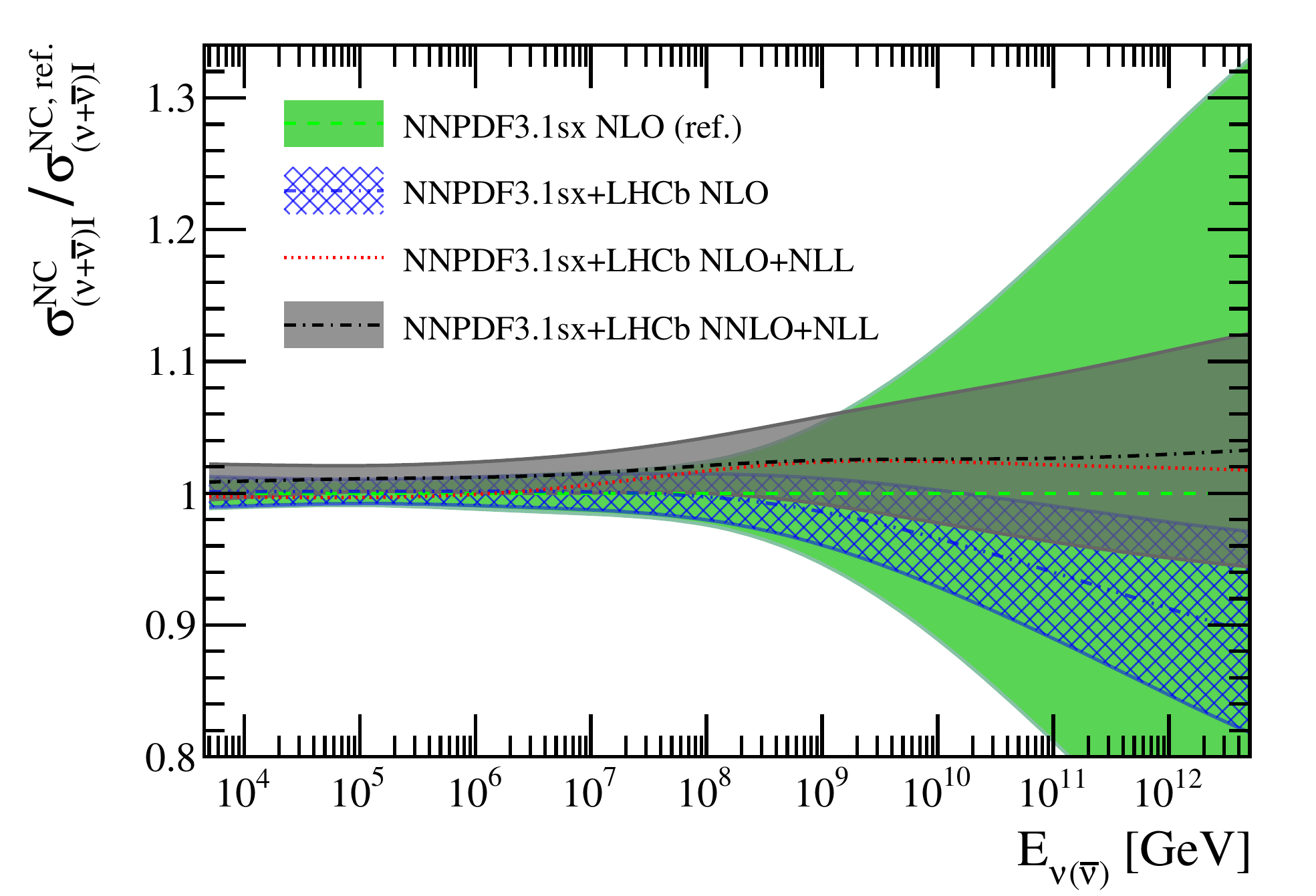}
\caption{\small The total CC (left) and NC (right plot) neutrino-isoscalar
  cross-section as a function of the energy $E_{\nu(\bar{\nu})}$ for
  different theoretical settings.
  Results are shown normalised to the central value of the NNPDF3.1sx NLO calculation
  and averaged over neutrino- and antineutrino-induced processes.
  The uncertainty bands correspond to the 1$\sigma$ PDF
  uncertainties.}
\label{fig:UHE_CC_ratio_Theory}
\end{figure}

In order to assess the perturbative stability of our calculation, we
compute the total cross-section in the $n_f=6$ FONLL scheme at NLO,
NLO+NLL$x$, and NNLO+NLL$x$ accuracy using the corresponding
NNPDF3.1sx+LHCb PDF sets presented in Sect.~\ref{sec:lhcb}.
The results are shown in Fig.~\ref{fig:UHE_CC_ratio_Theory}, where the
total cross-section for both the CC (left) and NC (right) scattering
is shown for the sum of neutrino- and antineutrino-induced processes, presented
as a function of the (anti)neutrino energy $E_{\nu(\bar{\nu})}$.
Predictions are normalised to the central value of those obtained with
the NNPDF3.1sx NLO set (\textit{i.e.} without the LHCb $D$-meson data)
and the quoted uncertainties indicate the 1$\sigma$ PDF uncertainty.
For the NLO+NLL$x$ predictions, we only show the central value, as the
results are almost identical to those obtained at NNLO+NLL$x$, both in
terms of central value and uncertainty.
Here we assume an isoscalar target without nuclear modifications.

Fig.~\ref{fig:UHE_CC_ratio_Theory} reveals that the impact of
the LHCb $D$-meson data is significant in the region
$E_{\nu(\bar{\nu})}\gsim 10^8$~GeV.
For instance, at $E_{\nu(\bar{\nu})}\simeq 10^{12}$~GeV the PDF
uncertainties decrease from about $30\%$ to below $10\%$.
The same qualitative behaviour is observed for both CC and NC
processes.
A further interesting observation is that the inclusion of the LHCb
$D$-meson data leads to a suppression of the central value of the
total UHE neutrino cross-section by around $10\%$ at the highest
energies.
Although this is shown here only for the NLO case, the LHCb data have
a similar impact when different theory settings, such as (N)NLO+NLL$x$, are adopted.
This is a consequence of the impact of the LHCb $D$-meson data on the behaviour
of the gluon PDF at small $x$. As shown in Fig.~\ref{fig:gPDF_rwgt}, the LHCb data
leads to relatively lower values of the gluon PDF in the small-$x$ region (irrespective of the theory settings).

Concerning the perturbative stability, Fig.~\ref{fig:UHE_CC_ratio_Theory} 
shows that the NNPDF3.1sx+LHCb predictions are consistent within the 
1$\sigma$ PDF uncertainties at all considered perturbative accuracies.
The central values of the NLO+NLL$x$ and NNLO+NLL$x$
calculations are remarkably consistent, in agreement
within 1\% for both CC and NC scattering
across the entire range of $E_{\nu(\bar{\nu})}$ values.
This difference is negligible as compared to PDFs uncertainties
and nuclear corrections (see Sect.~\ref{sec:nuclearPDFs}).

\begin{figure}[t]
\centering
\includegraphics[width=.49\linewidth]{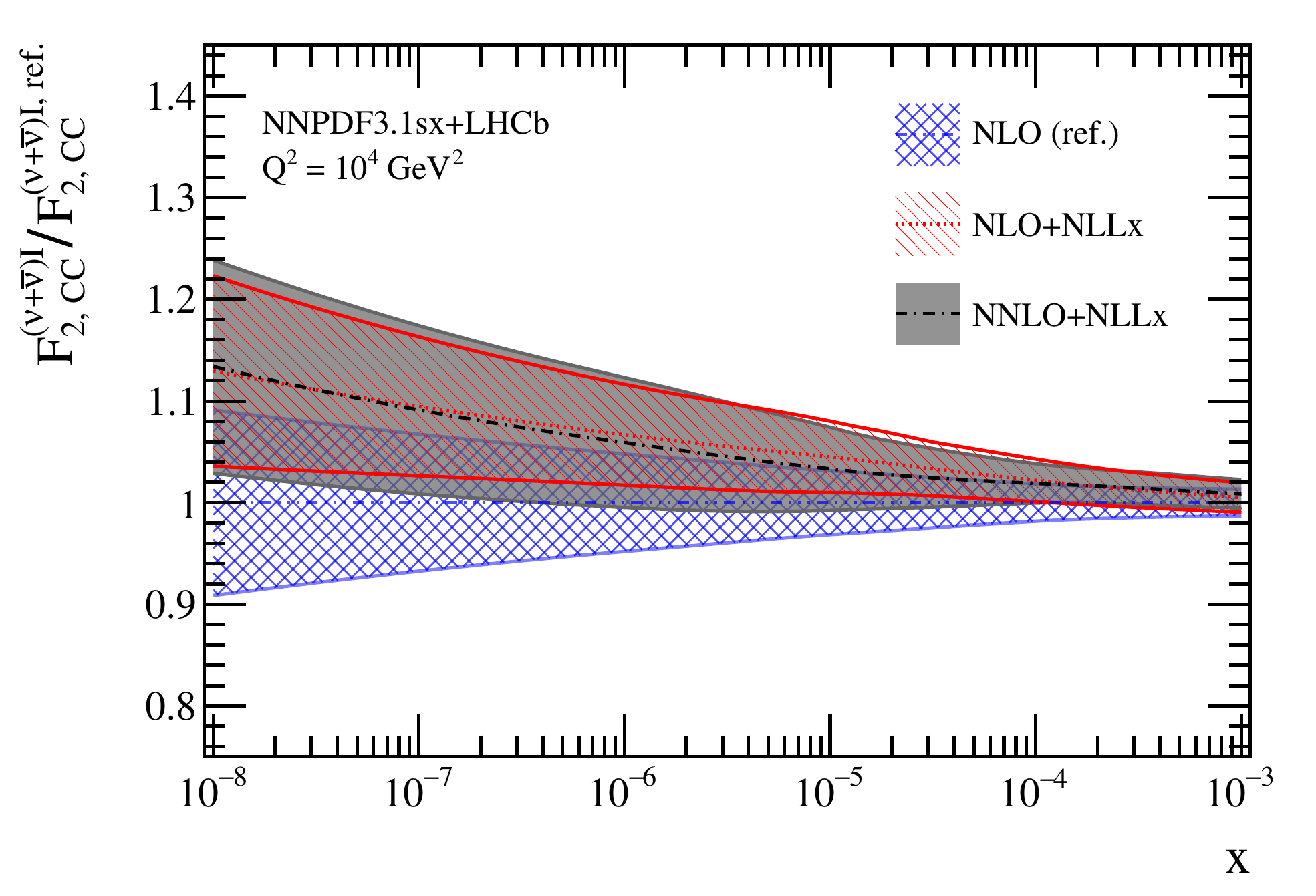}
\includegraphics[width=.49\linewidth]{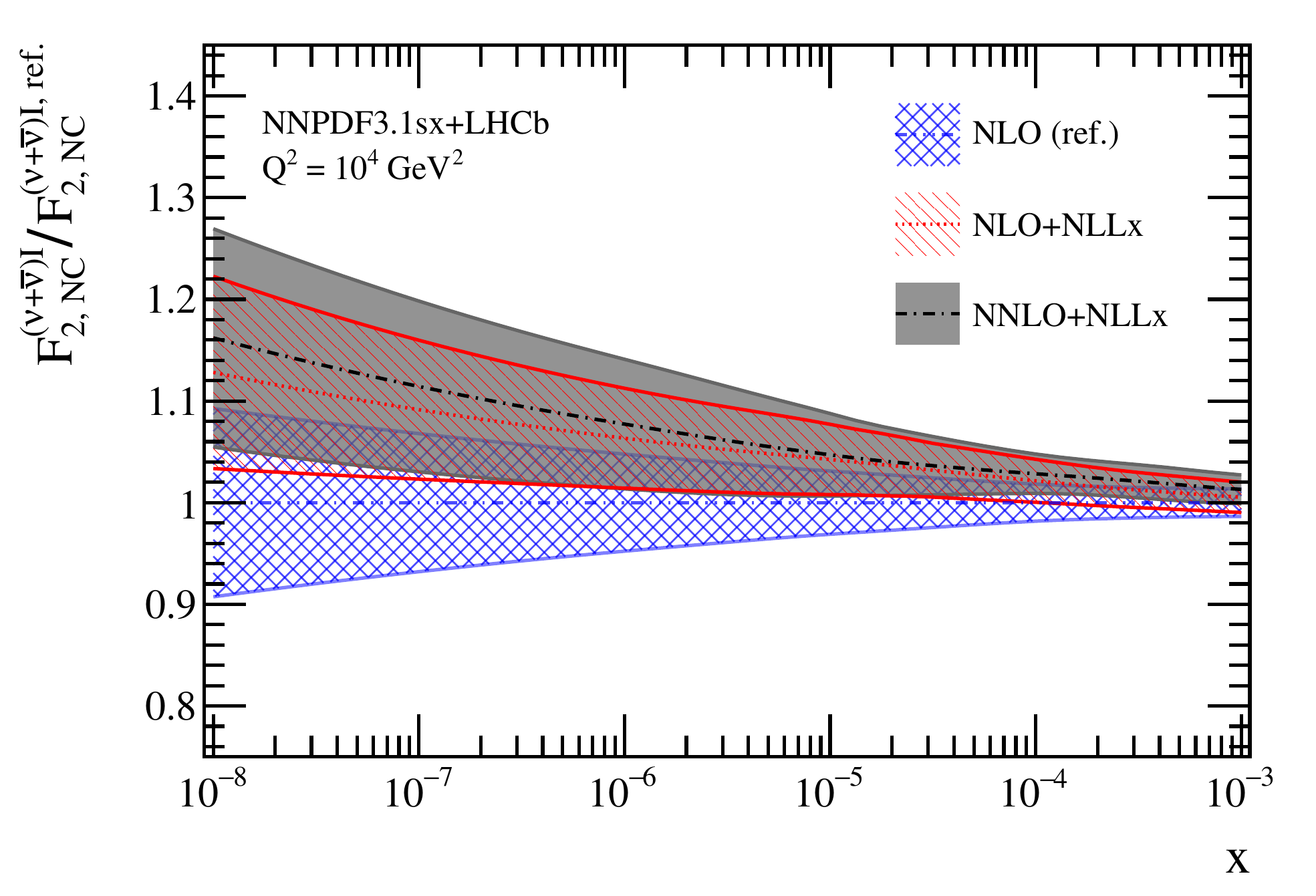}
\caption{\small The CC (left) and NC (right plot) structure function
  $F^{\nu+\bar{\nu}}_2(x,Q^2)$ computed for an isoscalar target
  at $Q^2 = 10^{4}$~GeV$^2$ with the NNPDF3.1sx+LHCb sets for three different
  theoretical settings: NLO, NLO+NLL$x$, and NNLO+NLL$x$.
  The structure functions are normalised
  to the central value the NLO calculation, with the
  bands indicating the $1\sigma$ PDF uncertainty.
}
\label{fig:F_smallx}
\end{figure}

The perturbative stability of the UHE cross-sections upon inclusion of
small-$x$ resummation effects is a direct consequence of the stability
of the structure functions.
To illustrate this point, in Fig.~\ref{fig:F_smallx} we show the
predictions for the $F^{\nu+\bar{\nu}}_2(x,Q^2)$ structure function
for both CC (left) and NC (right) scattering on an isoscalar target at
$Q = 100$~GeV, computed with the NNPDF3.1sx+LHCb sets for the three
different theoretical settings: NLO, NLO+NLL$x$, and NNLO+NLL$x$.
The central values of the NLO+NLL$x$ and NNLO+NLL$x$ structure
functions differ typically by around (1-2)\%, and 4\% at most for NC
scattering.
Differences between the fixed-order and the resummed calculations are
instead more pronounced and can be as large as $15\%$ at $x\simeq 10^{-8}$.
We also note that differences between the corresponding input PDF sets
for each of these three theoretical settings are typically larger (see Fig.~\ref{fig:gPDF_rwgt}).
Nonetheless, predictions for physical observables are in better
agreement due to the partial compensation between PDF evolution and
DIS coefficient functions.

The agreement between the NLO+NLL$x$ and NNLO+NLL$x$ predictions shown
in Figs.~\ref{fig:UHE_CC_ratio_Theory} and~\ref{fig:F_smallx}
demonstrates the excellent convergence of the perturbative expansion
at small $x$ after resummation effects are included. This indicates
that missing higher-order (MHO) corrections are likely to be small as
compared to other sources of theoretical uncertainty.
Altogether, we find that the combination of the constraints from the
LHCb $D$-meson data and the inclusion of NLL$x$ small-$x$ resummation
leads to robust predictions for CC and NC neutrino-nucleon
cross-section predictions up to the highest energies, with PDF
uncertainties below 10\% and negligible uncertainties due to MHO corrections.
At this level of precision, other sources of theoretical uncertainty,
such as nuclear corrections, cannot be neglected.

\subsection{Comparison to previous calculations} \label{sec:comparison}

As discussed in Sect.~\ref{sec:introduction}, predictions for the UHE
neutrino-nucleus cross-sections based on a variety of the different
theoretical setups have been provided in the past by a number of
groups~\cite{Gluck:1998js,CooperSarkar:2007cv,Gandhi:1995tf,Connolly:2011vc,CooperSarkar:2011pa,Gandhi:1998ri,Dicus:2001kb,Fiore:2005wf,Albacete:2015zra,Goncalves:2015fua,Arguelles:2015wba,Block:2013nia,JalilianMarian:2003wf}.
In the following, we provide a comparison of our results (labelled as BGR18) 
to a number of calculations of the UHE cross-sections present in the
literature.
The BGR18 predictions shown here correspond to the NNLO+NLL$x$
calculation in the $n_f=6$ FONLL scheme with the corresponding
NNPDF3.1sx+LHCb set as an input PDF set.
As nuclear corrections are absent from the selected benchmark
calculations, for the purpose of comparison we do not include them in our predictions.

\begin{figure}[t]
\centering
\includegraphics[width=.49\linewidth]{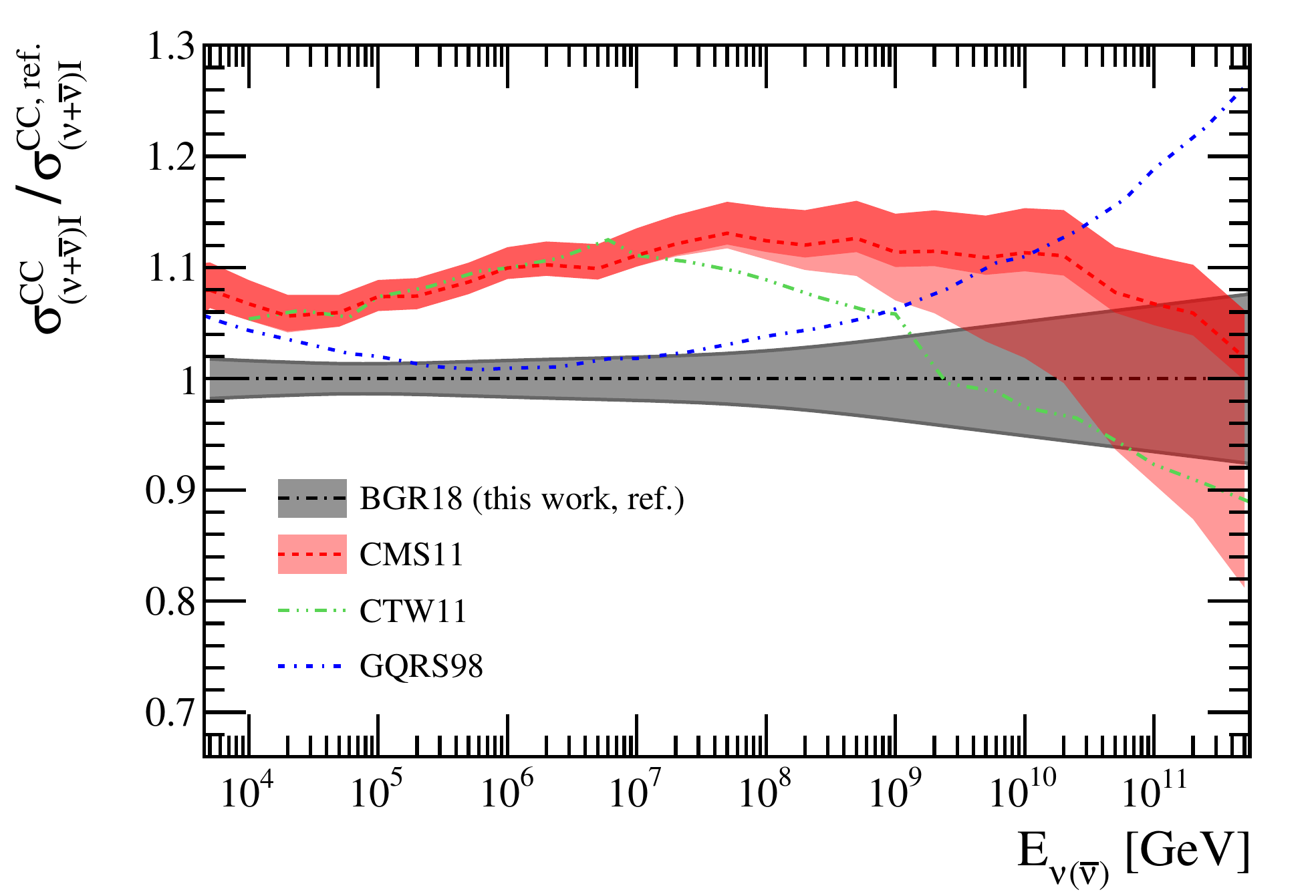}
\includegraphics[width=.49\linewidth]{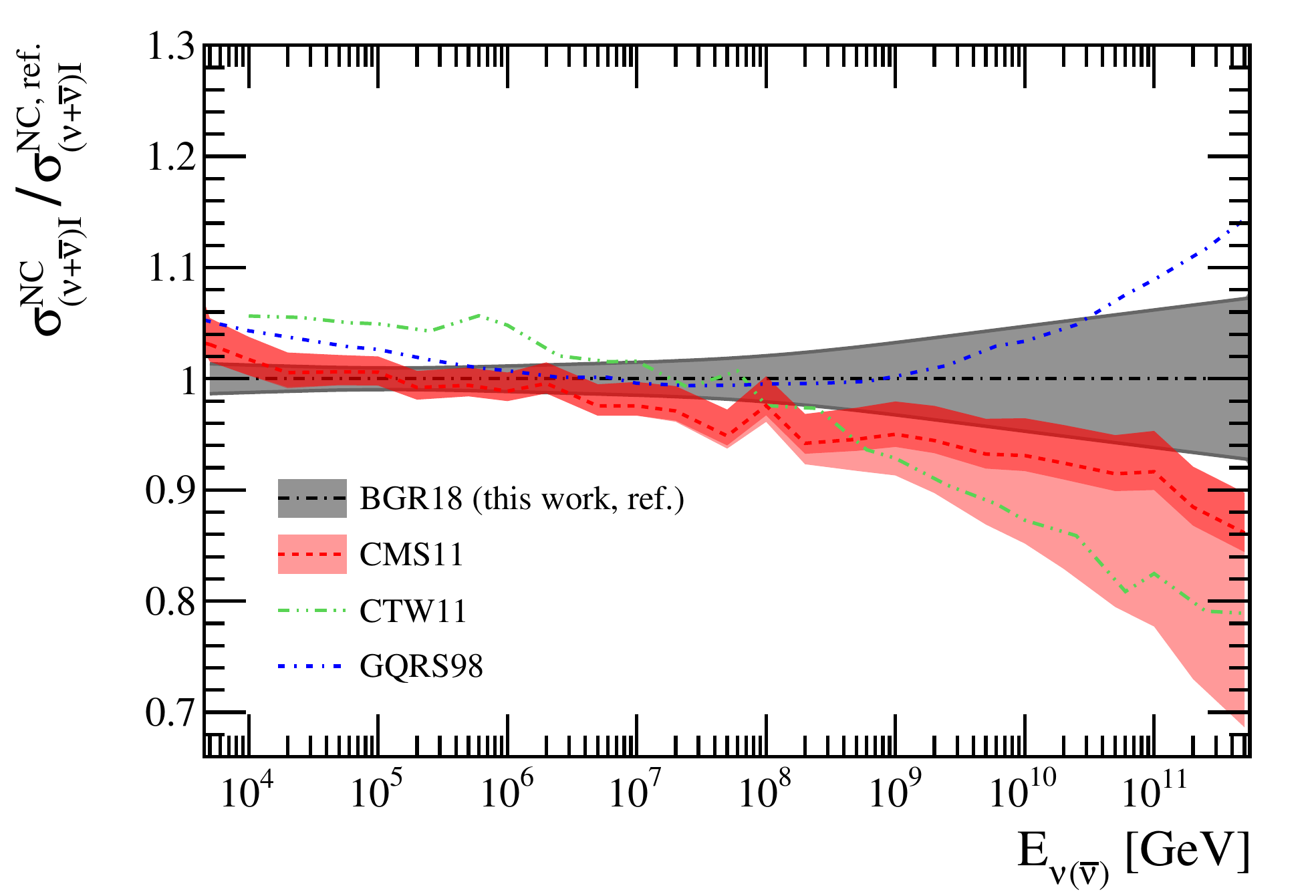}
\caption{\small Comparison of the results of the present work (BGR18) with
  previous calculations of the UHE neutrino-isoscalar cross-sections as a 
  function of $E_{\nu(\bar{\nu})}$ for CC (left) and NC (right plot)
  scattering, normalised to the central BGR18 value.
  The BGR18 predictions are provided in the $n_f=6$ FONLL scheme at NNLO+NLL$x$ accuracy with the
  corresponding NNPDF3.1sx+LHCb PDF set,
  and are compared to the GQRS98, CMS11, and CTW11 calculations.
  For both BGR18 and CMS11 the bands indicate the PDF uncertainty and
  in the CMS11 case the dark and light bands correspond to two
  alternative treatments of the PDF uncertainty, see text.}
\label{fig:UHE_xsec_comparison}
\end{figure}

We begin by comparing our results to the calculations from Gandhi {\it
  et al.}  (GQRS98)~\cite{Gandhi:1998ri}, Connolly, Thorne, and Waters
(CTW11)~\cite{Connolly:2011vc}, and Cooper-Sarkar, Mertsch, and Sarkar
(CMS11)~\cite{CooperSarkar:2011pa}, all of which have been obtained in
the framework of collinear factorisation. The GQRS98 and CTW11
predictions are obtained with LO-accurate coefficient functions with
CTEQ4M~\cite{Lai:1996mg} and MSTW08~\cite{Martin:2009iq} PDFs
respectively, while the CMS11 calculation is performed at NLO with the
HERAPDF1.5 PDFs~\cite{CooperSarkar:2010wm}.
The comparison for the sum of neutrino and antineutrino cross-sections
is presented in Fig.~\ref{fig:UHE_xsec_comparison} for both the CC
(left) and NC (right) processes.
For GQRS98 and CTW11 we only show the central values, while for BGR18
and CMS11 we also include the PDF uncertainties.
In the case of the CMS11 predictions we have added the central values
of the neutrino and antineutrino induced cross-sections, and show the relative
uncertainty of the neutrino induced process. The lighter and darker uncertainty bands 
of this prediction correspond to different prescriptions for estimating the PDF uncertainties. 
The darker band is obtained by excluding one particular member of the
HERAPDF1.5 PDF set.
This member dominates in the computation of the PDF uncertainty at
small $x$ and thus removing it results in a much smaller uncertainty
band, see Ref.~\cite{CooperSarkar:2011pa} for more details.

In general, there are marked differences between the BGR18 and the other calculations.
For the case of CC scattering, we find that the BGR18 and GQRS98
predictions are in agreement in the region
$10^5\,{\rm GeV}\lsim E_{\nu(\bar{\nu})}\lsim 10^{8}\,{\rm~GeV}$, but
differ significantly outside this range.
At $E_{\nu(\bar{\nu})}\simeq 10^{12}$~GeV the GQRS98 calculation is
larger by around $30\%$, which corresponds to a deviation of more than
3$\sigma$ in units of the PDF uncertainty of the BGR18 calculation.
The origin of this difference can be understood by considering that the CTEQ4M PDF
set used in the GQRS98 calculation must be extrapolated beyond its region
of validity, $x\in[10^{-5},1]$. Applying the extrapolation adopted in Ref.~\cite{Gandhi:1998ri}, we find 
the CTEQ4M gluon and quark PDFs evaluated at $x\simeq 10^{-7}$ and $Q\sim100$~GeV overshoot 
those of NNPDF3.1sx+LHCb by around 25\%.
While the CMS11 and CTW11 calculations are broadly
consistent with one another, we find that at intermediate energies
these predictions are approximately (8-10)\% larger than BGR18.
In the case of the CTW11 prediction, this can be partly attributed to the absence
of the $\mathcal{O}(\alpha_s)$ corrections to the coefficient functions, which are negative
and amount to (4-5)\% for $E_{\nu(\bar{\nu})}\in [10^4,10^8]$~GeV.
The origin of the difference with respect to the CMS11 prediction (which includes these corrections)  
is instead likely to originate from the treatment of top-quark production. Inspection of Fig.~11 (left) 
of Ref.~\cite{CooperSarkar:2011pa} suggests that the contribution to
the total cross-section from $b$-quark initiated diagrams (top-quark
production) amount to +10\% at $E_{\nu} = 10^6$~GeV. This calculation is performed in the ZM-VFNS where
heavy-quark mass effects are absent.
As discussed in Sect.~\ref{sec:heavyquarks}, our calculation of top-quark production for CC scattering 
includes heavy-quark mass correction to NLO. We find that at this energy the relative contribution of top-quark
production is below 1\% (see Fig.~\ref{fig:UHE_topquarkmass}).
At high (anti)neutrino energies, $E_{\nu(\bar{\nu})}\simeq 10^{11}$~GeV, the three calculations are
instead consistent within the BGR18 uncertainties, although the CTW11
central value is suppressed by around 10\% as compared to BGR18.

In the NC scattering case, the GQRS98 calculation agrees with BGR18 at
intermediate values of $E_{\nu(\bar{\nu})}$ but overshoots it at higher and lower energies.
The same behaviour was observed for CC scattering and these differences can again 
be primarily attributed to the behaviour of the input PDFs (see the discussion above).
We find reasonable agreement between BGR18 and the CMS11 and CTW11 results for NC
scattering. For the highest values of neutrino energies, both of these predictions tend to undershoot
the BGR18 predictions.
For instance, the CTW11 predictions are suppressed by around a factor
20\% at $E_{\nu}\simeq 10^{12}$ GeV as compared to the BGR18
calculation. This can be partly traced back to differences at the level of input
PDFs. However, when the PDF uncertainties of the CTW11 are accounted for (see Ref.~\cite{Connolly:2011vc}),
this behaviour is not significant.

\begin{figure}[t]
\centering
\includegraphics[width=.49\linewidth]{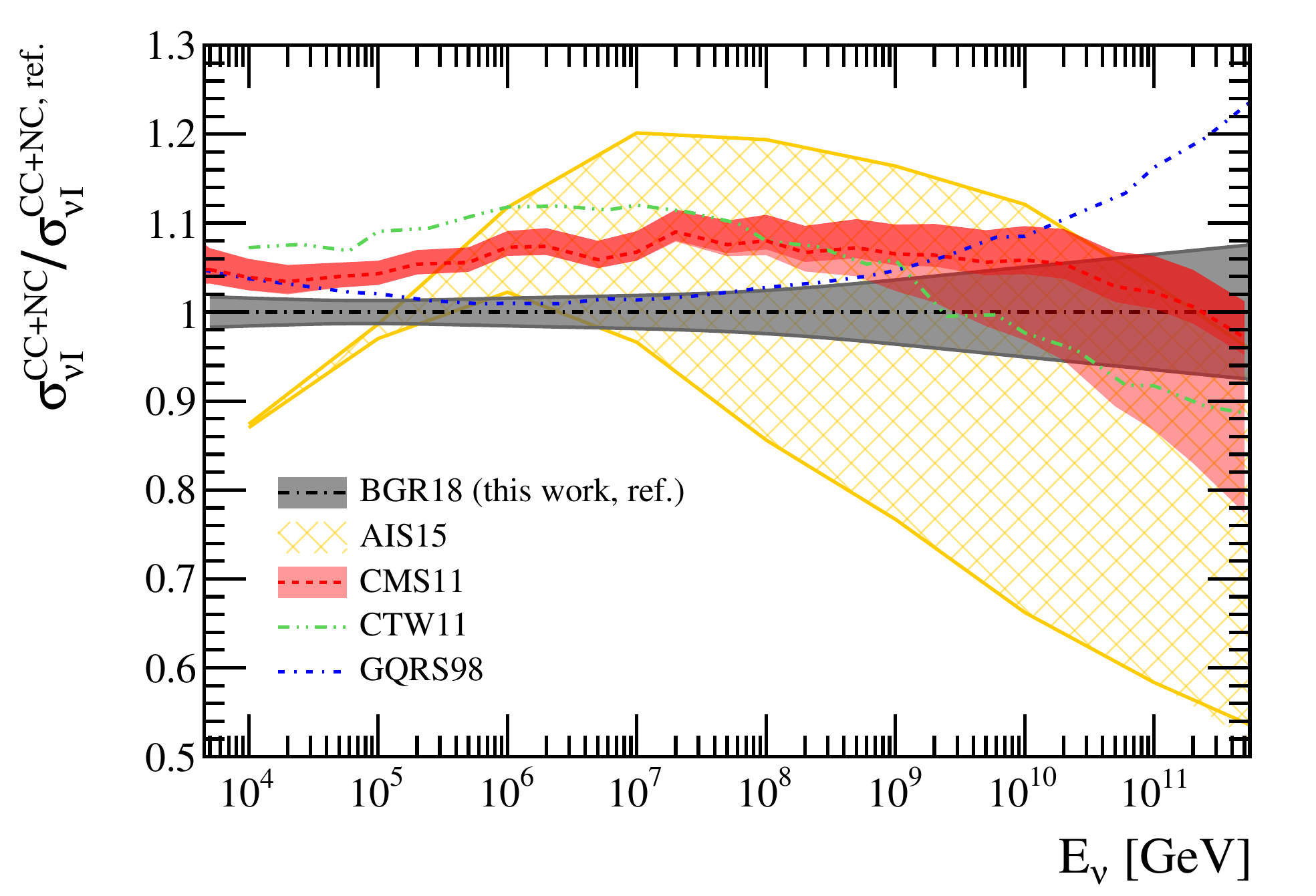}
\caption{\small Same as Fig.~(\ref{fig:UHE_xsec_comparison}), now for
  the sum of NC and CC neutrino-isoscalar cross-sections. In addition,
  the AIS15 calculation based upon the non-linear QCD (saturation)
  framework is included.}
\label{fig:UHE_xsec_comparison_tot}
\end{figure}

The calculations displayed in Fig.~\ref{fig:UHE_xsec_comparison} are
all based on collinear factorisation.
In order to assess the sensitivity of the UHE cross-sections to other
QCD theoretical frameworks, in Fig.~\ref{fig:UHE_xsec_comparison_tot}
we compare the total cross-section for neutrino-induced scattering
(the sum of CC and NC processes) with the predictions from Albacete
{\it et al.} (AIS15)~\cite{Albacete:2015zra}. This calculation is based on the
Balitsky-Kovchegov equation with running coupling, which incorporates
non-linear effects from gluon recombination (saturation).

In the AIS15 framework, the steep growth of the PDFs at small $x$ is
tamed by non-linear effects, leading to a suppression of the UHE
neutrino-nucleon cross-section. Indeed,
Fig.~\ref{fig:UHE_xsec_comparison_tot} shows that at the highest
energies the AIS15 calculation is in general suppressed by about $25\%$ as
compared to the BGR18 result.
However, for large neutrino energies the AIS15 prediction is affected by a large
theoretical uncertainty arising from the limited information on some
of the input parameters that enter the calculation.
Given the small theory uncertainties of the BGR18 calculation, a
possible suppression of the measured UHE cross-sections at high
energies as compared to our predictions may indicate the onset of
saturation effects.

\subsection{Comparison to IceCube data and impact of nuclear corrections}

\begin{figure}
\centering
\includegraphics[width=.49\linewidth]{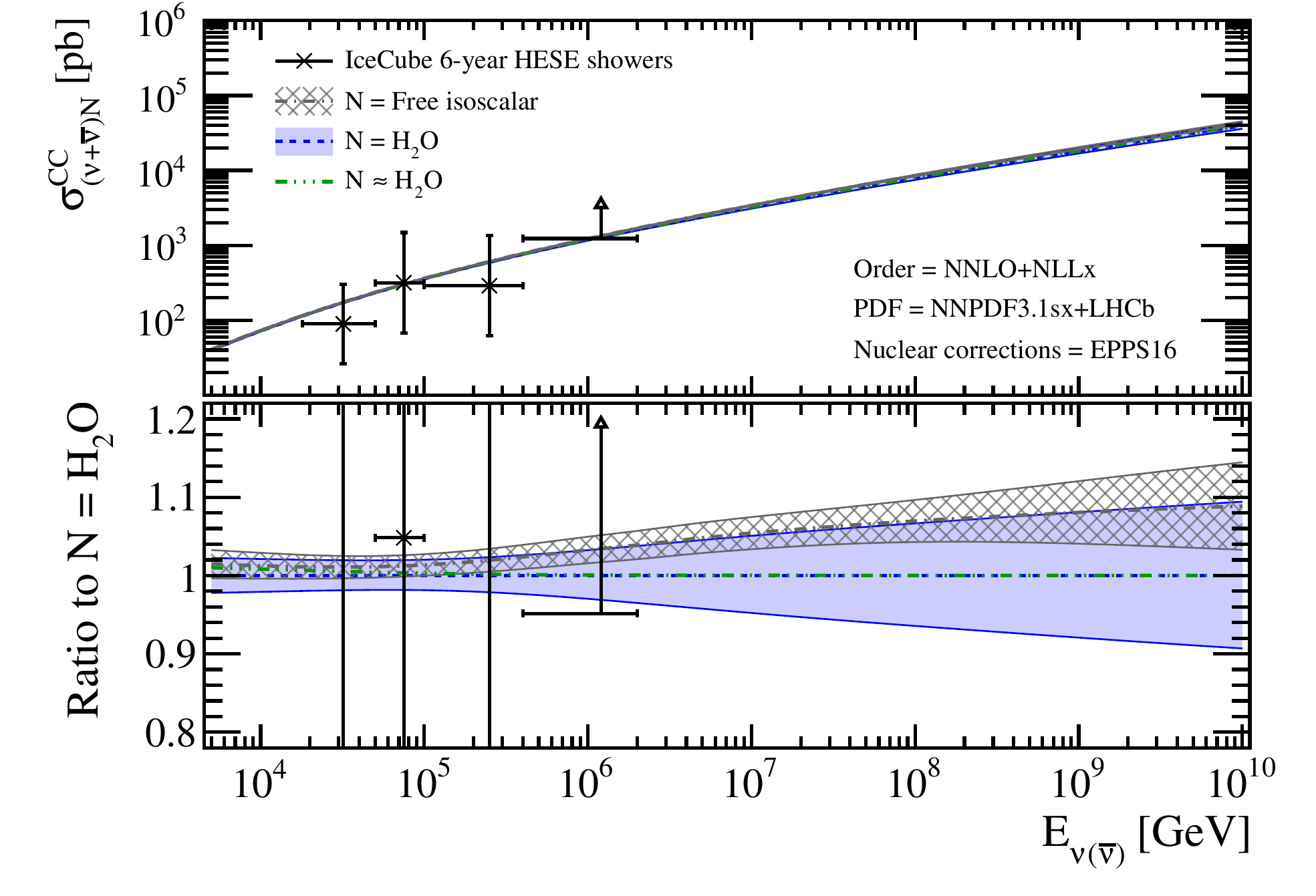}
\includegraphics[width=.49\linewidth]{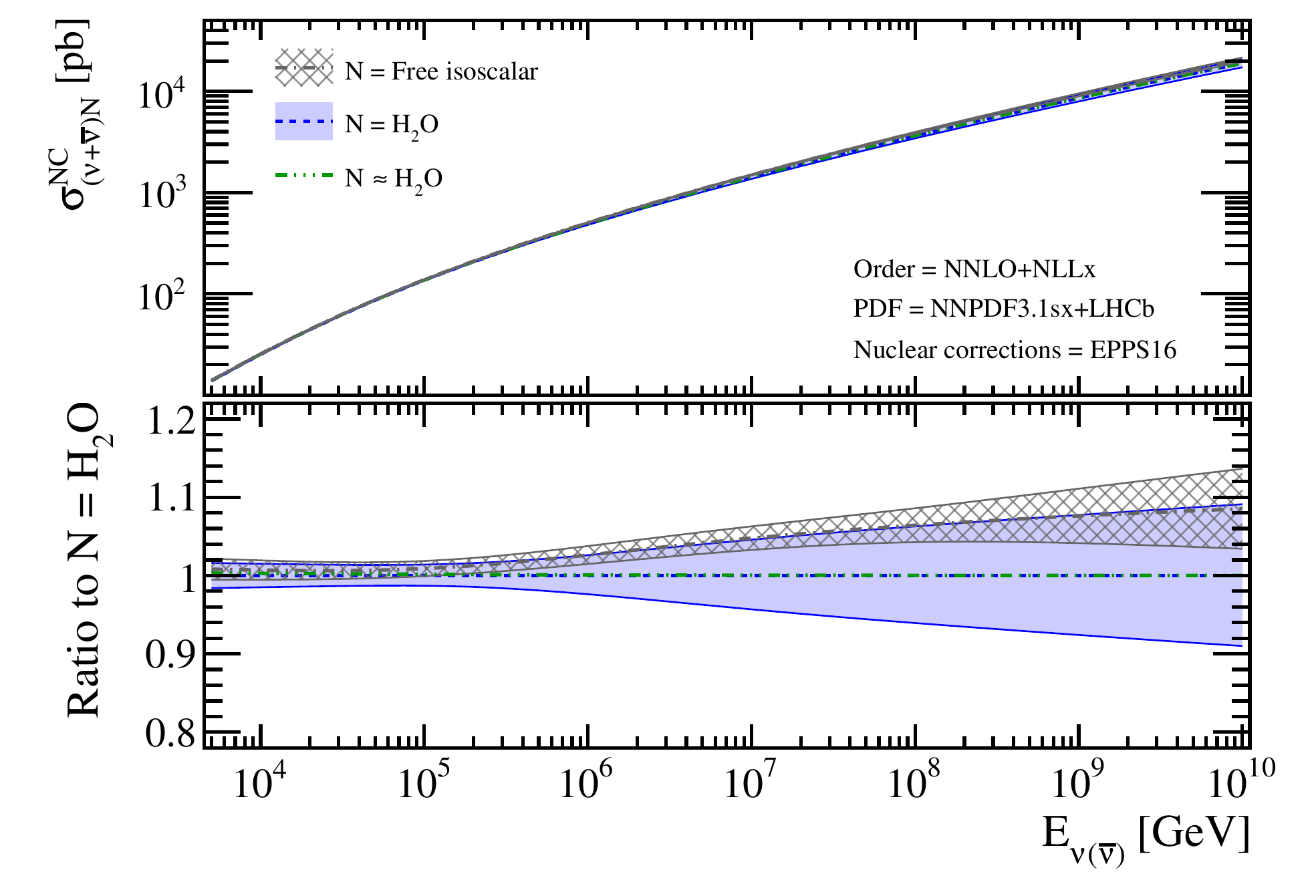}
\caption{\small 
The BGR18 predictions in the $n_f=6$ FONLL scheme at NNLO+NLL$x$ accuracy
as a function of $E_{\nu}$ in the CC (left)
and NC (right plot) cases for two different targets:
a free isoscalar nucleon (without nuclear modifications) and an
H$_2$O target (with the EPPS16 nuclear modifications).
In the lower panel, we show the same results normalised to the central value
of the BGR18 calculation for an H$_2$O target.
In the CC case, we also compare our predictions with recent IceCube
measurements based on the 6-year HESE shower
dataset~\cite{Aartsen:2017kpd}.  }
\label{fig:IceCube}
\end{figure}

To conclude the discussion of our results, in Fig.~\ref{fig:IceCube}
we show predictions for the sum of neutrino and antineutrino scattering
on different targets.
We consider the case of a free isoscalar target and that of H$_2$O
molecule, where nuclear modifications have been evaluated using the
EPPS16 nPDF set (see Sect.~\ref{sec:nuclearPDFs}).  Results are also
shown for an ``isoscalar'' H$_2$O molecule ($N \approx$ H$_2$O),
according to Eq.~\eqref{eq:nPDF_UHE}.
These predictions are obtained from the baseline NNLO+NLL$x$ accurate
results in the $n_f=6$ FONLL scheme and are 
presented as a function of $E_{\nu}$ for both CC (left) and NC (right) processes.
The upper panels of Fig.~\ref{fig:IceCube} display the absolute
cross-section, while in the lower panels the predictions are shown
normalised to the central value of the BGR18 calculation on H$_2$O.
In the CC case, we also show the recent IceCube measurements 
based on the 6-year HESE (high-energy showers) dataset~\cite{Aartsen:2017kpd}.
This dataset is based on high-energy starting events (or
contained-vertex events). The experimental uncertainties are the sum
in quadrature of the statistical and systematic errors.
Note that for the rightmost data point only a lower limit on the
magnitude of the cross-section can be derived.

The results in Fig.~\ref{fig:IceCube} highlight that, given the current
precision of free-nucleon calculations, effects due to the nuclear
modifications of the PDFs of nucleons bound inside a H$_2$O 
molecule are significant and cannot be neglected.
As already discussed in Sect.~\ref{sec:nuclearPDFs}, nuclear
modifications induce a suppression of the UHE cross-section by up to
10\% as compared to the free-nucleon case.
However, uncertainties associated to these nuclear modifications are large
and currently dominate over PDF and other theoretical uncertainties.
The fact that nuclear modifications are now the dominant source of
theoretical uncertainty on the UHE neutrino cross-section predictions 
provides a strong motivation to improve on the knowledge of nPDFs.
This goal could be achieved by exploiting the LHC $p$+$Pb$ collision data as
well as, in the near future, data collected at the EIC.
It is worth noting that the prediction for $N\approx$ H$_2$O provides an excellent
approximation for $E_{\nu}$ values. The difference between the full and approximate 
results at the lowest $E_{\nu}$ values is $\approx1\%$.

Focussing now on the CC scattering cross-section of Fig.~\ref{fig:IceCube}, 
we observe that the IceCube measurements extend up to neutrino 
energies of around $E_{\nu}\simeq 10^6$ GeV.
At the current level of precision, these measurements cannot discriminate
between the different theoretical predictions.
Nonetheless, future data based on a much larger sample from neutrino
telescopes, such as IceCube and KM3NET, and from other experiments
sensitive to very high-energy neutrinos should improve the
precision of the current measurements at intermediate energies and 
extend the measurement to larger $E_{\nu}$ values.
As discussed in Sect.~\ref{sec:comparison}, the different theoretical
predictions give rise to small differences at intermediate energies. The origin
of these differences are well understood, and are related to the perturbative accuracy
of the calculation.
A cross-section measurement at higher energies would instead probe QCD in the very
small-$x$ region providing a test of the assumptions of non-perturbative
information related to the distribution of quarks and gluons within bound nuclei, and may also 
test for the presence of saturation effects.

\section{Summary and outlook}
\label{sec:summary}

In this work, we have presented state-of-the-art predictions for the
cross-sections of high-energy neutrino scattering on nucleons, with particular
attention to a target material composed of H$_2$O molecules.
With respect to previous calculations, we have made a number of 
major improvements.

Firstly, we have extended the calculation of deep-inelastic structure
functions to NNLO matched with small-$x$ resummation corrections at
next-to-leading logarithmic (NLL$x$) accuracy.
We find that small-$x$ resummation corrections stabilise
the perturbative convergence of the UHE neutrino cross-sections. This
feature is highlighted by the similarity of the NLO+NLL$x$ and NNLO+NLL$x$ calculations.

The second main improvement is the inclusion of the $D$-meson
production data from LHCb in the determination of the PDF sets used in
our calculations. This dataset imposes a stringent constraint on the
PDFs at small $x$, with a consequent significant impact on the UHE neutrino cross-section.
This was achieved by producing dedicated PDF sets based on the
NNPDF3.1sx sets and including the LHCb data by means of Bayesian
reweighting. Due to the unavailability of small-$x$ resummation corrections
to the partonic cross-section for $D$-meson production, predictions for this process 
have instead been obtained at NLO+PS accuracy convoluted with PDFs determined
including small-$x$ resummation effects.
We have demonstrated that the LHCb measurements lead to a reduction of
the uncertainties on the UHE neutrino cross-sections related to PDFs
by up to a factor three.

As compared to previous studies, we have introduced further
improvements in our calculations.
One of these is the inclusion of charm-, bottom-, and top-quark mass
effects using the FONLL general-mass scheme for both CC and NC
structure functions.
We have also provided an estimate of the impact
on the UHE cross-sections related to nuclear effects.
We find that these corrections can be as large as $10\%$ at high
neutrino energies but are also affected by large uncertainties, which
in turn impact the quoted precision of the calculation.

We found notable differences between our predictions and previous
benchmark calculations, which can be traced back to differences in both the perturbative
and non-perturbative inputs to the calculation. For example, in the energy range of
$E_{\nu(\bar{\nu})} \in [6\times10^3,10^5]$~GeV we find a prediction which is $\approx5\%$
lower than the quoted SM prediction in the IceCube measurement~\cite{Aartsen:2017kpd}.
When more precise cross-section measurements are obtained at neutrino telescopes,
these effects, as well as a reliable estimate of the nuclear corrections, will eventually become 
relevant to interpret the data. For the moment, the recent data from IceCube are still affected
by large uncertainties and do not extend to large enough energies to discriminate between 
the different predictions.

We foresee that our calculations could be improved in at least two ways.
Firstly, we found that uncertainties attributed to nuclear effects
represent one of the dominant sources of theoretical uncertainty.
This uncertainty could be reduced by including in the nPDF fits measurements from
the LHC in $p$+$Pb$ collisions to constrain the distributions in small-$x$ region.
Secondly, small-$x$ resummation effects and NNLO corrections, once available, should also
be included in the partonic cross-section for $D$-meson production.
Our study indicates that these corrections are likely to be
small at the level of a normalised cross-section, but at the level of precision 
that the calculation has achieved it might be necessary to account for them.

To summarise, our analysis demonstrates how measurements of the
neutrino-nucleus cross-section will represent a unique probe to test
the strong interaction in an extreme regime where new dynamics are
expected to arise, such as BFKL or non-linear effects.
In this context, our calculations provide a robust building block for
the data analysis and interpretation of neutrino telescopes in the
coming years.
Our results should also be relevant for other phenomenological
applications, for instance to compute the attenuation of the
high-energy neutrino flux as they pass through the
Earth~\cite{Vincent:2017svp}.

\vspace{0.5cm}

\noindent
{\bf Acknowledgments}
\vspace{0.1cm}

\noindent
We are grateful to Aart Heijboer and Alfonso Garc\'ia for discussions
about the KM3NET experiment, to Marco Bonvini for discussions regarding
HELL, to Subir Sarkar for information about the CMS11 calculation, and
to Javier Albacete and Alba Soto for providing us with the AIS15 results.
We thank Hannu Paukkunen and Aleksander Kusina for providing us with
the EPPS16 and nCTEQ15 Oxygen nPDF grids.
We thank Rabah Abdul Khalek and Luca Rottoli for providing cross-checks 
relevant for some of the results presented in this paper.
V.~B. and J.~R. are supported by an European Research Council Starting
Grant ``PDF4BSM''.
J.~R. is also partially supported by the Netherlands Organization for
Scientific Research (NWO).
The research of R.~G. is funded by the ERC Advanced Grant ``MC@NNLO''
(340983).

\appendix
\section{The BGR18 UHE neutrino-nucleus cross-sections}
\label{sec:UHExsec}

In this appendix, we discuss the delivery of the results presented in
this paper. We provide a tabulation of the total UHE cross-sections
for a range of $E_{\nu}$ values together with the corresponding
theoretical uncertainties.
Results are obtained with our baseline theory settings, \textit{i.e.}
the FONLL scheme at NNLO with a maximum of $n_f=6$ active flavours
augmented by small-$x$ resummation corrections to NLL$x$ accuracy, and
based on the NNPDF3.1sx+LHCb NNLO+NLL$x$ PDF set.

\paragraph{Structure functions.}
In order to allow the users to reproduce our results for the
double-differential cross-sections, we provide predictions for the
neutrino structure functions $F_2$, $xF_3$, and $F_L$ both for NC and
CC scattering.
Structure functions are made available in the form of interpolation
grids in the {\tt LHAPDF6} format~\cite{Buckley:2014ana}.
Grids for NC and CC, neutrino and antineutrino structure functions are
provided for a free isoscalar target.
The grids contain the $N_{\rm rep}=40$ Monte Carlo replicas resulting
from the PDF reweighting analysis.\footnote{Note that some of the
  replicas are equivalent, as a consequence of the unweighting
  procedure~\cite{Ball:2011gg}.}
Mean and standard deviation of a structure function $F$ are obtained
according to
\begin{align}
  \langle F \rangle_{\rm rep} = \frac{1}{ N_{\rm rep}}\sum_{k=1}^{N_{\rm  rep}} F^{(k)} \,, \qquad
  \delta F = \sqrt{  \frac{\sum_{k=1}^{N_{\rm rep}}( \langle F \rangle_{\rm rep}
  - F^{(k)})^2}{N_{\rm rep}-1}} \,,
\label{eq:mcestimators}
\end{align}
where $F^{(k)}$ is the value of the structure function computed with
the $k$-th Monte Carlo replica.

\paragraph{Total cross-sections.}
The structure function grids discussed above allow constructing the
double-differential UHE neutrino cross-section with the associated PDF
uncertainties.
Predictions for the total cross-section can then be obtained by
integrating the double-differential cross-section by means of
Eq.~\eqref{eq:integrate}.

While in Eq.~(\ref{eq:d2sig}) we presented the explicit formula for
the CC cross-section, here we discuss in more detail the NC
cross-section. Throughout this paper we have implicitly assumed that
in our computations higher-order electroweak (EW) corrections can be
neglected. However, for the computation of the NC cross-section we
find advantageous to employ an ``improved'' scheme that includes part
of the higher-order EW corrections. A pure LO treatment of the EW
effects entails relations between the relevant parameters, such as
$G_F$, $M_W$, $M_Z$ and $\sin^2\theta_W$, that lead some of them to be
in strong disagreement with the measured values.
In order to overcome this limitation, we employ the prescription of
Ref.~\cite{Denner:1991kt} to include in our computation the leading
universal corrections, so that:
\begin{align}\label{eq:d2sigNC}
  \nonumber
  \frac{\rd^2 \sigma^{\rm NC}_{\nu(\bar{\nu})N} (x,Q^2,E_{\nu})}{\rd
  x\,\rd Q^2} = &\frac{8 G_{F}^2}{\pi x}\left[\frac{ ( M_Z^2 \rho  - M_W^2) M_W^2}{M_Z^2 Q^2  \rho^2( 2 - \rho )}\right]^2  \\
                &\times  \Bigg(Y_+ F_{2,{\rm NC}}^{\nu N}(x,Q^2)   \mp Y_- x F_{3,{\rm NC}}^{\nu N}(x,Q^2) - y^2 F_{L,{\rm NC}}^{\nu N}(x,Q^2)   \Bigg) \,,
\end{align}
where $\rho = 1 +\Delta\bar{\rho}$, with $\Delta\bar{\rho}$ given in
Eq.~(8.22) of Ref.~\cite{Denner:1991kt}. Note that, in contrast with
the CC case, the structure functions in Eq.~(\ref{eq:d2sigNC}) for
neutrino and antineutrino scattering are the same. Therefore, the only
difference between neutrino and antineutrino cross-sections is the
sign of the term proportional to $xF_3$.

In addition to the free-nucleon cross-sections, we also provide the
values of the nuclear correction factor $R_{\nu O}/A$ as
defined in Eq.~(\ref{eq:REPPS16}).
This allows one to evaluate the central value and the associated
uncertainty to the total cross-sections including the effects of
nuclear modifications.
Presenting the results in this format has the advantage that
predictions for scattering off a molecule of H$_2$O can be easily
updated once improved predictions for $R_{\nu O}$ become available,
see Sect.~\ref{sec:nuclearPDFs}.
Since proton PDF uncertainties and nPDF uncertainties are
uncorrelated, they can be combined by adding them in quadrature.

The BGR18 CC and NC (anti)neutrino total cross-sections,
$\sigma^{\rm CC}_{\nu(\bar{\nu}) I}$ and
$\sigma^{\rm NC}_{\nu(\bar{\nu}) I}$, as functions of the energy
$E_{\nu}$ are tabulated in Tables~\ref{tab:APP_CCxsecs}
and~\ref{tab:APP_NCxsecs}, respectively.
Results for the scattering off an isoscalar target without nuclear
effects are shown along with the corresponding PDF uncertainty
$\delta \sigma^{\rm CC}_{\nu(\bar{\nu})I}$.
The values for the nuclear correction factor $R_{\nu(\bar{\nu}) O}/A$,
Eq.~(\ref{eq:REPPS16}), and the corresponding nPDF uncertainty
$\delta R^{\rm CC}_{ \bar{\nu} ( \bar{\nu} ) O}$ are also provided.

The {\tt LHAPDF} structure function grids and an example code that
computes the total cross-sections tabulated in
Tables~\ref{tab:APP_CCxsecs} and~\ref{tab:APP_NCxsecs} are available
from the following web page:
\begin{center}
\tt https://data.nnpdf.science/BGR18/
\end{center}
along with a short set on instructions.

\renewcommand*{\arraystretch}{1.5}
\begin{table}[t]
  \footnotesize
	\centering
	\begin{tabular}{ c | c c | c c | c c | c c }
         &  \multicolumn{8}{c}{\large Charged-current neutrino scattering} \\
          \toprule
          $E_{\nu(\bar{\nu})}$ (GeV) &  
          $\sigma^{\rm CC}_{\nu I}$ (pb)  &
          $\delta\sigma^{\rm CC}_{\nu I}$ (\%) &  
          $\frac{1}{A}R^{\rm CC}_{\nu O}$ & 
          $\delta R^{\rm CC}_{\nu O} (\%)$ & 
          $\sigma^{\rm CC}_{\bar{\nu} I}$ (pb)  &
          $\delta\sigma^{\rm CC}_{\bar{\nu} I}$ (\%) &  
          $\frac{1}{A}R^{\rm CC}_{\bar{\nu} O}$ & 
          $\delta R^{\rm CC}_{\bar{\nu} O}$ (\%) \\
          \midrule
$5\times10^{3}$ &$25.5$ &$\pm 2.2$ & $1$ & $^{+1.2}_{-1.2}$ & $15.2$ &$\pm 1.4$ & $0.99$ & $^{+2.2}_{-2.1}$ \\
$1\times10^{4}$ &$44.6$ &$\pm 2$ & $1$ & $^{+1.2}_{-1.2}$ & $28.5$ &$\pm 1.3$ & $0.99$ & $^{+2.0}_{-1.9}$ \\
$2\times10^{4}$ &$73.8$ &$\pm 1.8$ & $1$ & $^{+1.2}_{-1.2}$ & $51.2$ &$\pm 1.2$ & $0.99$ & $^{+1.9}_{-1.8}$ \\
$5\times10^{4}$ &$133$ &$\pm 1.5$ & $0.99$ & $^{+1.3}_{-1.2}$ & $103$ &$\pm 1.3$ & $0.99$ & $^{+1.8}_{-1.6}$ \\
$1\times10^{5}$ &$199$ &$\pm 1.4$ & $0.99$ & $^{+1.5}_{-1.3}$ & $165$ &$\pm 1.3$ & $0.99$ & $^{+1.9}_{-1.6}$ \\
$2\times10^{5}$ &$287$ &$\pm 1.4$ & $0.98$ & $^{+1.8}_{-1.6}$ & $252$ &$\pm 1.4$ & $0.98$ & $^{+2.1}_{-1.8}$ \\
$5\times10^{5}$ &$453$ &$\pm 1.5$ & $0.97$ & $^{+2.5}_{-2.1}$ & $421$ &$\pm 1.6$ & $0.97$ & $^{+2.7}_{-2.3}$ \\
$1\times10^{6}$ &$628$ &$\pm 1.6$ & $0.97$ & $^{+3.1}_{-2.7}$ & $600$ &$\pm 1.7$ & $0.96$ & $^{+3.2}_{-2.9}$ \\
$2\times10^{6}$ &$859$ &$\pm 1.8$ & $0.96$ & $^{+3.7}_{-3.4}$ & $837$ &$\pm 1.7$ & $0.96$ & $^{+3.8}_{-3.5}$ \\
$5\times10^{6}$ &$1.28\times10^{3}$ &$\pm 1.9$ & $0.95$ & $^{+4.6}_{-4.3}$ & $1.27\times10^{3}$ &$\pm 1.8$ & $0.95$ & $^{+4.7}_{-4.3}$ \\
$1\times10^{7}$ &$1.71\times10^{3}$ &$\pm 2$ & $0.94$ & $^{+5.3}_{-4.9}$ & $1.71\times10^{3}$ &$\pm 1.9$ & $0.94$ & $^{+5.3}_{-5.0}$ \\
$2\times10^{7}$ &$2.27\times10^{3}$ &$\pm 2.1$ & $0.94$ & $^{+5.9}_{-5.5}$ & $2.28\times10^{3}$ &$\pm 2$ & $0.94$ & $^{+5.9}_{-5.5}$ \\
$5\times10^{7}$ &$3.26\times10^{3}$ &$\pm 2.3$ & $0.93$ & $^{+6.5}_{-6.2}$ & $3.28\times10^{3}$ &$\pm 2.3$ & $0.93$ & $^{+6.6}_{-6.3}$ \\
$1\times10^{8}$ &$4.25\times10^{3}$ &$\pm 2.5$ & $0.93$ & $^{+7}_{-6.7}$ & $4.29\times10^{3}$ &$\pm 2.5$ & $0.93$ & $^{+7.0}_{-6.7}$ \\
$2\times10^{8}$ &$5.51\times10^{3}$ &$\pm 2.8$ & $0.92$ & $^{+7.4}_{-7.1}$ & $5.56\times10^{3}$ &$\pm 2.8$ & $0.92$ & $^{+7.4}_{-7.1}$ \\
$5\times10^{8}$ &$7.69\times10^{3}$ &$\pm 3.3$ & $0.92$ & $^{+7.9}_{-7.6}$ & $7.76\times10^{3}$ &$\pm 3.3$ & $0.92$ & $^{+7.9}_{-7.6}$ \\
$1\times10^{9}$ &$9.82\times10^{3}$ &$\pm 3.7$ & $0.92$ & $^{+8.2}_{-8}$ & $9.93\times10^{3}$ &$\pm 3.7$ & $0.92$ & $^{+8.2}_{-8.0}$ \\
$2\times10^{9}$ &$1.25\times10^{4}$ &$\pm 4.1$ & $0.91$ & $^{+8.5}_{-8.3}$ & $1.26\times10^{4}$ &$\pm 4.1$ & $0.91$ & $^{+8.5}_{-8.3}$ \\
$5\times10^{9}$ &$1.7\times10^{4}$ &$\pm 4.7$ & $0.91$ & $^{+8.8}_{-8.6}$ & $1.72\times10^{4}$ &$\pm 4.7$ & $0.91$ & $^{+8.8}_{-8.6}$ \\
$1\times10^{10}$ &$2.14\times10^{4}$ &$\pm 5.1$ & $0.91$ & $^{+9}_{-8.8}$ & $2.17\times10^{4}$ &$\pm 5.1$ & $0.91$ & $^{+9.0}_{-8.8}$ \\
$2\times10^{10}$ &$2.69\times10^{4}$ &$\pm 5.6$ & $0.91$ & $^{+9.2}_{-9}$ & $2.72\times10^{4}$ &$\pm 5.6$ & $0.91$ & $^{+9.2}_{-9.0}$ \\
$5\times10^{10}$ &$3.6\times10^{4}$ &$\pm 6.1$ & $0.9$ & $^{+9.4}_{-9.3}$ & $3.64\times10^{4}$ &$\pm 6.1$ & $0.9$ & $^{+9.4}_{-9.3}$ \\
$1\times10^{11}$ &$4.47\times10^{4}$ &$\pm 6.6$ & $0.9$ & $^{+9.5}_{-9.4}$ & $4.52\times10^{4}$ &$\pm 6.6$ & $0.9$ & $^{+9.5}_{-9.4}$ \\
$2\times10^{11}$ &$5.54\times10^{4}$ &$\pm 7$ & $0.9$ & $^{+9.6}_{-9.5}$ & $5.61\times10^{4}$ &$\pm 7$ & $0.9$ & $^{+9.6}_{-9.5}$ \\
$5\times10^{11}$ &$7.32\times10^{4}$ &$\pm 7.6$ & $0.9$ & $^{+9.7}_{-9.6}$ & $7.41\times10^{4}$ &$\pm 7.6$ & $0.9$ & $^{+9.7}_{-9.6}$ \\
$1\times10^{12}$ &$9\times10^{4}$ &$\pm 8$ & $0.9$ & $^{+9.8}_{-9.7}$ & $9.12\times10^{4}$ &$\pm 8$ & $0.9$ & $^{+9.8}_{-9.7}$ \\
$2\times10^{12}$ &$1.1\times10^{5}$ &$\pm 8.4$ & $0.9$ & $^{+9.8}_{-9.7}$ & $1.12\times10^{5}$ &$\pm 8.4$ & $0.9$ & $^{+9.8}_{-9.7}$ \\
$5\times10^{12}$ &$1.44\times10^{5}$ &$\pm 8.9$ & $0.9$ & $^{+9.9}_{-9.8}$ & $1.45\times10^{5}$ &$\pm 8.9$ & $0.9$ & $^{+9.9}_{-9.8}$ \\
\bottomrule
\end{tabular}
\vspace{0.4cm}
\caption{\small The BGR18 charged-current neutrino total cross-sections $\sigma^{\rm CC}_{\nu(\nu) I}$
 as a function of the energy $E_{\nu}$.
We show the results for the scattering of a (anti)neutrino
off a free isoscalar target together with the percentage
proton PDF uncertainties, $\delta \sigma^{\rm CC}_{\nu(\bar{\nu})I}$.
We also list the values of
the nuclear correction factor $R_{\nu(\bar{\nu}) O}/A$, Eq.~(\ref{eq:REPPS16}),
computed with the EPPS16 set and the percentage
nPDF uncertainty, $\delta R^{\rm CC}_{ \bar{\nu} ( \bar{\nu} ) O}$.\label{tab:APP_CCxsecs}
}
\end{table}

\renewcommand*{\arraystretch}{1.7}
\begin{table}[t]
  \footnotesize
	\centering
	\begin{tabular}{ c | c c | c c | c c | c c }
          &  \multicolumn{8}{c}{\large Neutral-current neutrino scattering} \\
          \toprule
          $E_{\nu(\bar{\nu})}$ (GeV) &  
          $\sigma^{\rm NC}_{\nu I}$ (pb)  &
          $\delta\sigma^{\rm NC}_{\nu I}$ (\%) &  
          $\frac{1}{A}R^{NC}_{\nu O}$ & 
          $\delta R^{\rm NC}_{\nu O} (\%)$ & 
          $\sigma^{\rm NC}_{\bar{\nu} I}$ (pb)  &
          $\delta\sigma^{\rm NC}_{\bar{\nu} I}$ (\%) &  
          $\frac{1}{A}R^{\rm NC}_{\bar{\nu} O}$ & 
          $\delta R^{\rm NC}_{\bar{\nu} O}$ (\%) \\
          \midrule
$5\times10^{3}$ &$8.45$ &$\pm 1.7$ & $1$ & $^{+0.83}_{-0.84}$ & $5.52$ &$\pm 0.96$ & $0.99$ & $^{+1.4}_{-1.3}$ \\
$1\times10^{4}$ &$15.1$ &$\pm 1.5$ & $1$ & $^{+0.81}_{-0.82}$ & $10.4$ &$\pm 0.91$ & $0.99$ & $^{+1.3}_{-1.3}$ \\
$2\times10^{4}$ &$25.8$ &$\pm 1.4$ & $1$ & $^{+0.81}_{-0.79}$ & $19$ &$\pm 0.89$ & $0.99$ & $^{+1.2}_{-1.2}$ \\
$5\times10^{4}$ &$48.6$ &$\pm 1.1$ & $1$ & $^{+0.88}_{-0.79}$ & $38.9$ &$\pm 0.9$ & $0.99$ & $^{+1.2}_{-1.1}$ \\
$1\times10^{5}$ &$74.6$ &$\pm 1$ & $0.99$ & $^{+1}_{-0.89}$ & $63.5$ &$\pm 0.94$ & $0.99$ & $^{+1.3}_{-1.1}$ \\
$2\times10^{5}$ &$111$ &$\pm 0.96$ & $0.99$ & $^{+1.3}_{-1.1}$ & $99.5$ &$\pm 0.99$ & $0.99$ & $^{+1.5}_{-1.3}$ \\
$5\times10^{5}$ &$182$ &$\pm 1$ & $0.98$ & $^{+2}_{-1.7}$ & $170$ &$\pm 1.1$ & $0.98$ & $^{+2.1}_{-1.9}$ \\
$1\times10^{6}$ &$258$ &$\pm 1.1$ & $0.97$ & $^{+2.6}_{-2.3}$ & $248$ &$\pm 1.2$ & $0.97$ & $^{+2.7}_{-2.4}$ \\
$2\times10^{6}$ &$361$ &$\pm 1.2$ & $0.97$ & $^{+3.3}_{-3}$ & $352$ &$\pm 1.2$ & $0.96$ & $^{+3.4}_{-3.1}$ \\
$5\times10^{6}$ &$552$ &$\pm 1.3$ & $0.96$ & $^{+4.2}_{-3.9}$ & $545$ &$\pm 1.3$ & $0.96$ & $^{+4.3}_{-4.0}$ \\
$1\times10^{7}$ &$751$ &$\pm 1.4$ & $0.95$ & $^{+4.9}_{-4.6}$ & $746$ &$\pm 1.4$ & $0.95$ & $^{+4.9}_{-4.6}$ \\
$2\times10^{7}$ &$1.01\times10^{3}$ &$\pm 1.6$ & $0.94$ & $^{+5.5}_{-5.2}$ & $1.01\times10^{3}$ &$\pm 1.6$ & $0.94$ & $^{+5.5}_{-5.2}$ \\
$5\times10^{7}$ &$1.48\times10^{3}$ &$\pm 1.8$ & $0.94$ & $^{+6.2}_{-6}$ & $1.47\times10^{3}$ &$\pm 1.8$ & $0.94$ & $^{+6.3}_{-6.0}$ \\
$1\times10^{8}$ &$1.95\times10^{3}$ &$\pm 2$ & $0.93$ & $^{+6.7}_{-6.5}$ & $1.95\times10^{3}$ &$\pm 2$ & $0.93$ & $^{+6.7}_{-6.5}$ \\
$2\times10^{8}$ &$2.55\times10^{3}$ &$\pm 2.3$ & $0.93$ & $^{+7.2}_{-6.9}$ & $2.55\times10^{3}$ &$\pm 2.3$ & $0.93$ & $^{+7.2}_{-6.9}$ \\
$5\times10^{8}$ &$3.6\times10^{3}$ &$\pm 2.8$ & $0.92$ & $^{+7.6}_{-7.4}$ & $3.6\times10^{3}$ &$\pm 2.8$ & $0.92$ & $^{+7.7}_{-7.4}$ \\
$1\times10^{9}$ &$4.63\times10^{3}$ &$\pm 3.2$ & $0.92$ & $^{+8}_{-7.8}$ & $4.63\times10^{3}$ &$\pm 3.2$ & $0.92$ & $^{+8.0}_{-7.8}$ \\
$2\times10^{9}$ &$5.93\times10^{3}$ &$\pm 3.7$ & $0.92$ & $^{+8.3}_{-8.1}$ & $5.93\times10^{3}$ &$\pm 3.7$ & $0.92$ & $^{+8.3}_{-8.1}$ \\
$5\times10^{9}$ &$8.15\times10^{3}$ &$\pm 4.3$ & $0.91$ & $^{+8.6}_{-8.5}$ & $8.15\times10^{3}$ &$\pm 4.3$ & $0.91$ & $^{+8.6}_{-8.5}$ \\
$1\times10^{10}$ &$1.03\times10^{4}$ &$\pm 4.7$ & $0.91$ & $^{+8.9}_{-8.7}$ & $1.03\times10^{4}$ &$\pm 4.7$ & $0.91$ & $^{+8.9}_{-8.7}$ \\
$2\times10^{10}$ &$1.3\times10^{4}$ &$\pm 5.2$ & $0.91$ & $^{+9.1}_{-8.9}$ & $1.3\times10^{4}$ &$\pm 5.2$ & $0.91$ & $^{+9.1}_{-8.9}$ \\
$5\times10^{10}$ &$1.75\times10^{4}$ &$\pm 5.7$ & $0.91$ & $^{+9.3}_{-9.2}$ & $1.75\times10^{4}$ &$\pm 5.7$ & $0.91$ & $^{+9.3}_{-9.2}$ \\
$1\times10^{11}$ &$2.18\times10^{4}$ &$\pm 6.2$ & $0.91$ & $^{+9.4}_{-9.3}$ & $2.18\times10^{4}$ &$\pm 6.2$ & $0.91$ & $^{+9.4}_{-9.3}$ \\
$2\times10^{11}$ &$2.71\times10^{4}$ &$\pm 6.6$ & $0.9$ & $^{+9.6}_{-9.5}$ & $2.71\times10^{4}$ &$\pm 6.6$ & $0.9$ & $^{+9.6}_{-9.5}$ \\
$5\times10^{11}$ &$3.6\times10^{4}$ &$\pm 7.2$ & $0.9$ & $^{+9.7}_{-9.6}$ & $3.6\times10^{4}$ &$\pm 7.2$ & $0.9$ & $^{+9.7}_{-9.6}$ \\
$1\times10^{12}$ &$4.44\times10^{4}$ &$\pm 7.6$ & $0.9$ & $^{+9.7}_{-9.6}$ & $4.44\times10^{4}$ &$\pm 7.6$ & $0.9$ & $^{+9.7}_{-9.6}$ \\
$2\times10^{12}$ &$5.46\times10^{4}$ &$\pm 8.1$ & $0.9$ & $^{+9.8}_{-9.7}$ & $5.46\times10^{4}$ &$\pm 8.1$ & $0.9$ & $^{+9.8}_{-9.7}$ \\
$5\times10^{12}$ &$7.14\times10^{4}$ &$\pm 8.6$ & $0.9$ & $^{+9.8}_{-9.7}$ & $7.14\times10^{4}$ &$\pm 8.6$ & $0.9$ & $^{+9.8}_{-9.7}$ \\
     \bottomrule
        \end{tabular}
        \vspace{0.4cm}
        \caption{\small Same as Table~\ref{tab:APP_CCxsecs} for the neutral-current
        scattering cross-sections. \label{tab:APP_NCxsecs} }
\end{table}

\clearpage

\FloatBarrier

\phantomsection
\addcontentsline{toc}{section}{References}
\bibliography{UHEsig}

\end{document}